\documentclass[a4paper,11pt]{article}
\usepackage{a4wide}
\usepackage{amsmath,amssymb}
\usepackage{graphics,graphicx,hyperref}
\usepackage{changebar}
\usepackage{color}
\usepackage{cancel}
\usepackage[all,cmtip]{xy}

\def\ra{\rangle}
\def\la{\langle}

\def\bY{\mathbf{Y}}

\def\cX{\mathcal{X}}
\def\cL{\mathcal{L}}
\def\bQ{\mathbf{Q}}

\def\bK{\mathbf{K}}
\def\bZ{\mathbf{Z}}
\def\cM{\mathcal{M}}
\def\cH{\mathcal{H}}

\def\cC{\mathcal{C}}
\def\cG{\mathcal{G}}
\def\cF{\mathcal{F}}
\def\cB{\mathcal{B}}
\def\bQb{\overline{\bQ}}
\def\Kb{\overline{K}}
\def\cQ{\mathcal{Q}}
\def\cD{\mathcal{D}}
\def\cO{\mathcal{O}}
\def\cA{\mathcal{A}}
\def\cE{\mathcal{E}}
\def\Wb{\overline{W}}
\def\Fb{\overline{F}}
\def\Ab{\overline{A}}
\def\Bb{\overline{B}}
\def\cQb{\overline{\cQ}}
\def\cDb{\overline{\cD}}
\def\cOb{\overline{\cO}}

\def\zb{\overline{z}}
\def\xb{\overline{x}}
\def\ub{\overline{u}}

\def\qb{\overline{q}}
\def\phib{\overline{\phi}}
\def\psib{\overline{\psi}}

\def\cb{\overline{c}}
\def\Xb{\overline{X}}
\def\Jb{\overline{J}}
\def\Tb{\overline{T}}
\def\Omegab{\overline{\Omega}}
\def\etab{\overline{\eta}}
\def\chib{\overline{\chi}}
\def\betab{\overline{\beta}}

\def\cXb{\overline{\cX}}
\def\Psib{\overline{\Psi}}

\def\gammab{\overline{\gamma}}
\def\deltab{\overline{\delta}}
\def\alphab{\overline{\alpha}}
\def\thetab{\overline{\theta}}
\def\p{\partial}
\def\pb{\overline{\partial}}
\def\Db{\overline{D}}
\def\rhob{\overline{\rho}}
\def\ep{\epsilon}
\def\epb{\overline{\ep}}
\def\cEb{\overline{\cE}}
\def\GU{\text{U}}
\def\bq{\mathbf{q}}

\def\bv{\mathbf{v}}

\def\hb{\overline{h}}

\def\ft{\tilde{f}}

\def\Sigmat{\tilde{\Sigma}}
\def\Wt{\widetilde{W}}
\def\Ht{\widetilde{H}}
\def\Lt{\widetilde{L}}
\def\Jt{\widetilde{J}}
\def\Ft{\widetilde{F}}

\def\rank{{{\text{rank}}}}
\def\ch{{{\text{ch}}}}
\def\tot{{{\text{tot}}}}
\def\Hess{{{\text{Hess}}}}
\def\Hom{{{\text{Hom}}}}

\def\half{{{\frac12}}}

\def\CP{{{\mathbb{P}}}}
\def\C{{{\mathbb{C}}}}
\def\Z{{{\mathbb{Z}}}}

\numberwithin{equation}{section}

\begin{document}
\thispagestyle{empty}
\begin{flushright}
IPMU17-0188
\end{flushright}
\vspace{1cm}
\begin{center}
{\LARGE\bf Aspects of $(2,2)$ and $(0,2)$ hybrid models}
\end{center}
\vspace{8mm}
\begin{center}
{\large Marco Bertolini\footnote{{\tt marco.bertolini@ipmu.jp}}${}^{\dagger}$ and Mauricio Romo\footnote{{\tt mromoj@ias.edu}}${}^{\ddagger}$}
\end{center}
\vspace{6mm}
\begin{center}
${}^{\dagger}$ Kavli Institute for the Physics and Mathematics of the Universe (WPI),\\
The University of Tokyo, Kashiwa, Chiba 277-8583, Japan\\\vspace{2mm}
${}^{\ddagger}$ School of Natural Sciences, Institute for Advanced Study,\\ Princeton, NJ 08540, USA
\end{center}
\vspace{15mm}

\begin{abstract}
\noindent
In this work we study the topological rings of two dimensional (2,2) and (0,2) hybrid models.
In particular, we use localization to derive a formula for the correlators in both cases, focusing on the B- and $\frac{\mathrm{B}}{2}$-twists.
Although our methods apply to a vast range of hybrid CFTs, we focus on
hybrid models suitable for compactifications of the heterotic string. In this case, our formula provides
unnormalized Yukawa couplings of the spacetime superpotential.
We apply our techniques to hybrid phases of linear models, and we
find complete agreement with known results in other phases.
We also obtain a prediction for a certain class of correlators involving twisted operators in (2,2) Landau-Ginzburg orbifolds.
For (0,2) theories, our argument does not rely on the existence of a (2,2) locus.
Finally, we derive vanishing conditions concerning worldsheet instanton corrections
in (0,2) $\frac{\mathrm{B}}{2}$-twisted hybrid models.
\end{abstract}
\newpage
\setcounter{tocdepth}{2}
\tableofcontents
\setcounter{footnote}{0}

\section{Introduction}

Most of what is currently known about the structure of the moduli space of (2,2) and, more in general, (0,2) superconformal field theories (SCFTs)
is due to our ability to extract relevant structures at special loci/limits.
The most prominent example of this is probably the Calabi-Yau/Landau-Ginzburg (CY/LG) correspondence \cite{Greene:1988ut,Vafa:1988uu}.
For instance, the overwhelming evidence that various geometrical data of Calabi-Yau (CY) manifolds
is encoded in orbifolds of Landau-Ginzburg (LG) models has inspired the construction of the
gauged linear sigma model (GLSM) \cite{Witten:1993yc}. In turn, the GLSM has been one
of the main tools in exploring various aspects of the moduli space.
Another example is mirror symmetry, which in its early days
has been proved as a correspondence of orbifolds of Gepner models \cite{Greene:1990ud},
and later generalized to a vast collection of theories.

Despite these remarkable successes, the picture is far from complete. Even for (2,2) SCFTs,
generically the GLSM comprises only a subspace of the moduli space, and even in the realm of GLSMs,
a generic phase will not be described by a non-linear sigma model (NLSM) or a LG orbifold (LGO).
Moreover, (2,2) theories admit deformations which preserve only (0,2) superconformal symmetry.
This class of deformations is much less understood.
Although recently there has been progress in analyzing these models \cite{McOrist:2010ae}, questions such as, for example,
how to extend mirror symmetry to this larger moduli space of (0,2) deformations, have not yet found a complete answer \cite{Melnikov:2010sa,Melnikov:2012hk}.

Even more mysterious are (0,2) theories which do not admit a (2,2) locus.
For instance, the issue of whether instanton contributions to the spacetime superpotential
destabilize the vacuum of a heterotic compactification based on a (0,2) SCFT, which were believed
to be absent in the contest of (0,2) GLSMs \cite{Silverstein:1995re,Beasley:2003fx}, is still
unresolved even in this subset of theories \cite{Bertolini:2014dna}.

An important feature of SCFTs is the existence of chiral rings. In (2,2) theories there are two such rings, the
(a,c)/(c,c) rings (and their equivalent conjugates)\footnote{In this paper, we will often refer to the (a,c) ring as the A ring and to the (c,c) ring as the B ring.} which
are topological in nature and are independent of a set of parameters of the theory. In particular, they are exchanged under mirror symmetry.
In (0,2) theories, where we lack the definition of a left-moving chirality condition,
in favorable cases it is possible to show the existence of subrings of the (right-moving) chiral ring, which in particular generalize the A/B rings,
and reduce to these on the (2,2) locus (when available).
We refer to these, following the nomenclature in literature, as A/2 and B/2 rings.
Properties of such rings are still largely unknown.
For instance, do B/2 model correlators receive instanton corrections? If yes, is there a class of theories where they do not?
If yes, this class of theories would constitute a natural playground for improving our understanding of (0,2) mirror symmetry.

It should be evident at this point that one approach to tackle some of these issues is to deepen
our understanding of more general theories. This is the path we choose to follow in this work,
as we study the ring structure of B and B/2 rings in hybrid theories.
Loosely speaking, a hybrid model is a class of NLSMs with superpotential, where the theory can
be interpreted as a Landau-Ginzburg (orbifold) whose superpotential varies adiabatically over a compact and smooth manifold.
Such theories depend on a number of
parameters, namely the K\"ahler class and the complex structure of the target space,
a set of parameters defining a holomorphic bundle over it, and a set of parameters defining the superpotential.
While their existence has been known for over two decades \cite{Witten:1993yc,Aspinwall:1993nu}, and
both the mathematics \cite{Fan:2015ab} and physics \cite{Aspinwall:2009qy} literature exhibit instances of hybrid models,
it was only recently that a systematic study of their physical properties has begun \cite{Bertolini:2013xga,Bertolini:2017lcz}.
Besides \cite{Fan:2015ab}, there have also been other recent developments in mathematics
that study hybrid models closely such as \cite{2013arXiv1301.5530C,2015arXiv150602989C}.

An important property of (quasi-)topological theories is that
often they depend only on a subset of the parameters of the original theory.
In (2,2) hybrid theories, B-model correlators depend only on the parameters in the superpotential.
When a hybrid model describes a limit in the moduli space of a Calabi-Yau compactification,
these parameters determine a choice of complex structure on the Calabi-Yau.
In (0,2) models instead, this dependence is naturally enlarged to the set of parameters determining the
(0,2) superpotential. In a large radius interpretation,
this set of parameters determine a holomorphic bundle over the CY manifold.
Moreover, there could be in principle a dependence on worldsheet instantons.
As mentioned above, models for which this dependence can be ruled out are
particularly interesting.

When a hybrid model arises as a phase in a linear model, we can compare our results with the ones derived from GLSM techniques. In particular,
in such cases we can interpret the hybrid model as a fixed point along the RG flow between the GLSM (for a particular choice of the parameters) and the IR CFT.
The structure we find is quite interesting. First of all, we find that in some cases the hybrid model seems to allow instanton corrections
while the corresponding GLSM forbids them. We find that it is only possible to recover the vanishing result from the hybrid perspective by
constructing an appropriate compactification of the moduli space of worldsheet instantons along the lines of \cite{Sharpe:2006qd}.
We also find examples where the hybrid model forbids instanton corrections but the linear model does not rule them out.

Another interesting observation concerns the dependence of the correlators on the various parameters of the theory.
In a large class of GLSMs, B model correlators are independent of $E$ parameters -- which instead appear as parameters in A/2 model correlators --
and instead depend exclusively on $J$ parameters \cite{McOrist:2008ji}. This separation seems to disappear in the hybrid model, where $E$ parameters descend to
``bundle" parameters and $J$ parameters descend to superpotential parameters. While it is somehow expected that this splitting does not occur in the IR CFT,
it is interesting that it is not manifest in the UV hybrid theory as well.

The rest of this work is organized as follows. In section \ref{s:22hybrids} we review the construction of (2,2) hybrids.
In section \ref{s:chiralring} we describe the main object of interest, the B ring, and review the techniques
that allow to compute its elements.
In section \ref{s:22corrs}, we introduce the B-twisted version of the hybrid theory, and use localization
to derive a formula for the correlators and study its properties. In section \ref{s:22example}
we completely solve an example, and we check our results via the GLSM.
Finally, in section \ref{s:02hybrids} we derive an analogous formula
for correlators of B/2-twisted (0,2) hybrid models. We analyze the conditions
for instanton corrections to vanish, and apply our techniques to an example.
We conclude with some open questions in section \ref{s:outlook}.

\section{$\mathcal{N}=(2,2)$ hybrid models}
\label{s:22hybrids}

In this section we review the construction of \cite{Bertolini:2013xga}
of (2,2) hybrid models in order to set-up notation and to highlight the aspects relevant for the present work.
The results are valid for an arbitrary compact Riemann surface $\Sigma$.
We will work locally on an open patch $U\cong \mathbb{C}$ of $\Sigma$ with coordinates $(z,\bar{z})$.

A hybrid model is determined by the data $(\bY,W,\mathbb{C}^{*}_V)$, where $\bY$ is a K\"ahler manifold, $W\in H^{0}(\mathcal{O}_\bY)$,
i.e., $W$ is a holomorphic function on $\bY$, and $\mathbb{C}^{*}_V$ is a $\mathbb{C}^{*}$-action on $\bY$ defined by a Killing vector field $V$ on $\bY$
such that $\mathcal{L}_{V}W=W$. We will restrict ourselves to the case of $\bY$ being CY, i.e., its canonical class being trivial, $K_{\bY}\cong \cO_{\bY}$.
However, additional conditions on this data are necessary in order to be able to compute meaningful quantities.
We call the triple $(\bY,W,\mathbb{C}^{*}_{V})$ a good hybrid if $\bY$ has bundle structure with
a compact base\footnote{In all our applications we will restrict to $B$ being K\"ahler and smooth.}
$B$ and $V$ is a vertical Killing vector of $\bY$. We assume $\bY=\tot\left(X\rightarrow B\right)$ and $X=\oplus_i X_i$
is a decomposition into eigenspaces with respect to the $V$-action, that is,
\begin{align}
V(B)&=0~,		&V(X_i)=q_i X_i~,
\end{align}
where $q_i\in\mathbb{Q}_{>0}$, $i=1,\dots,n$. Although not a necessary condition to construct a well-defined model,
we will make the simplifying assumption that $X_i$ are line bundles over $B$.
Finally, the superpotential $W$ is chosen such that it satisfies the \textit{potential condition}
\begin{eqnarray}
dW^{-1}(0)=B\subset \bY~.
\end{eqnarray}
This is the class of models to which our methods apply. In the rest of this section we will construct the action
for the corresponding NLSM and study its symmetries.

\subsection{Action}

We work in $(2,2)$ superspace in Euclidean signature parametrized by $(z,\zb)$ and $(\theta^{\pm},\thetab^\pm)$, where the + ($-$) corresponds to the
right-moving (left-moving) sector. In this setting we define the supercharges
\begin{align}
\label{eq:rightsupercharges}
\cQ_+ &=  {\p \over \p \theta^+} + i\thetab^+ \pb_{\bar{z}}~,		&\cQb_+ &=  -{\p \over \p \thetab^+} - i\theta^+ \pb_{\bar{z}}~.
\end{align}
These anti-commute and satisfy the algebra $\{\cQ_+,\cQb_+\}=-2i\pb_{\zb}$, where we denote $\pb_{\bar{z}}:= \p/\p\zb$. We also define the
superderivatives
\begin{align}
\cD_+ &=  {\p \over \p \theta^+} - i\thetab^+ \pb_{\zb}~,		&\cDb_+ &=  -{\p \over \p \thetab^+} + i\theta^+ \pb_{\zb}~,
\end{align}
which anti-commute among each other and with the operators \eqref{eq:rightsupercharges}, and satisfy the algebra $\{\cD_+,\cDb_+\}=2i\pb_{\zb}$. There
is an equivalent structure on the left-moving sector of the theory, (anti-)commuting with the operators above
and which is obtained by replacing $\pb_{\zb}$ with $-\p_{z}:=-\p/\p_z$ and $(\theta^+,\thetab^+)$ with $(\theta^-,\thetab^-)$.

The field content of the theory is given by $d$ chiral (2,2) supermultiplets and their anti-chiral conjugates
\begin{align}
X^\alpha &= \cX^\alpha + \sqrt{2} \theta^- \Psi^\alpha + i\theta^-\thetab^- \p_{z} \cX^\alpha~, &
\Xb^{\alphab} & = \cXb^{\alphab} - \sqrt{2}\thetab^- \Psib^{\alphab} -i\theta^-\thetab^- \p_{z} \cXb^{\alphab}~,
\end{align}
where $\alpha = 1,\dots,d$, and where we define
\begin{align}
d&=\dim \bY~,		&b&=\dim B~,		&n&=\rank\ X~.
\end{align}
These are decomposed in terms of (0,2) bosonic and fermionic chiral supermultiplets $\cX^\alpha$ and $\Psi^\alpha$ respectively,
which have the following expansions
\begin{align}
\label{eq:02sfields22}
\cX^\alpha & = x^\alpha + \sqrt{2}\theta^+ \psi_+^\alpha -i \theta^+\thetab^+ \pb_{\bar{z}} x^{\alpha}~,&
\cXb^{\alphab} & = {\xb}^{\alphab} - \sqrt{2}\thetab^+ \psib_+^{\alphab} +i\theta^+\thetab^+ \pb_{\bar{z}} \xb^{\alphab}~, \nonumber\\
\Psi^\alpha & = \psi_-^\alpha - \sqrt{2}\theta^+ F^\alpha -i\theta^+\thetab^+ \pb_{\bar{z}} \psi_-^{\alpha}~,&
\Psib^{\alphab} & = \psib_-^{\alphab} - \sqrt{2}\thetab^+ \Fb^{\alphab} +i\theta^+\thetab^+ \pb_{\bar{z}} \psib_-^{\alphab}~.
\end{align}
The chirality conditions read
\begin{align}
\label{eq:chiralconds}
\cDb_\pm X^\alpha &=0~,		&\cDb_+ \cX^\alpha &=\cDb_+ \Psi^\alpha =0~,
\end{align}
and similarly for the conjugate anti-chiral fields. The lowest component $x^\alpha$ are coordinates on $\bY$, i.e., maps
\begin{align}
x^\alpha: \Sigma\rightarrow \bY~,
\end{align}
while $\psi_+^\alpha$ ($\psi_-^{\alpha}$) are
right-moving (left-moving) fermions on the worldsheet, i.e., $C^{\infty}$ sections of the tangent sheaf $T_{\bY}:=T^{(1,0)}\bY$, or more precisely
\begin{align}
\psi_-^\alpha &\in\Gamma(K_{\Sigma}^{\frac{1}{2}}\otimes x^{*}T_{\bY})~, &\psi_+^\alpha &\in\Gamma(K_{\Sigma}^{-\frac{1}{2}}\otimes x^{*}T_{\bY})~,
\end{align}
where $K_\Sigma$ is the anti-canonical bundle of the worldsheet $\Sigma$.
The action on $\Sigma$ in (2,2) superspace is given by\footnote{Here $m$ is a parameter with a dimension of mass and we will set it to one in the rest of the paper.}
\begin{align}
S[X]:=\int_{\Sigma}d^2z (\mathcal{L}_{K}+\mathcal{L}_{W})=\frac1{8\pi}\int_\Sigma d^2z \cD_+\cDb_+\cD_-\cDb_- K(X,\Xb) + \frac{m}{4\pi} \int_\Sigma d^2z \cD_+\cD_- W(X) + \text{c.c.}~,
\end{align}
where $K$ is a K\"ahler potential on $\bY$ with K\"ahler metric $g_{\alpha\betab}:=\p_\alpha \pb_{\betab}K$.
We can expand the action in components and we obtain
\begin{align}
\label{eq:compaction}
\mathcal{L}_{K}&=-g_{\alpha\betab}\left(\partial_{\mu} x^{\alpha}\partial^{\mu} \xb^{\betab}+\frac{i}{2}\langle{\psib}^{\betab},\gamma^{\mu}D_{\mu}\psi^\alpha\rangle
+\frac{i}{2}\langle\psi^\alpha,\gamma^{\mu}D_{\mu}{\psib}^{\betab}\rangle\right)\nonumber\\
&\quad-\frac{1}{4}R_{\alpha\betab\delta\gammab}\langle\psi^\alpha,\psi^\delta\rangle\langle\psib^{\betab},\psib^{\gammab}\rangle+
g_{\alpha\alphab}\left(F^\alpha-\frac{1}{2}\Gamma^\alpha_{\beta\delta}\langle\psi^\beta,\psi^\delta\rangle\right)\left(\overline{F}^{\alphab}+\frac{1}{2}\overline{\Gamma}^{\alphab}_{\betab\deltab}\langle\psib^{\betab},\psib^{\deltab}\rangle\right)~,\nonumber\\
\mathcal{L}_{W}&=\frac{1}{2}\left(F^\alpha\partial_\alpha W-\frac{1}{2}\partial_\alpha \partial_\beta W\langle\psi^\alpha,\psi^\beta\rangle+\overline{F}^\alpha\overline{\partial}_{\alphab}\overline{W}+\frac{1}{2}\overline{\partial}_{\alphab}\overline{\partial}_{\betab}\overline{W}\langle\bar{\psi}^{\alphab},\bar{\psi}^{\betab}\rangle\right)~.
\end{align}
The covariant derivatives act on the fermions as
\begin{eqnarray}
D_{\mu}\psi^{\alpha}=(\partial_{\mu}+\frac{1}{2}\omega_{\mu})\psi^{\alpha}+\partial_{\mu}x^{\beta}\Gamma^{\alpha}_{\beta\delta}\psi^{\delta}~,
\end{eqnarray}
where $\omega_{\mu}$ is the spin connection on $\Sigma$ and $\langle,\rangle$ denotes the frame invariant product (see appendix \ref{app:convents}).
The K\"ahler connection and the curvature are given by
\begin{align}
 \Gamma^\alpha_{\beta\gamma} &= g_{\gamma\betab,\beta} g^{\betab\alpha}~,		&R_{\alpha\betab\gamma}{}^\delta&= \Gamma_{\alpha\gamma,\betab}^\delta~.
\end{align}
Finally, it is possible to integrate out the auxiliary fields via the equations of motion
\begin{align}
F^{\alpha}&=-\frac{1}{2}g^{\alpha\betab}\partial_{\betab}\overline{W}-\Gamma_{\delta\gamma}^{\alpha}\psi^{\delta}_{+}\psi^{\gamma}_{-}~,
&\overline{F}^{\alphab}&=-\frac{1}{2}g^{\alphab\beta}\partial_{\beta}W-\overline{\Gamma}_{\deltab\gammab}^{\alphab}\bar{\psi}^{\deltab}_{+}\bar{\psi}^{\gammab}_{-}~.
\end{align}

\subsection{Supersymmetry transformations}

The action we defined in the previous section is by construction invariant under (2,2) supersymmetry.
For our study of the (c,c) ring in the following sections, we need the explicit supersymmetry transformations
induced by the supercharges $\cQb_+$ and $\cQb_-$.
For completeness, we present the supersymmetry transformations of the component fields for all the supercharges.

Let us define the operator $\bQ_\pm$ and $\bQb_{\pm}$ such that, acting on a superfield $A$, $[\ep_\mp \bQ_\pm , A]=\mp\frac1{\sqrt2} \ep_\mp \cQ_\pm A$, where
$\ep_\mp$ are anticommuting parameters, and similarly for the barred quantities.\footnote{The pairing $\ep_\mp \leftrightarrow \cQ_{\pm}$ 
corresponds to the frame invariant product defined in appendix \ref{app:convents}.}
With these conventions for the charges \eqref{eq:rightsupercharges} we find
\begin{align}
[\bQ_+,x^\alpha] &= -\psi_+^\alpha~,	&\{\bQ_+,\psib_+^{\alphab}\} &= -i\pb_{\bar{z}}\xb^{\alphab}~,	
&\{\bQ_+,\psi_-^\alpha\}&=F^\alpha~,	&[\bQ_+,\Fb^{\alphab}]&=-i\pb_{\bar{z}} \psib_-^{\alphab}~,	\nonumber\\
[\bQb_+,\xb^{\alphab}] &=\psib_+^{\alphab}~,		&\{\bQb_+,\psi_+^\alpha\} &= i\pb_{\bar{z}} x^\alpha~,
&\{\bQb_+,\psib_-^{\alphab}\}&=\Fb^{\alphab}~,	&[\bQb_+,F^{\alpha}]&=-i\pb_{\bar{z}} \psi_-^{\alpha}~,
\end{align}
as well as
\begin{align}
[\bQ_-,x^\alpha] &= \psi_-^\alpha~,	&\{\bQ_-,\psib_-^{\alphab}\} &= -i\p_{z}\xb^{\alphab}~,	
&\{\bQ_-,\psi_+^\alpha\}&=F^\alpha~,	&[\bQ_-,\Fb^{\alphab}]&= i\p_{z} \psib_+^{\alphab}~,	\nonumber\\
[\bQb_-,\xb^{\alphab}] &=-  \psib_-^{\alphab}~,		&\{\bQb_-,\psi_-^\alpha\} &=i \p_{z} x^\alpha~,
&\{\bQb_-,\psib_+^{\alphab}\}&=\Fb^{\alphab}~,	&[\bQb_-,F^{\alpha}]&=i\p_{z} \psi_+^{\alpha}~.
\end{align}
The $\mathcal{N}=(2,2)$ transformations can be written in a more covariant fashion as follows
\begin{align}
\label{22transfcov}
\delta x^\alpha&=\langle\epsilon,\psi^\alpha\rangle~,		 &\delta \xb^{\alphab}&=-\langle\bar{\epsilon},\psib^{\alphab}\rangle~,\nonumber\\
\delta \psi^\alpha&=i\gamma^{\mu}\partial_{\mu}x^\alpha\bar{\epsilon}+\epsilon F^\alpha~,
&\delta \psib^{\alphab}&=-i\gamma^{\mu}\partial_{\mu}\xb^{\alphab}\epsilon+\bar{\epsilon}\overline{F}^{\alphab}~,\nonumber\\
\delta F^\alpha&=i\langle \bar{\epsilon},\gamma^{\mu}\partial_{\mu}\psi^\alpha\rangle~,
&\delta \overline{F}^{\alphab}&=i\langle \epsilon,\gamma^{\mu}\partial_{\mu}\psib^{\alphab}\rangle~.
\end{align}

\subsection{R-symmetries and the low-energy limit}

The action \eqref{eq:compaction} at $W=0$ exhibits the chiral symmetries of the NLSM on $\bY$,
which act on the superfields \eqref{eq:02sfields22} as
\begin{align}
\xymatrix@C=5mm@R=0mm{
				&\theta^+		&\theta^-		&\cX^\alpha		&\Psi^\alpha\\
\GU(1)_L^0		&0			&1			&0				&-1\\
\GU(1)_R^0		&1			&0			&0				&0
}
\end{align}
These are however broken in the theory with a non-trivial superpotential.
A consequence of the $\C^\ast_V$ action induced by the Killing vector field $V$
is the fact that the superpotential satisfies a quasi-homogeneity condition.
Assuming that at least for a generic enough
superpotential $V$ is unique and that locally it can be written as $V=q_\alpha x^\alpha \p/\p x^\alpha$,
this condition reads
\begin{align}
W(x^{\alpha}e^{i\lambda q_{\alpha}})=e^{i\lambda}W(x^{\alpha})~.
\end{align}
It then follows that the classical action admits the symmetries
\begin{align}
\label{eq:Rcharges}
\xymatrix@C=5mm@R=0mm{
			&\theta^+		&\theta^-		&\cX^\alpha		&\Psi^\alpha\\
\GU(1)_L		&0			&1			&q_\alpha			&q_\alpha-1\\
\GU(1)_R		&1			&0			&q_\alpha			&q_\alpha
}
\end{align}
These are non-anomalous when $\bY$ is a Calabi-Yau manifold, which we assumed in our construction.
Moreover, the vertical property of $V$ implies that $q_{\alpha}=0$ for the base coordinates, while each
coordinate along the fiber component $X_i$ has charge $0<q_i\leq1/2$.

There is a particularly useful structure which we can use, together with known renormalization theorems, to relate UV data and the IR CFT.
Namely, we can construct a left-moving $\mathcal{N}=2$ superconformal algebra in $\bQb_+$-cohomology \cite{Silverstein:1994ih}.
For the class of models under study, the generators have been worked out in \cite{Bertolini:2013xga} and we report them here for convenience
\begin{align}
\label{eq:leftN2alg}
J_L &\equiv  (q_\alpha -1)\psi_-^\alpha\psib_{-,\alpha}- q_\alpha x^\alpha \rho_\alpha~,\nonumber\\
T &\equiv -\p_z x^\alpha\rho_\alpha- \half\left(\psib_{-,\alpha}  \p_z \psi_-^\alpha+ \psi_-^\alpha \p_z\psib_{-,\alpha}\right) -\half\p_z J_L~, \nonumber\\
G^+ &\equiv i\sqrt{2} \left[ \psib_{-,\alpha} \p_z x^\alpha - \p_z(q_\alpha \psib_{-,\alpha} x^\alpha)\right]~,\nonumber\\
G^- &\equiv i\sqrt{2} \psi_-^\alpha \rho_\alpha~,
\end{align}
where $\rho_\alpha \equiv g_{\alpha\alphab} \p_{z} \xb^{\alphab} + \Gamma^\delta_{\alpha\gamma}\psib_{-,\delta} \psi_-^\gamma$. In term of these fields,
the action reduces to a first-order system, with free fields OPEs
\begin{align}
\label{eq:OPEs}
x^\alpha(z) \rho_{\beta}(w) &\sim \frac{1}{z-w}\delta^\alpha_\beta~,
&\psi_-^\alpha(z)\psib_{-,\beta}(w)& \sim \frac{1}{z-w}\delta^\alpha_\beta~,
\end{align}
which define the left-moving $\mathcal{N}=2$ algebra in the full theory, while the non-trivial geometry is encoded in the
transformation properties across patches of the fields.
In this case, there is
plenty of evidence \cite{Bertolini:2013xga} that the theory flows under RG to a conformal fixed point characterized by the central charges
\begin{align}
c=\cb&=3\sum_\alpha \left( 1-2q_\alpha \right)~.
\end{align}
In the present work, we are concerned with the (c,c) ring of the theory, and as we will see, only a subset of the fields actively plays a role.
In particular, the field $\rho$ will not enter our discussion, thus we do not need to review its properties.
Nonetheless, this patch-wise free-fields first-order system description of a hybrid model
is particularly useful in performing explicit computations, as we will see in later sections.

\subsection{The orbifold and the twisted Jacobian algebra}

In application to string theory, the relevant objects are not quite hybrid models, but rather orbifolds thereof,
obtained by quotienting the theory by the discrete symmetry generated by $\exp(2\pi J_0)$,
where $J_0$ is the conserved charge associated to the $J_L$ current.\footnote{We remark that $\exp(2\pi i J_{0})$
acts on $x^{\alpha}$ and $\psi^{\alpha}_{\pm}$ with the same phase,
hence it is a flavor symmetry of the theory and the quotient theory is well-defined.}
Let $q_i=a_i/N_i$, where $a_i,N_i\in\mathbb{N}_{>0}$, then this discrete symmetry
is given by $\Gamma=\Z_N$ where $N=\text{lcm}(N_1,\dots,N_n)$.
As a consequence, all (NS,NS) states have integral charges under both $\GU(1)_L\times\GU(1)_R$
and the theory can be consistently completed to define a type II or heterotic string vacuum \cite{Gepner:1987vz}.
In particular, given the vertical property of $V$, the orbifold action is purely on the fiber $X$.
As mentioned above, this condition defines a good hybrid \cite{Bertolini:2013xga}.
When this fails to be the case, it has been shown \cite{Aspinwall:2009qy} that the theory develops a singularity at finite distance in the moduli space,
and the hybrid structure of a LGO fibered over a compact base breaks down.

From a purely mathematical point of view, the orbifold can be viewed as additional structure on the data defining the hybrid model.
In particular, given $W:\bY\rightarrow \C$, let the Jacobian algebra $\text{Jac}(W)$ be the finite-dimensional $\C$-algebra
defined by the cohomology of $\bQb_++\bQb_-$, which
we will study in detail in the following section.
This algebra is expected to have the structure of a Frobenius algebra \cite{Morrison:1994fr,Dubrovin:1994hc}.
The authors of \cite{2016arXiv160808962B} defined, in the case of orbifolds of an invertible polynomial $W$,
a $\Gamma$-twisted version of the Jacobian algebra, denoted $\text{Jac}'(W,\Gamma)$.
This object is further equipped with an orbifold residue pairing which defines an orbifold Jacobian algebra, i.e.,
the $\Gamma$-invariant subalgebra of $\text{Jac}'(W,\Gamma)$.
This in turn defines the structure of a $\Z_2$-graded (commutative) Frobenius algebra.

From this perspective, one of the results of this work is to derive a non-degenerate $\C$-bilinear form, which we can call the residue pairing, which
gives $(\bY,W,\mathbb{C}^{*}_{R})$ the structure of a (orbifold) Frobenius algebra.

\section{The B ring}
\label{s:chiralring}

Having reviewed the construction of the theories we wish to study, we now turn to the
characterization of the main object of interest in this first part of the work, namely, the (c,c) ring.
In generic (2,2) SCFTs, it is defined as the collection of
operators which satisfy the relations $h=q/2$ and $\hb=\qb/2$, where $h$($\hb$)
is the left(right)-moving weight and $q$($\qb$) the charge under
the left(right)-moving R-symmetry \cite{Lerche:1989uy}.
This is identified with the cohomology of the supercharges $\bQb_+$ and $\bQb_-$,\footnote{In (2,2) theories
there is another (in general) inequivalent ring, the (a,c) ring, defined by the cohomology of
the supercharges $\bQb_+$ and $\bQ_-$, or equivalently by the relations $h=-q/2$ and $\hb=\qb/2$.} or equivalently with the cohomology of the sum
\begin{align}
\bQb_{\text{(c,c)}}:=\bQb_+ +\bQb_-~.
\end{align}
A simple consequence of the $\mathcal{N}=2$ superconformal algebra is that in a compact SCFT the
number of such elements is finite. The ring structure is provided by the OPE between these operators,
which is non-singular as a consequence of the unitarity bounds.
This also implies that the three-point functions $\langle \cO_1(z_1) \cO_2(z_2)\cO_3(z_3) \rangle$
of such operators
in a suitable twisted theory (the B model) are independent
of the insertion points $z_{1,2,3}$. The computation of these correlators in hybrid theories is the main result of this work.

Before we proceed to review the techniques, developed in \cite{Bertolini:2013xga}, to compute the elements of the ring,
a comment is in order.
While it is customary in the literature to use $\bQb_{\text{(c,c)}}$ as the BRST charge in the B model \cite{Vafa:1990mu,Witten:1991zz},
this choice has the disadvantage that the corresponding representatives of the cohomology classes, i.e., the elements of the ring,
do not, in general, admit a well-defined $\GU(1)_L\times\GU(1)_R$-action.
In other words, such representatives are not eigenvectors of the chiral symmetries \eqref{eq:Rcharges}.
This issue is avoided in the context of (2,2) LG models as the cohomology is localized at one
particular position in the (dual) Koszul complex, the ring assumes the usual form $\C[x_1,\dots,x_N]/\la \p_1W,\dots,\p_N W \ra$, and
the representatives have well-defined charges and weights.
This is not the case for a more general hybrid, as the description of the ring generally involves, as we will see,
some particular linear combinations of the Fermi fields which do not admit a well-defined action under the chiral symmetries.

For this reason, we find it more convenient to consider
the cohomology of both $\bQb_+$ and $\bQb_-$ separately.
Specifically, we observe that there exists a deformation
of the $\bQb_{\text{(c,c)}}$-cohomology
defined by the supercharge $\bQb_\zeta := \bQb_++\zeta\bQb_-$.
We will argue that $\bQb_\zeta$-cohomology is equivalent to $\bQb_{\text{(c,c)}}$-cohomology.
It is clear that $\bQb_\zeta$-cohomology will not depend on $\zeta$
as long as $\zeta\neq0$, while at $\zeta=0$ obviously $\bQb_{\zeta=0}=\bQb_+$.
A look at \eqref{22transfcov} shows that
\begin{align}
\label{eq:newsupch}
[\bQb_\zeta,\xb^{\alphab}] &= \psib_+^{\alphab}-  \zeta \psib_-^{\alphab}~,		&\{\bQb_\zeta,\psi_+^\alpha\} &=i \pb_{\bar{z}} x^\alpha~,\nonumber\\
\{\bQb_\zeta,\psi_-^\alpha\} &=\zeta i \p_{z} x^\alpha~,
&\{\bQb_\zeta,\psib_{-,\alpha} +\zeta \psib_{+,\alpha}  \}&=(1+\zeta^{2})W_{\alpha}~.
\end{align}
At first order in $\zeta$, the only additional condition is that $\psi_-^\alpha$ are no longer exact. This implies that
$\bQb_\zeta$-cohomology is equivalent, up to $\bQb_-$-exact terms, to the cohomology of an operator $\bQb$ which acts as
\begin{align}
\label{eq:newsupch2}
[\bQb,\xb^{\alphab}] &= \psib_+^{\alphab}~,		&\{\bQb,\psi_+^\alpha\} &=i \pb_{\bar{z}} x^\alpha~,
&\{\bQb,\psi_-^\alpha\} &= i \p_{z} x^\alpha~,
&\{\bQb,\psib_{-,\alpha}  \}&=W_{\alpha}~.
\end{align}
In particular, we can split $\bQb = \bQb_0+\bQb_W$,
where $\bQb_0:=\bQb|_{W=0}$ and $\bQb_W$ contains all the dependence on $W$. These satisfy $\bQb_0^2=\bQb_W^2=\{\bQb_0,\bQb_W\}=0$,
and the non-trivial action of these operators is represented by
\begin{align}
\label{eq:chirringcohm}
[\bQb_0,\xb^{\alphab}] &=\psib_+^{\alphab}~,		&\{\bQb_0,\psi_+^\alpha\} &=i\pb_{\bar{z}} x^\alpha~,
&\{\bQb_0,\psi_-^\alpha\} &= i\p_{z} x^\alpha~, &\{\bQb_W,\psib_{-,\alpha}  \}&=W_{\alpha}~.
\end{align}
In order to simplify notation it is convenient to redefine the fields as follows
\begin{align}
\label{eq:newfermi}
\etab^{\alphab}& :=  \psib_+^{\alphab}~,		&\chib_\alpha&:=\psib_{-,\alpha}~.
\end{align}
Now, we can rewrite \eqref{eq:chirringcohm} as
\begin{align}
\label{eq:chirringcohmnewfermi}
[\bQb_0,\xb^{\alphab}] &=  \etab^{\alphab}~,		&\{\bQb_0,\psi_+^\alpha\} &=i\pb_{\bar{z}} x^\alpha~,
&\{\bQb_0,\psi_-^\alpha\} &=i \p_{z} x^\alpha~, &\{\bQb_W,\chib_\alpha \}&=W_{\alpha}~.
\end{align}
From \eqref{eq:chirringcohmnewfermi} we see that up to (anti-)holomorphic derivatives, $\psi_\pm$ are $\bQb_0$-exact and the candidates for
elements in $\bQb$-cohomology are the local operators
\begin{align}
\label{eq:states}
\cO(\omega) = \omega(x,\xb)_{\betab_1\dots\betab_s}^{\alpha_1\dots\alpha_r}\etab^{\betab_1}\cdots\etab^{\betab_s}\chib_{\alpha_1}\cdots\chib_{\alpha_r}~.
\end{align}
The action of the supercharges on the states \eqref{eq:states} is given by
\begin{align}
\label{eq:Q0action}
\bQb_0 : \cO(\omega) \mapsto (\pb \omega)(x,\xb)_{\gammab \betab_1\dots\betab_s}^{\alpha_1\dots\alpha_r}\etab^{\gammab}\etab^{\betab_1}\cdots\etab^{\betab_s}\chib_{\alpha_1}\cdots\chib_{\alpha_r}~,
\end{align}
where $(\pb\omega)_{\gammab}\equiv \p\omega/\p \xb^{\gammab}$, and
\begin{align}
\label{eq:QWaction}
\bQb_W : \cO(\omega) \mapsto (-1)^s \omega(x,\xb)_{\betab_1\dots\betab_s}^{\alpha_1\dots\alpha_r}W_{\alpha_1}
\etab^{\betab_1}\cdots\etab^{\betab_s}\chib_{\alpha_2}\cdots\chib_{\alpha_r}~.
\end{align}
We now specialize this structure to the hybrid geometry $\bY=\tot\left( X \xrightarrow{\pi} B\right)$.
Let $\{ U_{a}\}$ be an open cover of $B$,
then $\{\pi^{-1}U_{a}\}$ is an open cover of $\bY$.
Consider an open set $\pi^{-1}U_{a}\cong U_{a}\times \mathbb{C}^{n}$ parametrized by the coordinates
\begin{align}
x^{\alpha}=(y^{I},\phi^{i})\in U_{a}\times \mathbb{C}^{n}~,
\end{align}
where $I=1,\dots,b$ and $i=1,\dots,n$.
On this patch we can identify the operators (\ref{eq:states}) as sections of the sheaf
\begin{align}
\left(\bigoplus_{s_{1}+s_{2}=s}\Omega^{0,s_{1}}(U_a)\otimes\Omega^{0,s_{2}}(\mathbb{C}^{n})\right)\otimes \wedge^r T_{\bY}~,
\end{align}
where all the products $\otimes$ are over the ring of $C^{\infty}$ functions on $\pi^{-1}U_{a}$.
We take this ring to be $C^{\infty}$ functions on $U_{a}$ with at most polynomial growth
along the fiber directions. This condition does not affect the cohomology,
as we show in appendix \ref{app:fibcohom} using the results of \cite{Babalic:2016mbw}.

As shown in \cite{Bertolini:2013xga}, upon specialization to $\bY$, in order to compute $\overline{\mathbf{Q}}$-cohomology
we can restrict our attention to operators that are horizontal forms and independent of $\bar{\phi}$.
That is, we can interpret $\omega$ in (\ref{eq:states}) as a $(0,s)$-horizontal forms valued in $\wedge^r T_{\bY}$.
We identify this vector space as
\begin{align}
\label{eq:wedgeTYdef}
\wedge^r_s T_{\bY}&:=\Omega^{0,s}(\bY,\wedge^r T_{\bY})~,
\end{align}
that is, $(0,s)$-forms in $\bY$ valued in $\wedge^r T_{\bY}$ with at most polynomial growth along the fiber directions. 
The supercharges act on \eqref{eq:wedgeTYdef} as
\begin{align}
\label{eq:supchactionsdg}
\bQb_0 &: \wedge^r_s T_{\bY} \rightarrow \wedge^r_{s+1} T_{\bY}~,		&\bQb_W &: \wedge^r_s T_{\bY} \rightarrow \wedge^{r+1}_s T_{\bY}~.
\end{align}
The space of operators \eqref{eq:states} constitutes a double graded complex $K^{r,s}$
\begin{equation}
\begin{matrix}\vspace{5mm}\\E_0^{r,s} \equiv K^{r,s}:\end{matrix}
\begin{xy}
\xymatrix@C=5mm@R=5mm{
0 & 0 & \cdots & 0 & 0 & 0 &0\\
0 &\wedge^d_b T_{\bY} & \cdots &\wedge^2_b T_{\bY} & \wedge^1_b T_{\bY}  &   \wedge^0_b T_{\bY} & 0\\
\vdots &\vdots & \ddots &\vdots	&\vdots & \vdots &\vdots\\
0 &\wedge^d_1T_{\bY} & \cdots &\wedge^2_1 T_{\bY} & \wedge^1_1T_{\bY}  &  \wedge^0_1 T_{\bY} & 0\\
0 &\wedge^d T_{\bY} & \cdots &\wedge^2 T_{\bY} & T_{\bY}  &  \cO_{\bY} & 0
}
\save="x"!LD+<-6mm,0pt>;"x"!RD+<40pt,0pt>**\dir{-}?>*\dir{>}\restore
\save="x"!LD+<82mm,-3mm>;"x"!LU+<82mm,2mm>**\dir{-}?>*\dir{>}\restore
\save!RD+<13mm,-3mm>*{r}\restore
\save!CL+<85mm,28mm>*{s}\restore
\end{xy}
\end{equation}
and \eqref{eq:supchactionsdg} implies that $\bQb_0$ and $\bQb_W$
act as the vertical and horizontal differentials, respectively. Thus, the total cohomology $\bQb$ is computed by
a spectral sequence determined by this data. In particular, the first stage is obtained as $E_1^{r,s}=H^s_{\bQb_0}(\bY,K^{r,\bullet})$, which yields
\begin{equation}
\label{eq:22spetrseq}
\begin{matrix}\vspace{5mm}\\E_1^{r,s} :\end{matrix}
\begin{xy}
\xymatrix@C=10mm@R=5mm{
0 \ar[r]&H^b(\bY,\wedge^dT_{\bY}) \ar[r]^-{\bQb_W}& \cdots \ar[r]^-{\bQb_W}& H^b(\bY,T_{\bY}) \ar[r]^-{\bQb_W} &   H^b(\bY,\cO_{\bY})\ar[r]& 0\\
\vdots &\vdots & \ddots &\vdots	&\vdots & \vdots\\
0 \ar[r]&H^1(\bY,\wedge^dT_{\bY}) \ar[r]^-{\bQb_W}& \cdots\ar[r]^-{\bQb_W} & H^1(\bY,T_{\bY})  \ar[r]^-{\bQb_W}&   H^1(\bY,\cO_{\bY})\ar[r]& 0\\
0 \ar[r]&H^0(\bY,\wedge^dT_{\bY}) \ar[r]^-{\bQb_W}& \cdots \ar[r]^-{\bQb_W}& H^0(\bY,T_{\bY}) \ar[r]^-{\bQb_W} &   H^0(\bY,\cO_{\bY}) \ar[r]& 0\\
}
\save="x"!LD+<-6mm,0pt>;"x"!RD+<20pt,0pt>**\dir{-}?>*\dir{>}\restore
\save="x"!LD+<85mm,-3mm>;"x"!LU+<85mm,2mm>**\dir{-}?>*\dir{>}\restore
\save!RD+<03mm,-3mm>*{r}\restore
\save!CL+<87mm,24mm>*{s}\restore
\end{xy}
\end{equation}
The next stage is given by taking cohomology with respect to the horizontal map $\bQb_W$
according to \eqref{eq:QWaction}. This is the hybrid generalization of the Koszul complex familiar from Landau-Ginzburg models \cite{Kawai:1994np,Melnikov:2009nh}.
Higher differentials are constructed from \eqref{eq:supchactionsdg} using the standard zig-zag procedure \cite{Bott:1982df}.
An important fact is that the spectral sequence defined above is ensured to converge, since $s\leq b=\dim B$.

An apparent issue in computing the spectral sequence is that, due to the non-compactness of $\bY$,
the cohomology groups $H_{\bQb_0}^\bullet(\bY,\wedge^r T_{\bY})$ are generically infinite dimensional.
This issue can be circumnavigated
due to the fact that $\bQb$-cohomology commutes with the left-moving $\mathcal{N}=2$ algebra \eqref{eq:leftN2alg},
and we can use the generators $J_L$ and $T$ to introduce additional gradings on
the space of operators \eqref{eq:states}.
This is why it is convenient to choose cohomology class representatives that admit well-defined charges.
In particular, as we discussed above, the left-chirality condition $2h=q$ is already automatically imposed by $\bQb$-cohomology, therefore it
suffices to restrict our attention to the grading $\bq$ corresponding to $J_L$.
In practice, this means that we can compute the spectral sequence at a fixed value of $\bq$, and
the groups $H_{\bq}^\bullet(\bY,\wedge^r T_{\bY})$ are then finite-dimensional. A prescription
for how to compute these graded cohomology groups in terms of cohomology groups on the base
is given in appendix C of \cite{Bertolini:2013xga}.

\section{B-twisted $\mathcal{N}=(2,2)$ hybrid models and $S^{2}$ correlators}
\label{s:22corrs}

In this section we turn to the study of the B model for hybrid theories.
The vector and axial R-charges act on the component fields as
\begin{align}
\xymatrix@C=5mm@R=0mm{
			&\phi^{\alpha}		&\psi_{+}^{\alpha}		&\psi_{-}^{\alpha}	&F^{\alpha}\\
\GU(1)_V		&q^\alpha_{V}		&q^\alpha_{V}-1		&q^\alpha_{V}-1	&q^\alpha_{V}-2\\
\GU(1)_A		&q^\alpha_{A}		&q^\alpha_{A}-1		&q^\alpha_{A}+1	&q^\alpha_{A}
}
\end{align}
These are related to the left- and right-moving R-charges by
\begin{align}
\GU(1)_{V}&=\GU(1)_{L}+\GU(1)_{R}~,	 &\GU(1)_{A}&=-\GU(1)_{L}+\GU(1)_{R}~.
\end{align}
From \eqref{eq:Rcharges} it follows that in the models under study, $q^\alpha_{A}=0$ and $q^\alpha_V=2q_\alpha$.
The B-twist \cite{Witten:1991zz} of the theory
amounts to twisting the Euclidean rotation group $U(1)_{E}$ by $U(1)_{A}$.
There are two options for performing such a twist, namely
\begin{align}
\label{eq:Btwistdef}
B_{(+)}&:U(1)_{E}'=U(1)_{E}+\frac{1}{2}U(1)_{A}~, 	&B_{(-)}&:U(1)_{E}'=U(1)_{E}-\frac{1}{2}U(1)_{A}~,
\end{align}
and for definitiveness we choose the twist denoted $B_{(+)}$.
Under this choice, the spinors $\epb_{\pm}$ become scalars,\footnote{The $B_{(-)}$-twist is equivalent and, in such case, $\epsilon_{\pm}$ become scalars.}
and the matter fermions transform as sections of the bundles
\begin{align}
\psi^{\alpha}_{+}&\in \Gamma(\overline{K}_{\Sigma}\otimes x^{*}(T_{\bY}))~,	&\psi^{\alpha}_{-}&\in\Gamma(K_{\Sigma}\otimes x^{*}(T_{\bY}))~,\nonumber\\
{\psib}^{\alphab}_{+}&\in \Gamma(x^{*}(\overline{T}_{\bY}))~,				&\psib^{\alphab}_{-}&\in \Gamma(x^{*}(\overline{T}_{\bY}))~.
\end{align}
There exists a family of nilpotent operators $\delta_{\zeta}$, parametrized by a phase
$\zeta$,\footnote{In \cite{Hori:2000ck}, $\zeta$ is denoted $e^{i\beta}$ and fixed to $\beta=\pi$.
$\delta_\zeta$ is also the same differential denoted $\overline{\partial}_{f}$ in \cite{Li:2013kja}.} which is defined by setting
\begin{eqnarray}
\delta_{\zeta}=\delta|_{\bar{\epsilon}_{+}=\zeta\bar{\epsilon}_{-}}~.
\end{eqnarray}
This differential corresponds to the operator $\bQb_\zeta$ in \eqref{eq:newsupch}. Since we showed in the previous section that the cohomology of $\bQb_\zeta$
does not depend on $\zeta$ as long as $\zeta\neq0$, we choose in this section to restrict our attention to $\zeta$ being a phase.
The SUSY transformations generated by $\delta_{\zeta}$ acquire a particularly simple form if we redefine the fields as follows\footnote{The Fermi field $\rho^\alpha$ is not
to be confused with the field $\rho_\alpha$ in \eqref{eq:OPEs}. As mentioned before, the latter will make no significant appearance in our computations.}
\begin{align}
\label{eq:22newfields}
\kappa^{\bar{\alpha}}&=\psib^{\alphab}_{+}-\zeta \psib^{\alphab}_{-}~,
&\theta^{\alphab}&=\psib^{\alphab}_{+}+\zeta\psib^{\alphab}_{-}~,
&\rho^{\alpha}&=\psi^{\alpha}_{\bar{z}}d\bar{z}+\zeta^{-1}\psi^{\alpha}_{z}dz~,\nonumber\\
F'^{\alpha}&=\zeta^{-1} F^{\alpha}~,  		&\overline{F}'^{\bar{\alpha}}&=\zeta \overline{F}^{\bar{\alpha}}~.
\end{align}
In terms of these, from \eqref{22transfcov}, we obtain
\begin{align}
\delta_{\zeta}x^{\alpha}&=0		&\delta_{\zeta} \bar{x}^{\bar{\alpha}}&=\bar{\epsilon}_{-}\kappa^{\bar{\alpha}}\nonumber\\
\delta_{\zeta}\rho^{\alpha}_{\mu}&=2i\bar{\epsilon}_{-}\partial_{\mu}x^{\alpha}
&\delta_{\zeta}\theta^{\bar{\alpha}}&=2\bar{\epsilon}_{-}\overline{F}'^{\bar{\alpha}}\nonumber\\
\delta_{\zeta}F'^{\alpha}&=2i\bar{\epsilon}_{-}\varepsilon^{\mu\nu}\partial_{\mu}\rho^{\alpha}_{\nu}~,
&\delta_{\zeta}\overline{F}'^{\bar{\alpha}}&=0~,\nonumber\\
\delta_{\zeta}\kappa^{\bar{\alpha}}&=0~,
\end{align}
where we defined the symbol $\varepsilon^{\mu\nu}$ as $\varepsilon^{z\bar{z}}=-\varepsilon^{\bar{z}z}=1$.
The $B_{(+)}$-twisted Lagrangian, in terms of the fields \eqref{eq:22newfields}, reads
\begin{align}
\mathcal{L}_{K}&=-g_{\alpha\bar{\beta}}\left(h^{\mu\nu}\partial_{\mu}x^{\alpha}\partial_{\nu}\bar{x}^{\bar{\beta}}+\frac{i}{2}h^{\mu\nu}\rho^{\alpha}_{\mu}D_{\nu}\kappa^{\bar{\beta}}
+i\varepsilon^{\mu\nu}\partial_{\nu}\rho^{\alpha}_{\mu}\theta^{\bar{\beta}}\right)+\frac{1}{4}\varepsilon^{\mu\nu}R_{\alpha\bar{\alpha}\beta\bar{\beta}}\rho^{\alpha}_{\mu}\rho^{\beta}_{\nu}\kappa^{\bar{\alpha}}\theta^{\bar{\beta}}\nonumber\\
&\quad+g_{\alpha\bar{\alpha}}F'^{\alpha}\overline{F}'^{\bar{\alpha}}+\frac{1}{2}g_{\alpha\bar{\alpha}}\Gamma^{\alpha}_{\beta\gamma}\varepsilon^{\mu\nu}\rho^{\beta}_{\mu}\rho^{\gamma}_{\nu}\overline{F}'^{\bar{\alpha}}
+\frac{1}{2}g_{\alpha\bar{\alpha}}\overline{\Gamma}^{\bar{\alpha}}_{\bar{\beta}\bar{\gamma}}F'^{\alpha}\kappa^{\bar{\beta}}\theta^{\bar{\gamma}}\nonumber\\
&\quad+\frac{1}{4}g_{\alpha\bar{\alpha}}\Gamma^{\alpha}_{\beta\gamma}\overline{\Gamma}^{\bar{\alpha}}_{\bar{\beta}\bar{\gamma}}\varepsilon^{\mu\nu}\rho^{\beta}_{\mu}\rho^{\gamma}_{\nu}\kappa^{\bar{\beta}}\theta^{\bar{\gamma}}~, \nonumber\\
\mathcal{L}_{W}&=\frac{1}{2}\left(\zeta F'^{\alpha}\partial_{\alpha}W+\frac{\zeta}{2}\partial_{\alpha}\partial_{\beta}W\varepsilon^{\mu\nu}\rho^{\alpha}_{\mu}\rho^{\beta}_{\nu}+\zeta^{-1}\overline{F}'^{\bar{\alpha}}\overline{\partial}_{\bar{\alpha}}\overline{W}+\frac{\zeta^{-1}}{2}\overline{\partial}_{\bar{\alpha}}\overline{\partial}_{\bar{\beta}}\overline{W}\kappa^{\bar{\alpha}}\theta^{\bar{\beta}}\right)~.
\end{align}
From the above expressions it is easy to see that the net effect of keeping the phase $\zeta$ arbitrary is equivalent to a rescaling of the superpotential
\begin{align}
W&\rightarrow \zeta W~, 	&\overline{W}&\rightarrow \zeta^{-1} \overline{W}~.
\end{align}
This is a good point to comment on the geometric interpretation of the differential $\delta_{\zeta}$. The local operators that are candidates to be $\delta_{\zeta}$-closed are
\begin{eqnarray}
\label{eq:opalt}
\mathcal{O}(\omega)=\omega^{\alpha_{1},\ldots,\alpha_{s}}_{\bar{\beta}_{1},\ldots,\bar{\beta}_{r}}\kappa^{\bar{\beta}_{1}}\cdots\kappa^{\bar{\beta}_{r}}\theta_{\alpha_{1}}\cdots\theta_{\alpha_{s}}~,
\end{eqnarray}
where we defined $\theta_{\alpha}:=g_{\alpha\alphab}\theta^{\alphab}$.
While the structure is precisely the same as in \eqref{eq:states},
here we are using a different basis which does not have a well defined vector R-charge.
The operators $\mathcal{O}(\omega)$ can be identified with global sections of the sheaf of polyvector fields
\begin{eqnarray}
PV:=\bigoplus_{s,r}PV^{s,r}=\bigoplus_{s,r}\Omega^{0,r}\otimes\wedge^{s} T_{\bY}=\bigoplus_{s,r}\wedge^{s}_{r} T_{\bY}~,
\end{eqnarray}
via the mapping
\begin{align}
\kappa^{\alphab}&\rightarrow d\xb^{\alphab}~, 		&\theta_{\alpha}&\rightarrow \partial_{\alpha}~.
\end{align}
Then, we identify the operators \eqref{eq:opalt} with
\begin{eqnarray}
\mathcal{O}(\omega)\rightarrow\omega^{s}_{r}:=\omega^{\alpha_{1},\ldots,\alpha_{s}}_{\bar{\beta}_{1},\ldots,\bar{\beta}_{r}}d\bar{x}^{\bar{\beta}_{1}}\cdots d\bar{x}^{\bar{\beta}_{r}}\otimes \partial_{\alpha_{1}}\cdots\partial_{\alpha_{s}}\in PV(\bY):=\Gamma(\bY,PV)~.
\end{eqnarray}
Upon acting on these, the differential $\delta_{\zeta}$ is identified with
\begin{eqnarray}
\delta_{\zeta}\rightarrow \overline{\partial}-\zeta\iota_{dW}~,
\end{eqnarray}
where the operator $\iota_{dW}$ acts as
\begin{eqnarray}
\iota_{dW}\circ\omega^{s}_{r}&=&\sum_{l=1}^{s}(-1)^{r+l+1}
\partial_{\alpha_{l}}W\omega^{\alpha_{1},\ldots,\alpha_{s}}_{r}\partial_{\alpha_{1}}\wedge\cdots\wedge \widehat{\partial_{\alpha_{l}}}\wedge\ldots\partial_{\alpha_{s}}\nonumber\\
&=&s(-1)^{r}\partial_{\alpha}W\omega^{\alpha,\alpha_{1},\ldots,\alpha_{s-1}}_{r}\partial_{\alpha_{1}}\wedge\ldots\wedge\partial_{\alpha_{s-1}}~.
\end{eqnarray}
The charge orthogonal to $\delta_\zeta$ has instead the following interpretation
\begin{eqnarray}
\tilde{\delta}_{\zeta}:=\delta|_{\bar{\epsilon}_{+}=-\zeta\bar{\epsilon}_{-}}\quad \rightarrow\quad[\Lambda,\overline{\partial}-\zeta\iota_{dW}]~.
\end{eqnarray}
The operator $\Lambda$ is defined as the contraction with the inverse of the K\"ahler (symplectic) form on $\bY$, and it acts on $\omega^{s}_{r}$ as
\begin{eqnarray}
\Lambda\circ \omega^{s}_{\bar{\alpha}_{1},\ldots,\bar{\alpha}_{r}}d\bar{x}^{\bar{\alpha}_{1}}\wedge\cdots\wedge d\bar{x}^{\bar{\alpha}_{r}}
=r(-1)^{r-1}g^{\alpha\bar{\alpha}}\omega^{s}_{\bar{\alpha},\bar{\alpha}_{2},\ldots,\bar{\alpha}_{k}}d\bar{x}^{\bar{\alpha}_{2}}\wedge\cdots\wedge d\bar{x}^{\bar{\alpha}_{k}}~.
\end{eqnarray}
The space $PV(\bY)=\oplus_{q,p}PV^{q,p}(\bY)$ admits a $\mathbb{Z}$-grading (see for example \cite{Li:2013kja}), given by
\begin{eqnarray}
\mathrm{deg}(\omega^{s}_{r})=r-s~,	\qquad \omega^{s}_{r}\in \wedge^{s}_{r}T_{\bY}~.
\end{eqnarray}
In particular, $\delta_{\zeta}$ shifts $\mathrm{deg}(\omega^{r}_{s})$ by $+1$.

\subsection{Localization on $S^{2}$}

We now turn to the derivation of the formula for the closed B-twisted correlators of local fields
via $S^{2}$ localization. This can be interpreted as the hybrid generalization of the LG correlators from \cite{Vafa:1990mu}.
We remark that this result has been first derived in \cite{Herbst:2004ax} -- although in a slightly different fashion than the one we will present here --
as well as in \cite{Guffin:2008kt}. In the latter work, however, the result is valid
only for the non-degenerate case, as it involves determinants of the Hessian of $W$.
This is not the case for the models we study in this work whenever $\bY$ is non-compact, that is,
whenever $X$ and $W$ are not trivial.
Nevertheless, we find instructive and useful to re-derive this result from our perspective.
This will also pay off in section \ref{s:02hybrids} when we will perform an analogous localization computation for (0,2) models,
where our result is instead new.

First, let us note that the kinetic term $\mathcal{L}_{K}$ in the Lagrangian is $\delta_{\zeta}$-exact, as
\begin{eqnarray}
\bar{\epsilon}_{-}\mathcal{L}_{K}=\delta_{\zeta}\left(\frac{i}{2}g_{\alpha\bar{\alpha}}h^{\mu\nu}\rho^{\alpha}_{\mu}\partial_{\nu}\bar{x}^{\bar{\alpha}}
+\frac{1}{2}g_{\alpha\bar{\alpha}}\left(F'^{\alpha}+\frac{1}{2}\varepsilon^{\mu\nu}\Gamma^{\alpha}_{\beta\gamma}\rho^{\beta}_{\mu}\rho^{\gamma}_{\nu}\right)\theta^{\bar{\alpha}}\right)~.
\end{eqnarray}
In particular, the term
\begin{eqnarray}\label{exactred}
\delta_{\zeta}\left(\frac{i}{2}g_{\alpha\bar{\alpha}}h^{\mu\nu}\rho^{\alpha}_{\mu}\partial_{\nu}\bar{x}^{\bar{\alpha}}\right)=-\bar{\epsilon}_{-}g_{\alpha\bar{\alpha}}\left(h^{\mu\nu}\partial_{\mu}x^{\alpha}\partial_{\beta}\bar{x}^{\bar{\alpha}}+\frac{i}{2}h^{\mu\nu}\tilde{\rho}^{\alpha}_{\mu}D_{\nu}\kappa^{\bar{\alpha}}\right)
\end{eqnarray}
is positive-definite, and we can use it to localize the action
by regarding, at a first stage, the fields $F'^{\alpha},\overline{F}'^{\bar{\alpha}}$ and $\theta^{\bar{\alpha}}$ as background fields.
This means that we first localize regarding only the fields $\{\kappa^{\bar{\alpha}},x^{\alpha},\bar{x}^{\bar{\alpha}},\rho^{\alpha}\}$ as dynamical.
The solution to the saddle point equations
\begin{align}
\delta_{\zeta}\kappa=\delta_{\zeta}\rho=0
\end{align}
on $\Sigma=S^{2}$ is simply given by $x^{\alpha}=\mathrm{const}$.
The classical action evaluated at this locus reduces to
\begin{align}
S_{0}[F,\theta]:=\int_{\Sigma}d^2z(\mathcal{L}_{K}+\mathcal{L}_{W})|_{x=\mathrm{const}}=\int_{\Sigma}d^2z\left(g_{\alpha\bar{\alpha}}F'^{\alpha}\overline{F}'^{\bar{\alpha}}+\frac{1}{2}\left(\zeta F'^{\alpha}\partial_{\alpha}W+\zeta^{-1}\overline{F}'^{\bar{\alpha}}\overline{\partial}_{\bar{\alpha}}\overline{W}\right)\right)~.
\end{align}
The 1-loop determinant coming from the expansion of (\ref{exactred}) over non-zero modes is actually a numerical constant that we can ignore.
While the zero modes of $x^{\alpha}$ are weighted by $S_{0}[F,\theta]$,
we have to be careful with the zero modes of the fermions $\kappa^{\bar{\alpha}}$ (on $S^{2}$, $\rho$ has no zero modes).
The usual trick (see for example \cite{Aspinwall:1991ce})
is to use the classical action evaluated at the zero modes of $\kappa^{\bar{\alpha}}_{0}$ to absorb them.
Thus, the path integral reads
\begin{eqnarray}\label{eq:pathint22}
\int \mathcal{D}F'\mathcal{D}\overline{F}'\mathcal{D}\theta\int_{\bY}d^2x\prod_{\alpha}\int d\kappa^{\bar{\alpha}}_{0} e^{S[F,\theta]}~,
\end{eqnarray}
and it is weighted by
\begin{align}
S[F,\theta]:=\int_{\Sigma}d^2z(\mathcal{L}_{K}+\mathcal{L}_{W})|_{(x=\mathrm{const},\kappa_{0}^{\bar{\alpha}})}=S_{0}[F,\theta]+\int_{\Sigma}d^2z\left(\frac{\zeta^{-1}}{4}\overline{\partial}_{\bar{\alpha}}\overline{\partial}_{\bar{\beta}}\overline{W}
\kappa^{\bar{\alpha}}_{0}\theta^{\bar{\beta}}\right)~.
\end{align}
The integral over $\mathcal{D}\theta$ is to be interpreted as an integral over the $\theta^{\alphab}$ zero modes,
and the integration over $F'^{\alpha},\Fb'$ can be performed by means of a change of variables.
The result, using the geometric interpretation outlined in the previous section, is the following formula for the correlators
\begin{align}
\label{eq:formula}
\langle\mathcal{O}(\omega)\rangle_{S^{2}}:=\int_{\bY}d^{2}x\int(\prod_{\alpha} d\theta_{\alpha}^{0}) (\prod_{\alpha}d\kappa^{\bar{\alpha}}_{0}) \exp\left(-\frac{\mathbf{v}}{4}\| dW \|^{2}+\mathbf{v}\frac{\zeta^{-1}}{4}\overline{\nabla}_{\bar{\alpha}}\overline{\partial}_{\bar{\beta}}\overline{W}
\kappa^{\bar{\alpha}}_{0}\theta^{\bar{\beta}}_{0}\right)\mathcal{O}(\omega)~,
\end{align}
where $\mathbf{v}$ is the worldsheet volume and $\| dW \|^{2}:=g^{\alpha\alphab}\partial_{\alpha}W\bar{\partial}_{\alphab}\Wb$.
The last step is the integration over the fermion zero modes. If $\omega \in PV^{r,s}(\bY)$, we have the equality
\begin{eqnarray}
\int_{\bY}d^{2}x\int(\prod_{\alpha} d\theta_{\alpha}^{0}) (\prod_{\alpha}d\kappa^{\bar{\alpha}}_{0}) \mathcal{O}(\omega)=\int_{\bY}\Omega_{\bY}\wedge(\Omega_{\bY}\lrcorner \omega^{s}_{r})~,
\end{eqnarray}
where $\Omega_{\bY}\lrcorner$ is defined as
\begin{eqnarray}
\Omega_{\bY}\lrcorner\omega^{s}_{r}:=(\Omega_{\bY})_{\alpha_{1}\cdots \alpha_{s}\alpha_{s+1}\cdots\alpha_{d}}\omega^{\alpha_{1}\cdots \alpha_{s}}_{\bar{\beta}_{1}\cdots \bar{\beta}_{r}}d\bar{x}^{\bar{\beta}_{1}}\wedge\cdots\wedge d\bar{x}^{\bar{\beta}_{r}}\wedge dx^{\alpha_{s+1}}\wedge\cdots\wedge dx^{\alpha_{d}}~.
\end{eqnarray}
Some comments about this formula are in order. The contributions of the Fermi fields to the measure are identified with sections of the bundles
\begin{align}
&\prod_{\alpha} d\theta_{\alpha}^{0}\in \Gamma(K_{\bY})~, &&\prod_{\alpha}d\kappa^{\bar{\alpha}}_{0}\in \Gamma(\overline{K}_{\bY}^{*})~.
\end{align}
The condition for the measure to be well-defined
is then $K_{\bY}\otimes\overline{K}_{\bY}^{*}\cong K_{\bY}\otimes  K_{\bY}\cong \mathcal{O}_{\bY}$
since $\mathcal{O}_{\bY}$ always has a nowhere vanishing global section. This condition was also obtained in \cite{Sharpe:2006qd}.
For the case at hand, where we want to consider $\bY$ being non-compact,
the holomorphic volume form $\Omega_{\bY}$
is defined only up to a multiplicative
non-vanishing holomorphic function on $\bY$.
We remedy this ambiguity by requiring
\begin{align}
\Omega_{\bY}\wedge\overline{\Omega}_{\bY}&=\widehat{K}^{n}~, 		&\widehat{K}&=ig_{\alpha\bar{\alpha}}dx^{\alpha}\wedge d\bar{x}^{\bar{\alpha}}~.
\end{align}
Finally, the $S^2$ correlators can be written as
\begin{eqnarray}
\label{eq:22S2corrs}
\langle\mathcal{O}(\omega)\rangle_{S^{2}}=\int_{\bY}\Omega_{\bY}\wedge(\Omega_{\bY}\lrcorner(e^{\hat{L}} \omega^{s}_{r}))~,
\end{eqnarray}
where we introduced the operator
\begin{align}
\hat{L}&:=-\frac{\mathbf{v}}{4}\| J\|^{2}+\mathbf{v}\frac{\zeta^{-1}}{4}\overline{\partial}\overline{J}^{\alpha}\partial_{\alpha}~,
&\overline{J}^{\alpha}\partial_{\alpha}&:=g^{\alpha\alphab}\overline{\partial}_{\alphab}\overline{W}\partial_{\alpha}\in PV^{1,0}(\bY)~,
\end{align}
whose exponential acts by the usual wedge product on polyvector fields
\begin{eqnarray}
\label{eq:expexpLh}
e^{\hat{L}}=e^{-\frac{\mathbf{v}}{4}\| J \|^{2}}\sum_{r}\frac{\zeta^{-r}\mathbf{v}^{r}}{4^{r}r!}\left(\overline{\partial}\overline{J}^{\alpha}\partial_{\alpha}\right)^{r}~.
\end{eqnarray}
We remark that the correlator \eqref{eq:22S2corrs}, when $\bY$ is compact and smooth (thus $\hat{L}\equiv 0$), reduces
to the well known expression for a NLSM with target the CY manifold $\bY$ (see for example section 16 of \cite{Hori:2003ic}).

Although the formula \eqref{eq:22S2corrs} is formally a one-point function,
it is of most interest when interpreted as a three-point function, that is,
by expressing the insertion as the product $\cO(\omega)=\cO(\omega_1)\cO(\omega_2)\cO(\omega_3)$,
where $\cO(\omega_{1,2,3})\in H^\bullet_{\bQb}(\bY,\wedge^\bullet T_{\bY})$.
In particular, the correlators must be invariant under the chiral symmetries.
From the expression \eqref{eq:formula}, it follows that the measure has charge $-c/3$ under $\GU(1)_L$, thus the
correlators vanish unless
\begin{align}
\bq(\cO(\omega_1))+\bq(\cO(\omega_2))+\bq(\cO(\omega_3))=\frac{c}3~.
\end{align}
Invariance of the formula \eqref{eq:formula} under $\GU(1)_A$, under which the measure has charge 0, instead implies that $\cO(\omega)\in \oplus_{p=0}^b PV^{p,p}(\bY)$.

\subsection{Properties of $S^{2}$ correlators}

The formula for the $S^{2}$ correlators we derived in the previous section enjoys a series of properties that we will argue for in this section.
We remark that some of these properties have already been shown in \cite{Babalic:2016mbw}
and \cite{2015arXiv150802769C}.

Let us start by showing that \eqref{eq:22S2corrs} is independent of the choice of representatives. Consider the following integral on $S^{2}$
\begin{align}
\langle\delta_{\zeta}\mathcal{O}(\omega)\rangle'_{S^{2}}&:=\int_{\bY}\Omega_{\bY}\wedge (\Omega_{\bY}\lrcorner \delta_{\zeta}\omega)~, 	&\omega&\in PV^{p,q}_{c}(\bY)~.
\end{align}
In order for the integral to be well-defined, we take $\mathcal{O}(\omega)$ to be a compactly supported (we only need it to be compactly supported along $X$),
nonsingular, homogenous (in degree) polyvector field.
Since $\overline{\partial}\omega\in PV^{p,q+1}(\bY)$ and $\iota_{J}\omega\in PV^{p-1,q}(\bY)$,
then $\langle\delta_{\zeta}\mathcal{O}(\omega)\rangle'_{S^{2}}=0$ unless $q+1=p=d$ or $p-1=q=d$.
If $p=d+1$, $\omega$ vanishes identically, while in the former case we are left with
\begin{eqnarray}
\langle\delta_\zeta\mathcal{O}(\omega)\rangle'_{S^{2}}=\int_{\bY}\Omega_{\bY}\wedge (\Omega_{\bY}\lrcorner \overline{\partial}\omega)=\int_{\bY}d\Omega_{\bY}\wedge (\Omega_{\bY}\lrcorner\omega)=0~,
\end{eqnarray}
where the last equality follows from our assumptions that $\omega$ has no poles on $\bY$ and that the integral is convergent.
By a direct computation one can show that
\begin{eqnarray}
[\delta_{\zeta},e^{\hat{L}}]=\delta_{\zeta}(e^{\hat{L}})+e^{\hat{L}}\delta_{\zeta}-e^{\hat{L}}\delta_{\zeta}=0~.
\end{eqnarray}
Hence
\begin{eqnarray}
\label{eq:indreprs}
\langle\delta_{\zeta}\mathcal{O}(\omega)\rangle_{S^{2}}=\langle\delta_{\zeta}(e^{\hat{L}}\mathcal{O}(\omega))\rangle'_{S^{2}}=0~,
\end{eqnarray}
for any polyvector field $\omega$, concluding the proof that the correlators do not depend on the choice of representatives.

A second property of $\langle\mathcal{O}(\omega)\rangle_{S^{2}}$ is that it is independent of small variations of $\mathbf{v}$ whenever $\omega$ is in $\delta_{\zeta}$-cohomology.
This is expected for a topological field theory and it follows from the identity
\begin{eqnarray}
\exp\left(-\frac{\mathbf{v}+\delta \mathbf{v}}{4}\| J \|^{2}+\zeta^{-1}\frac{(\mathbf{v}+\delta \mathbf{v})}{4}\overline{\partial}\overline{J}^{\alpha}\partial_{\alpha} \right)=e^{\hat{L}}+\delta_{\zeta}(\tilde{\alpha})\wedge e^{\hat{L}}~,
\end{eqnarray}
where
\begin{eqnarray}
\tilde{\alpha}=\delta \mathbf{v}\frac{\zeta^{-1}}{4}\overline{J}^{\alpha}\partial_{\alpha}~.
\end{eqnarray}
Now, since all the dependence on $\mathbf{v}$ is contained in $\hat{L}$, it follows that
\begin{eqnarray}
\langle \mathcal{O}(\omega)\rangle_{S^{2}}|_{\mathbf{v}+\delta\mathbf{v}}&=&\langle \mathcal{O}(\omega)\rangle_{S^{2}}|_{\mathbf{v}}+\langle\delta_{\zeta}(\mathcal{O}(\tilde{\alpha}))\wedge e^{\hat{L}}\mathcal{O}(\omega)\rangle'_{S^{2}}\nonumber\\
&=&\langle \mathcal{O}(\omega)\rangle_{S^{2}}|_{\mathbf{v}}+\langle\delta_{\zeta}(\mathcal{O}(\tilde{\alpha})\wedge e^{\hat{L}}\mathcal{O}(\omega))\rangle'_{S^{2}}\nonumber\\
&=&\langle \cO(\omega) \rangle_{S^{2}}|_{\mathbf{v}}~,
\end{eqnarray}
where the last equality follows from \eqref{eq:indreprs}.

Next, we are going to show that $\langle \mathcal{O}(\omega)\rangle_{S^{2}}$ is independent of variations $\delta g_{\alpha\bar{\beta}}$ of the metric on $\bY$
such that $\delta g_{\alpha\beta}=\delta g_{\bar{\alpha}\bar{\beta}}=0$, i.e.,
our B model correlators are independent of variations of the K\"ahler moduli of $\bY$, as expected.
Let, as before, $\omega\in PV^{s,s}(\bY)$ be a homogeneous polyvector field in $\delta_{\zeta}$-cohomology. Then
\begin{align}
\label{eq:corrindg}
\Omega_{\bY}\lrcorner(e^{\hat{L}}\omega )&=e^{-\frac{\mathbf{v}}{4}\| J\|^{2}}\zeta^{-r}\left(\frac{\mathbf{v}}{4}\right)^{r}\frac{(-1)^{rs+\frac{r(r-1)}{2}}}{r!} \nonumber\\
&\quad\times (\Omega_{\bY})_{\gamma_{1}\ldots \gamma_{d}}\overline{\partial}_{\betab_1}\overline{J}^{\gamma_{1}}\wedge\cdots\wedge\overline{\partial}_{\betab_{r}}\overline{J}^{\gamma_{r}}\omega^{\gamma_{r+1}\cdots \gamma_{d}}_{\bar{\beta}_{r+1}\cdots\bar{\beta}_{d}}d\bar{z}^{\bar{\beta}_{1}}\wedge\cdots\wedge d\bar{z}^{\bar{\beta}_{d}}~,
\end{align}
where $s=d-r$.
Moreover, notice that the only dependence of  $\hat{L}$ on the metric is in the form of the inverse metric $g^{\alpha \betab}$.
The variation $\delta g^{\alpha \bar{\beta}}$ in $\exp(\hat L)$ can be split in two pieces, namely the variation of the multiplicative exponent
\begin{eqnarray}
\delta(e^{-\frac{\mathbf{v}}{4}\| J\|^{2}})=-\frac{\mathbf{v}}{4}\delta g^{\alpha \bar{\beta}}\overline{J}_{\bar{\beta}}J_{\alpha}e^{-\frac{\mathbf{v}}{4}\| J\|^{2}}~,
\end{eqnarray}
and the variation of the $\overline{\partial}\overline{J}^{\alpha}\partial_{\alpha}$ factors
\begin{eqnarray}
\delta(\overline{\partial}\overline{J}^{\alpha}\partial_{\alpha})=\overline{\partial}(\delta g^{\alpha\bar{\beta}}\overline{J}_{\bar{\beta}})\partial_{\alpha}~.
\end{eqnarray}
The full variation of \eqref{eq:corrindg}, after a slightly lengthy computation, is given by
\begin{align}
\delta(\Omega_{\bY}\lrcorner(e^{\hat{L}}\omega ))&=\frac{\zeta^{-1}\mathbf{v}}{4}\Omega_{\bY}\lrcorner\left(e^{\hat{L}}\delta_{\zeta}(\Theta)\wedge\omega \right) \nonumber\\
&=\frac{\zeta^{-1}\mathbf{v}}{4}\Omega_{\bY}\lrcorner\left(e^{\hat{L}}\delta_{\zeta}(\Theta\wedge\omega) \right)~,
\end{align}
where
\begin{eqnarray}
\Theta:=\delta g^{\alpha\bar{\beta}}\overline{J}_{\bar{\beta}}\partial_{\alpha}\in PV^{(1,0)}(\bY)~.
\end{eqnarray}
Thus, the variation of the full correlator is
\begin{align}
\delta \la \cO(\omega) \ra_{S^2} = \frac{\zeta^{-1}\mathbf{v}}{4}  \la \delta_\zeta \cO(\Theta\wedge\omega) \ra_{S^2} =0~.
\end{align}

Finally, we want to comment on the independence of the parameters in $\Wb$.
While we expect $\langle \mathcal{O}(\alpha)\rangle_{S^{2}}$ to depend holomorphically on the superpotential parameters,
\eqref{eq:22S2corrs} is certainly not explicitly holomorphic.
Let us consider a variation
$\overline{J}\rightarrow\overline{J}+\delta \overline{J}$. A direct computation shows
\begin{eqnarray}
e^{\hat{L}}\big|_{\overline{J}+\delta \overline{J}}=e^{\hat{L}}+\delta_{\zeta}(\varpi)\wedge e^{\hat{L}}~,
\end{eqnarray}
where
 \begin{eqnarray}
\varpi=\frac{\zeta^{-1}\mathbf{v}}{4}\delta \overline{J}^{\alpha}\partial_{\alpha}\in PV^{(1,0)}(\bY)~.
\end{eqnarray}
Thus, this variation produces a $\delta_{\zeta}$-exact form, which gives a vanishing contribution to the correlator, and the
correlators do not depend on the anti-holomorphic parameters in $\Wb$.

\subsection{Reduction to integral over $B$ and residue formula}

In this section we are going to analyze more closely the integral arising in the $S^{2}$ correlators.
Our goal is to show that such integral reduces to an integral over $B$, and we will compare it with the residue formula given in \cite{2015arXiv150802769C}.
Given the factor $\exp(-\mathbf{v}\| dW\|^2/4)$ in \eqref{eq:22S2corrs} and the assumption of polynomial growth along $X$,
the integration over the fiber coordinates is absolutely convergent. Thus, it should be possible to perform such integration
and obtain $\langle \mathcal{O}(\omega)\rangle_{S^{2}}$ as an integral over $B$.
In this section we will push this line of reasoning a bit further to derive a residue formula for \eqref{eq:22S2corrs}.

Let us consider $\omega=\sum_{p=0}^{d}\omega_{p}~,$ where $\omega_{p}\in PV^{p,p}(\bY)$,
as appropriate for a non-vanishing integral. This is an element in $\delta_{\zeta}$-cohomology if and only if
\begin{align}
\overline{\partial}\omega_{p}&=\zeta \iota_{J}\omega_{p+1}~,	&p&=0,\ldots,d-1~.
\end{align}
Let us define the collection of functions
\begin{eqnarray}
f_{r}:=(-1)^{r}\frac{d^{r-1}}{d a^{r-1}}a^{-1}e^{-a\frac{\mathbf{v}}{4}\| J \|^{2}}\Big|_{a=1}~,
\end{eqnarray}
for $r\in\Z_{>0}$. These satisfy the following properties
\begin{align}
\overline{\partial}f_{r}&=\left(\frac{\mathbf{v}}{4}\right)^{r}e^{-\frac{\mathbf{v}}{4}\| J \|^{2}}\| J \|^{2r-2}\overline{\partial}(\| J \|^{2})~,
&f_{r+1}&=rf_{r}-\left(\frac{\mathbf{v}}{4}\right)^{r}\| J \|^{2r}e^{-\frac{\mathbf{v}}{4}\| J \|^{2}}~.
\end{align}
Let us also define the collection of polyvector fields $A_{s-1}\in PV^{s,s-1}(\bY)$ by
\begin{eqnarray}
A_{s-1}:=\left(\frac{\overline{J}^{\alpha}\partial_{\alpha}}{\| J \|^{2}}\right)\left(\frac{\overline{\partial}\overline{J}^{\alpha}\partial_{\alpha}}{\| J \|^{2}}\right)^{s-1}~,
\end{eqnarray}
for $s=1,\dots,d-1$, which instead satisfy
\begin{align}
\overline{\partial}A_{s-1}&=\iota_{J}A_{s}~,		&\overline{\partial}A_{d-1}&=0~.
\end{align}
It is possible to show that
\begin{align}
\left(\frac{\mathbf{v}}{4}\right)^{p}\frac{1}{p}e^{-\frac{\mathbf{v}}{4}\| J \|^{2}}(\overline{\partial}\overline{J}^{\alpha}\partial_{\alpha})^{p}=\overline{\partial}f_{p}A_{p-1}+\left(\frac{\mathbf{v}}{4}\right)^{p}\frac{1}{p}
e^{-\frac{\mathbf{v}}{4}\| J \|^{2}}\| J \|^{2p}\overline{\partial}A_{p-1}~.
\end{align}
With this property we can write
\begin{align}
\left(\frac{\mathbf{v}}{4}\right)^{p}\frac{1}{p}e^{-\frac{\mathbf{v}}{4}\| J \|^{2}}(\overline{\partial}\overline{J}^{\alpha}\partial_{\alpha})^{p}\omega_{d-p}&=
\overline{\partial}\left(f_{p}A_{p-1}\omega_{d-p}\right)-f_{p}\overline{\partial}A_{p-1}\omega_{d-p}\nonumber\\
&\quad+f_{p}A_{p-1}\overline{\partial}\omega_{d-p}+\left(\frac{\mathbf{v}}{4}\right)^{p}\frac{1}{p}
e^{-\frac{\mathbf{v}}{4}\| J \|^{2}}\| J \|^{2p}\overline{\partial}A_{p-1}\omega_{d-p}\nonumber\\
&=\overline{\partial}\left(f_{p}A_{p-1}\omega_{d-p}\right) -\frac{1}{p}f_{p+1}\iota_{J}A_{p}\omega_{d-p}\nonumber\\
&\quad+\zeta f_{p}A_{p-1}\iota_{J}\omega_{d-p+1}~.
\end{align}
Now, let us define
\begin{align}
B_{p}&:=-\frac{\zeta^{-p}}{p!}f_{p+1}\iota_{J}A_{p}\omega_{d-p}~, 	&C_{p}&:=\frac{\zeta^{-p+1}}{(p-1)!}f_{p}A_{p-1}\iota_{J}\omega_{d-p+1}~,
\end{align}
for $p=0,\ldots,d$, with $C_{0}:=0$ and $B_{d}:=0$. It is possible to show that\footnote{Here we use the following property:
given $\alpha\in PV^{p,q}(\bY)$ and $\beta\in PV^{r,s}(\bY)$ in an explicit basis 
in terms of fermions, $\alpha ^{[p]}_{[q]}\bar{\eta}^{[q]}\bar{\chi}_{[p]}$, etc., it is easy to see that 
$\iota_{J}(\alpha \beta)=\iota_{J}(\alpha)\beta+(-1)^{p+q}\alpha\iota_{J}\beta$.}
\begin{eqnarray}
B_{p}+C_{p+1}=0~,\qquad  p=0,\dots,d-1~.
\end{eqnarray}
Putting all together, we obtained the expression
\begin{eqnarray}
e^{\hat{L}}\omega=\overline{\partial}\left(\sum_{p=1}^{d}\frac{\zeta^{-p}}{(p-1)!}f_{p}A_{p-1}\omega_{d-p}\right)~.
\end{eqnarray}
Thus, a generic correlator can be written as
\begin{eqnarray}
\label{eq:corrsoverB}
\langle \mathcal{O}(\omega)\rangle_{S^{2}}=\int_{\bY\setminus B}d\left(\Omega_{\bY}\wedge\left(\Omega_{\bY}\lrcorner \Xi(\omega)\right)\right)~,
\end{eqnarray}
where the polyvector
\begin{eqnarray}	
\Xi(\omega):=\sum_{p=1}^{d}\frac{\zeta^{-p}}{(p-1)!}f_{p}A_{p-1}\omega_{d-p}\in PV^{d,d-1}(\bY)
\end{eqnarray}
is meromorphic, and \eqref{eq:corrsoverB} is an integral over $\bY\setminus B$,
where $B$ is the zero section of $\bY$ and is the singular loci of $\Xi$.
By Stokes' theorem, the integral reduces to an integral over $M=\partial(\bY\setminus B)$,
where $M$ has the structure of a fiber bundle over $B$, say $M:S\rightarrow B$, and the fiber is the sphere $S=S^{2d-2b-1}$. Hence
\begin{eqnarray}
\label{eq:residueform}
\langle \mathcal{O}(\omega)\rangle_{S^{2}}=\int_{B}\pi_{*}(\Omega_{\bY}\wedge(\Omega_{\bY}\lrcorner \Xi(\omega)))~,
\end{eqnarray}
where $\pi_{*}(\Omega_{\bY}\wedge(\Omega_{\bY}\lrcorner \Xi(\omega)))$ is defined by integration over the fiber coordinates by
\begin{eqnarray}
\pi_{*}(\Omega_{\bY}\wedge(\Omega_{\bY}\lrcorner \Xi(\omega)))\big|_{p\in B}=\int_{\pi^{-1}(p)}\Omega_{\bY}\wedge(\Omega_{\bY}\lrcorner \Xi(\omega))~.
\end{eqnarray}
We can interpret \eqref{eq:residueform} as our residue formula, and it coincides with the one presented in \cite{2015arXiv150802769C}.
As a consistency check we can see that in the case of a pure LG model, where $B=\mathrm{pt.}$~and $\bY=\C^d$,
\eqref{eq:residueform} reduces to an integral over $S^{2d-1}$. By a well-known theorem
in residue theory (see, for example, chapter 5 of \cite{GHbook}) \eqref{eq:residueform} can be
expressed as an integral of a holomorphic form in $\mathbb{C}^{d}\setminus \cup_{\alpha}\{J_{\alpha}=0\}$
over the torus $\{|J_{\alpha}|=\varepsilon_{\alpha}\}$ (recall that, for a LG model, the $\delta_{\zeta}$-cohomology
collapses to degree $(0,0)$, i.e., the relevant $\omega$'s are just holomorphic functions).
This is exactly the formula derived in \cite{Vafa:1990mu}. A detailed derivation of this can also be found in \cite{Li:2013kja},
where our function $f_{d}$ at large $\mathbf{v}$ plays the role of the cut-off function introduced in that work.

As a final comment, we were not able to reduce our formula to an integration over a cycle in $\bY\setminus B$ (and neither are the authors of \cite{2015arXiv150802769C}).
More importantly perhaps, it seems challenging to implement \eqref{eq:residueform} for explicit calculations.
For the purpose of actually finding the value of the integrals in specific examples,
or to be more precise, the dependence of the correlators on the parameters in $W$, we present in the next section a proposal for a transformation law.

\subsection{A transformation law for hybrid integrals}
\label{ss:transflaw}

The transformation law for local residues is well known (see chapter 5 of \cite{GHbook}).
This property is very useful for computing correlators in $\mathcal{N}=(2,2)$ LG models,
since they can be expressed in terms of residue integrals \cite{Vafa:1990mu}.
Let us review how this transformation law can be used to compute $S^{2}$ B-twisted
correlators in LG models, from the formula of \cite{Vafa:1990mu}.
Let us consider the LG model $(\mathbb{C}^{n},W_{\text{LG}})$ (this considerations
also apply to LG orbifolds $(\mathbb{C}^{n},W_{\text{LG}},\Gamma)$ where $\Gamma$ is a finite group),
where $\phi_1,\dots,\phi_n$ are coordinates on $\C^n$.
Then, there exists a $n\times n$ matrix $\mathcal{B}_{i}^{ \ j}$, whose entries are polynomials in the $\phi_i$'s, such that
\begin{align}
\mathcal{B}_{i}^{ \ j}J_{j}&=T_{i}~,		&T_{i}&:=\phi_{i}^{a_{i}}~,\qquad a_{i}\in\mathbb{N}_{> 0}~,
\end{align}
where we defined $J_{j}:=\partial_{j}W_{LG}$. The matrix $\mathcal{B}$ is not unique (and neither are the integers $a_{i}$),
but the residue is not affected by this choice. By construction, $\mathcal{B}$
contains all the dependence on the parameters in $W_{\text{LG}}$, which we collectively denote $\xi$. Then we can write
\begin{eqnarray}\label{expandB}
\mathrm{det}(\mathcal{B})=\sum_{t}h_{t}(\xi)m_{t}~,
\end{eqnarray}
where $h_{t}(\xi)$ are rational functions in $\xi$ and $m_{t}$ are monomials in the $\phi_i$'s.
The local operators spanning the (c,c) ring of the theory  are given by $\C[\phi_1,\dots,\phi_n]/<J>$,
where $<J>$ is the Jacobian ideal \cite{Lerche:1989uy}.
Let $\mathcal{O}(\omega)\in PV^{0,0}(\mathbb{C}^{n})$
be an element of the (c,c) ring.
Then, using the local transformation law, LG $S^{2}$ correlators can be written as
\begin{eqnarray}
\langle\mathcal{O}(\omega)\rangle^{LG}_{S^{2}}=\sum_{t}h_{t}(\xi)\mathrm{Res}\left( \frac{m_{t}\omega}{\prod_{j}T_{j}}\right)=\sum_{t}h_{t}(\xi)\langle \mathcal{O}(m_{t}\omega)\rangle^{T}_{S^{2}}~,
\end{eqnarray}
where we denoted $\langle \mathcal{O}(m_{t}\omega)\rangle^{T}_{S^{2}}$ the correlators obtained by replacing $J_{i}$ with $T_{i}$.
These do not contain any dependence on $\xi$ and are straightforward to evaluate.
The proof of this transformation property for the residue formula, given in \cite{GHbook}, relies on the
property that the residue integral is explicitly expressible as a holomorphic integral over a cycle.
As we pointed out above, we do not know how to generalize this property to the hybrid case,
despite the fact that formally the hybrid correlator has no dependence on the
anti-holomorphic parameters.
However, we conjecture that a similar property also holds in the more general hybrid setting.
In the remainder of this section we are going to provide
a non-rigourous argument in favor of the existence of such a property for the general hybrid $S^{2}$ correlator.

Let $\omega\in PV^{p,p}(\bY)$,
and we do not need to require that $\omega$ is an element in $\delta_{\zeta}$-cohomology for the following argument to hold.
This specific form of the insertion selects one term in the expansion \eqref{eq:expexpLh}, and the correlator \eqref{eq:22S2corrs} reads
\begin{eqnarray}
\label{eq:corrtranflaw1}
\langle\mathcal{O}(\omega)\rangle_{S^{2}}=\left(\frac{\mathbf{v}\zeta^{-1}}{4}\right)^{d-p}\int_{\bY}\Omega_{\bY}\wedge\left(\Omega_{\bY}\lrcorner\left(e^{-\frac{\mathbf{v}}{4}\| J \|^{2}}  (\overline{\partial}\overline{J}^{\alpha}
\partial_{\alpha})^{d-p}\omega\right)\right)~.
\end{eqnarray}
Let us now assume that there exists $\mathcal{B}\in \Gamma\left(\Hom(T_{\bY}^{*},T_{\bY}^{*})\right)=\Gamma\left(T_{\bY}\otimes T_{\bY}^{*}\right)$ such that
\begin{eqnarray}
T_{\alpha}=\mathcal{B}_{\alpha}^{ \ \beta}J_{\beta}\in\Gamma(T^\ast_{\bY})
\end{eqnarray}
does not depend on the parameters $\xi$ of $W$ and $B=\{T_{\alpha}=0\}\subset \bY$,
that is the potential condition holds for $T$ as well.
We can interpret the inverse of the K\"ahler metric $g^{-1}$ to be a section of $\Hom(\overline{T}^{*}_{\bY},T_{\bY})$ and write
\begin{eqnarray}
\| J \|^{2}=J(g^{-1}\overline{J})~.
\end{eqnarray}
Let us define the metric $h$ on $\bY$ by
\begin{eqnarray}
\label{eq:defhmetric}
g^{-1}=\mathcal{B}^{t}h^{-1}\overline{\mathcal{B}}~,
\end{eqnarray}
where $\mathcal{B}^{t}\in\Gamma\left(\mathrm{Hom}(T_{\bY},T_{\bY})\right)$ denotes the transpose of $\cB$
(and $\overline{\cB}\in\Gamma\left(\Hom(\Tb^\ast_{\bY},\Tb^\ast_{\bY})\right)$). We can then express
the operator $\hat{L}$ in terms of the metric $h$ as
\begin{eqnarray}
\hat{L}=-\frac{\mathbf{v}}{4}\| T \|^{2}_{h}+\mathbf{v}\frac{\zeta^{-1}}{4}(\mathcal{B}^{t})^{\beta}_{ \ \alpha}\overline{\partial}\overline{T}^{\alpha}\partial_{\beta}~,
\end{eqnarray}
where we emphasized that $\| J \|^{2}=\| T \|^{2}_{h}$ is contracted using the metric $h$ and $ \overline{T}^{\alpha}=h^{\alpha\bar{\alpha}}\overline{T}_{\bar{\alpha}}$.
Since $\mathcal{B}^{t}\in\Gamma(T^\ast_{\bY}\otimes T_{\bY})$, the wedge product $\mathcal{B}^{t}\wedge \mathcal{B}^{t}$
is well-defined, as is the tensor
\begin{eqnarray}
\mathcal{M}(\mathcal{B}):=\Omega_{\bY}\lrcorner \underbrace{\mathcal{B}^{t}\wedge\ldots\wedge\mathcal{B}^{t}}_{(d-p)\text{-times}}\in \Gamma(\bY,\wedge^{d}T^{*}_{\bY})~.
\end{eqnarray}
Finally, by plugging in these definitions in \eqref{eq:corrtranflaw1} we obtain
\begin{eqnarray}
\label{minorcorr}
\langle\mathcal{O}(\omega)\rangle_{S^{2}}=\left(\frac{\mathbf{v}\zeta^{-1}}{4}\right)^{d-p}\int_{\bY}\Omega_{\bY}\wedge\left(\mathcal{M}(\mathcal{B})\lrcorner\left(e^{-\frac{\mathbf{v}}{4}\| T\|^{2}}  (\overline{\partial}\overline{T}^{\alpha}
\partial_{\alpha})^{d-p}\omega\right)\right)~.
\end{eqnarray}
Although the above expression appears to be merely a rewriting of the original correlator \eqref{eq:corrtranflaw1}, notice that all the $\xi$ dependence
is now contained in the metric $h$ and the tensor $\cM(\cB)$.  Next, we wish to make use of the property that
the $S^2$ correlators do not depend on variations of the K\"ahler metric, as we showed above. This implies that we can choose $g$ in \eqref{eq:defhmetric}
such that $h$ has no dependence on $\xi$. In particular, we can choose $h$ to be diagonal and constant over $\bY$.
The tensor $\mathcal{M}(\mathcal{B})$ can be written in terms of minors of $\mathcal{B}$.
Let us make this more explicit. Let us define the matrix $\mathcal{M}_{\alphab}^{ \ \beta}:=\overline{\partial}_{\alphab}\overline{J}^{\beta}$
and let $\mathcal{M}^{\alphab_{1},\dots,\alphab_{p}}_{\beta_{1},\ldots,\beta_{p}}$ be the completely skew-symmetric tensor
where each component is given by the determinant of the minor of $\mathcal{M}$ obtained by removing
the columns $\bar{\alpha}_{1},\dots,\bar{\alpha}_{p}$ and the rows $\beta_{1},\ldots,\beta_{p}$.
This can be expressed as a sum of products
\begin{eqnarray}
\label{eq:Mminasdets}
\mathcal{M}^{\bar{\alpha}_{1},\dots,\bar{\alpha}_{p}}_{\beta_{1},\ldots,\beta_{p}}=\sum_{[\gamma]}(\overline{\partial}\overline{T})^{\bar{\alpha}_{1},\dots,\bar{\alpha}_{p}}_{\gamma_{1},\ldots,\gamma_{p}}(\mathcal{B}^{t})^{\gamma_{1},\ldots,\gamma_{p}}_{\beta_{1},\ldots,\beta_{p}}~.
\end{eqnarray}
In a patch, we can write $(\Omega_{\bY})_{\alpha_{1},\ldots,\alpha_{d}}=f_{\bY}\varepsilon_{\alpha_{1},\ldots,\alpha_{d}}$, with $f_{\bY}$ a nowhere vanishing holomorphic function. Then (\ref{minorcorr}) can be written more explicitly as
\begin{eqnarray}\label{xidepen}
\langle\mathcal{O}(\omega)\rangle_{S^{2}}=\left(\frac{\mathbf{v}\zeta^{-1}}{4}\right)^{d-p}\sum_{[\gamma]}\int_{\bY}d^{2d}x f_{\bY}^{2}e^{-\frac{\mathbf{v}}{4}\| T\|^{2}} (\mathcal{B}^{t})^{\gamma_{1},\ldots,\gamma_{p}}_{\beta_{1},\ldots,\beta_{p}}(\overline{\partial}\overline{T})^{\bar{\alpha}_{1},\dots,\bar{\alpha}_{p}}_{\gamma_{1},\ldots,\gamma_{p}}
\omega^{\beta_{1},\ldots,\beta_{p}}_{\bar{\alpha}_{1},\dots,\bar{\alpha}_{p}}~.
\end{eqnarray}
The determinants $(\mathcal{B}^{t})^{\gamma_{1},\ldots,\gamma_{p}}_{\beta_{1},\ldots,\beta_{p}}$
can be expanded, similarly to \eqref{expandB}, as
\begin{align}
(\mathcal{B}^{t})^{\gamma_{1},\ldots,\gamma_{p}}_{\beta_{1},\ldots,\beta_{p}} = \sum_t h_t(\xi) \mathsf{M}_t~,
\end{align}
where $\mathsf{M}_t$ are monomials in the local coordinates of $\bY$,
and the coefficients $h_t(\xi)$ are rational functions of $\xi$.
Thus, (\ref{xidepen}) is expressed as a sum
\begin{eqnarray}\label{eq:xidepen2}
\langle\mathcal{O}(\omega)\rangle_{S^{2}}= \sum_t h_t(\xi) \la \mathsf{M}_t\cO(\omega) \ra_{S^2}^T~,
\end{eqnarray}
where each of the correlators $\la \mathsf{M}_t\cO(\omega) \ra_{S^2}^T$ is computed with respect to $T$ and do not depend on the parameters $\xi$.
We have arrived, at least formally, at an analogous situation as in the LG case above.

We conclude by commenting where our argument fails to provide a rigorous proof of the conjectured transformation law.
Despite the fact that \eqref{minorcorr} appears to be simply a rewriting of \eqref{eq:corrtranflaw1},
it implicitly assumes that the section $\cB$ exists and that the map \eqref{eq:defhmetric} is invertible, but
$\mathrm{det}(\mathcal{B})$ is in general a non-trivial function on $\bY$, hence it can vanish.
Nevertheless, we believe that a rigorous proof should exist, and that the argument presented here is sufficiently close to it for our purposes.

\section{Main example: the octic hybrid}
\label{s:22example}

In this section we apply the techniques developed above to a non-trivial example.
Let us consider the hybrid model defined by the geometric data $\bY=\tot\left(\cO(-2)\oplus\cO^{\oplus3}\rightarrow \CP^1\right)$
and $\C^\ast_V$-action with weights $q_i=\frac14$, $i=1,\dots,4$, acting on the fiber coordinates and fixing the $\CP^1$ base.
The orbifold action $\Gamma=\Z_4$ introduces 3 additional twisted sectors, which will not be relevant for our analysis.
The most general superpotential compatible with this structure is given by
\begin{align}
W = \sum_{t=0}^4 S_{[2i]}F_{[4-t]}(\phi^{a})(\phi^1)^t~,
\end{align}
where $S_{[p]}\in  H^0(\CP^1, \cO(p))$ and $F_{[q]}$ is a homogeneous polynomial of degree $q$ in the variables $\phi^a$, $a=2,3,4$.
This model inherits its name from the geometric phase of the corresponding 2-parameter GLSM, which
in the large radius phase describes an octic hypersurface in the toric resolution of $\CP^{4}_{11222}$.

Specifically, we choose a one-dimensional parameter subspace of the above superpotential
\begin{align}
\label{eq:22supoct}
W&=\frac{1}{8}(x_1^8+x_2^8)(\phi^1)^4 + \frac{1}{4}(\phi^2)^4+\frac{1}{4}(\phi^3)^4+\frac{1}{4}(\phi^4)^4-\psi x_1x_2\phi^1\phi^2\phi^3\phi^4~,		
\end{align}
where $[x_1:x_2]$ are homogeneous coordinates on $B=\CP^1$. Let $u=x_2/x_1$ and $v=x_1/x_2$ be local coordinates on the standard cover $U_1=\{x_1\neq0\}$
and $U_2=\{x_2\neq0\}$ respectively. Just in this section we slightly alter our notation to $\alpha=0,\dots,d-1=4$, so that the base index assumes the value $I=0$
and the fiber coordinates have indices $i=1,\dots,4$.
Thus, in the $U_1$ patch we have the (0,2) superpotential\footnote{This terminology will become clear when we study (0,2) hybrids. Here $J^{u,v}_{i}:=\partial_{i}W|_{u,v}$
for the fiber coordinates, while $J^{u}_0=\partial_{u}W|_{u}$ and $J^{v}_0=\partial_{v}W|_{v}$ for the local base coordinates.}
\begin{align}
\label{eq:02supoctu}
J_0^u&=u^7(\phi^1_u)^4-\psi \phi_u^1\phi_u^2\phi_u^3\phi_u^4~,	&J_1^u&=\half(u^8+1)(\phi_u^1)^3-\psi u \phi_u^2\phi_u^3\phi_u^4~,\nonumber\\
J_2^u&= (\phi_u^2)^3 - \psi u\phi_u^1\phi_u^3\phi_u^4~,			&J_3^u&= (\phi_u^3)^3 - \psi u\phi_u^1\phi_u^2\phi_u^4~,	
&J_4^u&= (\phi_u^4)^3 - \psi u\phi_u^1\phi_u^2\phi_u^3~,
\end{align}
and similarly in the $U_2$ patch
\begin{align}
\label{eq:02supoctv}
J_0^v&=v^7(\phi_v^1)^4-\psi \phi_v^1\phi_v^2\phi_v^3\phi_v^4~,	&J_1^v&=\half(v^8+1)(\phi_v^1)^3-\psi v \phi_v^2\phi_v^3\phi_v^4~,\nonumber\\
J_2^v&= (\phi_v^2)^3 - \psi v\phi_v^1\phi_v^3\phi_v^4~,			&J_3^v&= (\phi_v^3)^3 - \psi v\phi_v^1\phi_v^2\phi_v^4~,	
&J_4^v&= (\phi_v^4)^3 - \psi v\phi_v^1\phi_v^2\phi_v^3~.
\end{align}
Here, $\phi^i_{u,v}=\phi^i|_{U_{1,2}}$ indicate the restrictions to the respective patches of the appropriate sections over the base, which transform according to
\begin{align}
\phi_u^1 &= v^2\phi_v^1~,		&\phi_u^a = \phi_v^a~.
\end{align}
It is then easy to verify that \eqref{eq:02supoctu} and \eqref{eq:02supoctv} transform as a section of $T^\ast_{\bY}$.  In particular, we can make
use of the fact that the splitting of the geometry $\bY=\cO^{\oplus3}\oplus\bY'$, where $\bY'=\tot\left( \cO(-2)\rightarrow \CP^1\right)$,
induces a similar splitting of the cotangent bundle
\begin{align}
T^\ast_{\bY}=\cO^{\oplus3}\oplus T^\ast_{\bY'}~,
\end{align}
that is,
\begin{align}
\begin{pmatrix} J_0^u \\ J^u_1 \end{pmatrix} &=
\begin{pmatrix}
-v^{2} & 2v \phi^{1}_v\\
0 & v^{-2}
\end{pmatrix}
\begin{pmatrix} J_0^v \\ J_1^v \end{pmatrix}~,		&J_a^u&=J_a^v~.
\end{align}

\subsection*{The chiral ring}

The dimension of the chiral ring for the $\mathbb{Z}_4$ orbifold of this hybrid model has been computed in \cite{Bertolini:2013xga}.
However, for our purposes, we need an explicit representation of the elements of the ring.
In this section we will achieve this by following the prescription outlined in section \ref{s:chiralring}.
Before we delve into the computation, we make a couple of observations which will simplify considerably our task.
First, the GSO projection is onto integral charges, $\bq\in\Z$, and unitarity bounds further restrict $0\leq \bq\leq c/3=3$.
Since we can compute the spectral sequence at fixed value of $\bq$,
we only need to analyze the cases $\bq=0,1,2,3$.
Second, it is well known that the dimension of the ring does not depend on the particular form of the superpotential
as long as it does not lead to a singular model. In particular, such choice only affects the
representatives of each cohomology class. Since, as we showed above, our formula
is independent of representatives, we can compute the correlators for any non-singular $W$.
The simplest choice is to set $\psi=0$ in \eqref{eq:02supoctu}. With this set-up, we are ready to describe the elements of the ring.

At $\bq=0$ the spectral sequence is trivial except at $H^0_{\bq=0}(\bY,\cO_{\bY})=\C$. This operator has charges $(0,0)$
and therefore it can be interpreted as the identity element in the ring.

At $\bq=1$, the spectral sequence at the first stage is
\begin{equation}
\label{eq:statesoctq1}
\begin{matrix}\vspace{5mm}\\E_1^{r,s} :\end{matrix}
\begin{xy}
\xymatrix@C=10mm@R=5mm{
\C^3		&0\\
\C^{22}	\ar[r]^-{\bQb_W}	&\C^{105}
}
\save="x"!LD+<-6mm,0pt>;"x"!RD+<20pt,0pt>**\dir{-}?>*\dir{>}\restore
\save="x"!LD+<28mm,-3mm>;"x"!LU+<28mm,2mm>**\dir{-}?>*\dir{>}\restore
\save!RD+<-7mm,-3mm>*{0}\restore
\save!RD+<-25mm,-3mm>*{1}\restore
\save!RD+<03mm,-3mm>*{r}\restore
\save!CL+<30mm,10mm>*{s}\restore
\end{xy}
\end{equation}
To show how we obtained this, let us look first at the bottom row ($s=0$).
A generic section of $H^0_{\bq=1}(\bY,T_{\bY})$ is parametrized by the operators
\begin{align}
&\begin{pmatrix}
T_u \chib_0 & Y_u\phi^1 \chib_1
\end{pmatrix}~,		&&c_{ab}\phi^b\chib_a~,	&&S^a_{[2]}\phi^1\chib_a~,
\end{align}
where in the patch $U_1$ we have (see appendix \ref{app:sectbundles})
\begin{align}
T_u&=a_0+a_1u+a_2u^2~,
&Y_u&=(b_0-2a_1)  -2a_2u ~,	&S_{[2]}&=d_0+d_1u+d_2u^2~,
\end{align}
and $c_{ab}\in\C$. Counting the number of parameters
\begin{align}
\#a + \#b + \#c + \#d = 3+1+9+3\cdot3=22~,
\end{align}
leads to the value in \eqref{eq:statesoctq1}.
Sections of $H^0_{\bq=1}(\bY,\cO_{\bY})$ are simply
\begin{align}
S_{[2t]}F_{[4-t]}(\phi^a)(\phi^1)^t~,		\qquad\quad t=0,\dots,4~,
\end{align}
where $S_{[2t]}\in H^0(B,\cO(2t))$ and $F_{[d]}(\phi^a)$ is a generic polynomial of degree $d$ in $\phi^a$. It is easy to count those, as
\begin{align}
\sum_{t=0}^4 (2t+1) \binom{6-t}{4-t} = 105~.
\end{align}
The first row ($s=1$) is much simpler, and the only non-vanishing $\pb$-closed polyvectors $PV^{r,1}(\bY)$ have the form
\begin{align}
\label{eq:firstrowq1}
S_{[-2],\overline0}^a \phi^a\chib_1\etab^{\overline0}~,
\end{align}
where
\begin{align}
S_{[-2],\overline0}^a\in H^1_{\bq=0}(\bY,\pi^\ast\cO(-2))=H^1(B,\cO(-2))=\C~,
\end{align}
which therefore yield three elements in $\pb$-cohomology at $r=s=1$.
To compute the second stage of the spectral sequence,
for the bottom row we can write explicitly the action of $\bQb_W$ on the states \eqref{eq:statesoctq1} as
\begin{align}
\label{eq:actionQWq1}
\bQb_W \begin{pmatrix}
T_u \chib_0 & Y_u\phi^1 \chib_1
\end{pmatrix}&=
\begin{pmatrix}
T_u & Y_u\phi^1
\end{pmatrix}
\begin{pmatrix} J_0 \\ J_1 \end{pmatrix}  = \half \left[ (b_0-2a_1)-2a_2u  +2a_0 u^7 +b_0 u^8  \right] (\phi^1)^4~, \nonumber\\
\bQb_W(c_{ab}\phi^b+S_{[2]}\phi^1)\chib_a &=(c_{ab}\phi^b+S_{[2]}\phi^1)J_a = (c_{ab}\phi^b+S_{[2]}\phi^1) (\phi^a)^3 ~.
\end{align}
While it is easy to verify that this map has a trivial kernel, we would like to have explicit representatives of its cokernel. As mentioned
 above, these will be good representatives for the cohomology classes even after we turn on the $\psi$ deformation.
First, the operators
\begin{align}
\label{eq:cokerbot}
S_{[2t_1]}(\phi^1)^{t_1}(\phi^2)^{t_2}(\phi^3)^{t_3}(\phi^4)^{t_4}~,	\qquad	t_1\leq3,\ t_{2,3,4}\leq2~,
\end{align}
are clearly not in the image of \eqref{eq:actionQWq1}. Since $t_1+t_2+t_3+t_4=4$, we count 78 of them.
When $t_1=4$, a little more carefulness is required, and the cokernel of the first map in \eqref{eq:actionQWq1} can be parametrized by the operators
\begin{align}
u^t (\phi^1)^4~,		\qquad 2\leq t\leq 6~,
\end{align}
accounting for additional 5 operators, which give rise to the second stage of the spectral sequence
\begin{equation}
\label{eq:E2q122oct}
\begin{matrix}\vspace{5mm}\\E_2^{r,s} :\end{matrix}
\begin{xy}
\xymatrix@C=10mm@R=5mm{
\C^3		&0\\
0		&\C^{83}
}
\save="x"!LD+<-6mm,0pt>;"x"!RD+<20pt,0pt>**\dir{-}?>*\dir{>}\restore
\save="x"!LD+<28mm,-3mm>;"x"!LU+<28mm,2mm>**\dir{-}?>*\dir{>}\restore
\save!RD+<-5mm,-3mm>*{0}\restore
\save!RD+<-21mm,-3mm>*{1}\restore
\save!RD+<7mm,-3mm>*{r}\restore
\save!CL+<30mm,10mm>*{s}\restore
\end{xy}
\end{equation}
However, we are not quite done yet, as the first row operators \eqref{eq:firstrowq1} are manifestly not $\bQb_W$-closed.
This can be fixed by requiring
\begin{align}
\label{eq:firstrowq1clos}
\bQb\left(S^a_{[-2]\overline0} \phi^a \chib_1 \etab^{\overline0} + R^a \phi^a J_1 \right)=S^a_{[-2]\overline0} \phi^a J_1 \etab^{\overline0} + \pb_{\ub} R^a \phi^a J_1\etab^{\overline0}=0~,
\end{align}
which has solution when $R^a\in \Gamma(\cO_{\bY})$ such that $\pb_{\ub} R^a= -S^a_{[-2]\overline0}$.
Explicitly, the representatives for the cohomology classes corresponding to the $PV^{1,1}$ operators \eqref{eq:firstrowq1} are given by
\begin{align}
\label{eq:firstrowins1}
- {1 \over (1+u \ub)^2} \phi^a \chib_1 \etab^{\overline0} -\half {u^7-\ub \over  1+u\ub}(\phi^1)^3\phi^a~, \qquad a=2,3,4~.
\end{align}
The states \eqref{eq:E2q122oct} have a geometrical interpretation
when the $\Z_4$-orbifold of the hybrid theory is employed for heterotic or type II compactifications.\footnote{From a spacetime
point of view, these consist of the internal part of the vertex operators associated to the emission
of a scalar field in the $\mathbf{10}_{1}$ component of the $\overline{\mathbf{27}}$ of $\text{E}_6$.} These 86 states correspond to
complex structure deformations of the K3-fibered CY manifold obtained by blowing up the $c=\cb=6$ LG fiber. The 83 states from the bottom row correspond to
polynomial deformations, that is, deformations of the equation defining the hypersurface. The 3 extra states \eqref{eq:firstrowins1} correspond to
non-polynomial complex structure deformations \cite{Aspinwall:2010ve,Baumgartl:2012uh}, that is, deformations that are not related to parameters
in the action \eqref{eq:compaction}. This fact makes them hard to study with GLSM techniques, as one lacks the UV description for those, or
with LG techniques, where these parameters appear in twisted sectors. In our hybrid model, while technically slightly more challenging due to the $PV^{1,1}$ component,
they appear essentially at the same footing as the polynomial deformations, and we are able to compute correlators in the ring including these operators.

For the operators at $\bq=2$ we have
at the first stage
\begin{equation}
\begin{matrix}\vspace{5mm}\\E_1^{r,s} :\end{matrix}
\begin{xy}
\xymatrix@C=10mm@R=5mm{
\C^{18}	\ar[r]^-{\bQb_W}	&\C^3		&0\\
\C^{126}	\ar[r]^-{\bQb_W}	&\C^{868}		\ar[r]^-{\bQb_W} &\C^{825}
}
\save="x"!LD+<-6mm,0pt>;"x"!RD+<20pt,0pt>**\dir{-}?>*\dir{>}\restore
\save="x"!LD+<28mm,-3mm>;"x"!LU+<28mm,2mm>**\dir{-}?>*\dir{>}\restore
\save!RD+<-46mm,-3mm>*{2}\restore
\save!RD+<-26mm,-3mm>*{1}\restore
\save!RD+<-6mm,-3mm>*{0}\restore
\save!RD+<03mm,-3mm>*{r}\restore
\save!CL+<30mm,10mm>*{s}\restore
\end{xy}
\end{equation}
The second stage of this spectral sequence has been computed in \cite{Bertolini:2013xga} and we simply report it here
\begin{equation}
\begin{matrix}\vspace{5mm}\\E_2^{r,s} :\end{matrix}
\begin{xy}
\xymatrix@C=10mm@R=5mm{
\C^3		&0\\
0		&\C^{83}
}
\save="x"!LD+<-6mm,0pt>;"x"!RD+<20pt,0pt>**\dir{-}?>*\dir{>}\restore
\save="x"!LD+<28mm,-3mm>;"x"!LU+<28mm,2mm>**\dir{-}?>*\dir{>}\restore
\save!RD+<-21mm,-3mm>*{1}\restore
\save!RD+<-5mm,-3mm>*{0}\restore
\save!RD+<5mm,-3mm>*{r}\restore
\save!CL+<30mm,10mm>*{s}\restore
\end{xy}
\end{equation}
which agrees as expected with \eqref{eq:E2q122oct}.
Let us list the explicit representatives.
The bottom row consists again of states of the form \eqref{eq:cokerbot}, where now instead $t_1+t_2+t_3+t_4=8$,
yielding $26$ states. The missing elements can be identified as the cokernel of the map
\begin{align}
\bQb_W \begin{pmatrix}
T_u \phi^t \chib_0 & Y_u\phi^{t+1} \chib_1
\end{pmatrix}F_{[4-t]}(\phi^a)&=
\half \left[ (b_0-2a_1)+\cdots+(b_1-2a_{2t+1})u^{2t}-2a_{2t+2}u^{2t+1} \right.\nonumber\\
&\quad \ \left.+2a_0 u^7 +b_0 u^8+\cdots+b_{2t}u^{8+2t}  \right] (\phi^1)^{4+t}F_{[4-t]}(\phi^a)~,
\end{align}
for $t=0,\dots,4$, where $F_{[4-t]}$ has degree $4-t$ in the $\phi^a$ coordinates and is subject to the condition that each variable is allowed to appear to a power not greater than 2.
Explicitly,
\begin{align}
\xymatrix@R=0mm@C=10mm{
t		&\text{operator}							&					&\#\\
0		&u^{\tilde t} (\phi^1)^4F_{[4]}(\phi^a)~,		&2\leq {\tilde t}\leq 6~,	&30\\
1		&u^{\tilde t} (\phi^1)^5F_{[3]}(\phi^a)~,		& 4\leq {\tilde t}\leq 6~,	&21\\
2		&u^{\tilde t} (\phi^1)^6F_{[2]}(\phi^a)~,		&{\tilde t}=6~,			&6
}
\end{align}
while there are no states for $t=3,4$. In total, we produced 83 operators, as expected. The 3 operators from the first row can be written as
\begin{align}
\label{eq:firstrowins2}
- {1 \over (1+u \ub)^2} (\phi^a)^2(\phi^b)^2\phi^c \chib_1 \etab^{\overline0} -\half {u^7-\ub \over  1+u\ub}(\phi^1)^3(\phi^a)^2(\phi^b)^2\phi^c~,\qquad\quad a\neq b\neq c = 2,3,4~.
\end{align}
Finally, at $\bq=3$ we only find one operator in cohomology, which corresponds to $\det \Hess W$, and has the expression
\begin{align}
\label{eq:detHessW}
\det \Hess W = u^6 (\phi^1)^6(\phi^2)^2(\phi^3)^2(\phi^4)^2~.
\end{align}
The cohomology is empty for $\bq>3$, as expected from unitarity bounds, thus this concludes our discussion of the (untwisted) (c,c) ring.

\subsection{Correlators}

In this section we will completely solve the example by evaluating the map \eqref{eq:formula}.
As we have seen above, we can qualitatively distinguish between two types of elements in the (c,c) ring, that is
{\it bottom row} and {\it first row operators}.\footnote{We saw in \eqref{eq:firstrowins1} and \eqref{eq:firstrowins2} that elements involving $PV^{1,1}(\bY)$ operators
are accompanied by appropriate $PV^{0,0}(\bY)$ tails. The term {\it first row} operators refers to the full cohomology classes given by the combination of both operators.}
It turns out that it is technically easier to distinguish correlators by the numbers of ``first row insertions".

As discussed in section \ref{ss:transflaw}, we are going to evaluate the map \eqref{eq:formula} by
implementing the proposed hybrid transformation law.
The key ingredient for achieving this
is a section $\cB\in\Gamma(T_{\bY}\otimes T^\ast_{\bY})$ such that
\begin{align}
\label{eq:transflawrem}
\cB_\alpha{}^\beta J_\beta = T_{\alpha} \in \Gamma(T^\ast_{\bY})~,
\end{align}
such that $T$ is independent of the parameter $\psi$ and $T^{-1}(0)=B$, that is, $T$ satisfies the potential condition as well.
An appropriate such section is given by
\begin{align}
\label{eq:Texplexpr}
T=\begin{pmatrix} u^{25}\phi_1^{13} & \half(u^{26}+1)\phi_1^{12} & \phi_2^{13} & \phi_3^{13} & \phi_4^{13}   \end{pmatrix}^{\top}~.
\end{align}
The expression of the section $\cB$ such that \eqref{eq:transflawrem} holds is rather bulky and unilluminating,
thus we relegate it to appendix \ref{app:transflawoct}.

\subsubsection{Bottom row correlators}

Let $\alpha=\cO_1\cO_2\cO_3\in \Gamma(\cO_{\bY})$, where $\cO_{1,2,3}$ are (c,c) elements of the bottom row type such that
$\bq(\cO_1)+\bq(\cO_2)+\bq(\cO_3)=3$.
According to the discussion regarding the B ring for this example, $\alpha$ will be of the general form
\begin{align}
\label{eq:geninsertion}
\alpha = u^{t_0} (\phi^1)^{t_1} (\phi^2)^{t_2} (\phi^3)^{t_3} (\phi^4)^{t_4} ~,	\qquad\qquad t_\alpha\geq0~,
\end{align}
such that $t_1+\cdots+t_4=12$ and $t_0\leq 2t_1$.
From \eqref{eq:Texplexpr} we compute
\begin{align}
\label{eq:octicpbTb}
\det \pb \Tb= -2028 (19 \ub^{26}-150)\ub^{24}(\phib^1)^{24}(\phib^2)^{12}(\phib^3)^{12}(\phib^4)^{12}~.
\end{align}
Thus, the correlators now read
\begin{align}
\label{eq:correlatorbottrow}
\la \alpha \ra_{S^2} &= \int \Omega_{\bY}\wedge \Omegab_{\bY} \exp^{-\bv/4 ||T||^{2}} \alpha \det \cB \det (\pb \Tb)~,
\end{align}
where $||T||^2 =T_\alpha \delta^{\alpha\betab}\Tb_{\betab}=T_\alpha \Tb^{\alpha}$, that is, we used the independence of the correlators from the metric to set
$h_{\alpha\betab}=\delta_{\alpha\betab}$.

Solving this type of integrals exactly proves to be a daunting challenge,
which we are not able to fully overcome.
It is possible, however, to perform the integral over the various phases $\arg(x^\alpha)$.
This will turn out to be enough to determine
the entire dependence of the correlators on the parameter $\psi$, up to an unknown constant.
While this is easily done for the fiber coordinates, we have to be careful in considering the exponential factor,
as it is not phase invariant with respect to the base coordinate $\arg(u)$.
Let us consider the argument of the exponential
\begin{align}
T_\alpha \Tb^\alpha =|u|^{50}|\phi^1|^{26}+\frac14 (|u|^{52}+u^{26}+\ub^{26}+1)|\phi^1|^{24}+ |\phi^2|^{26}+|\phi^3|^{26}+|\phi^4|^{26}~.
\end{align}
We note that the only term which is not phase invariant is $\frac14 (\ub^{26}+u^{26})|\phi_1|^{24}$.
Now, using the fact that \eqref{eq:correlatorbottrow} is independent of $\bv$, we can take the limit $\bv\rightarrow0$ and consider the expansion
\begin{align}
e^{-\bv \Tb^\alpha T_\alpha} = \sum_{n=0}^\infty \frac{(-\bv)^n}{n!} (\Tb^I T_I)^n= \sum_{n=0}^\infty F_n(|\phi|,|u|) (\ub^{26}+u^{26})^n
= \sum_{n=0}^\infty G_n(|\phi|,|u|) (\ub^{26n}+u^{26n})~,
\end{align}
where $F_n,G_n$ only depend on the magnitudes $|x^\alpha|$ of the coordinates.
By incorporating the full $\ub$ dependence in \eqref{eq:correlatorbottrow} we obtain the expansion
\begin{align}
\label{eq:antiholmdepub}
e^{-\mathbf{v} \Tb^\alpha T_\alpha}  \left(\ub^{26}-\frac{150}{19}\right)\ub^{24} =  \sum_{n=0}^\infty \left( H_n(|\phi|,|u|) \ub^{26n+24} +L_n(|\phi|,|u|)u^{26n-50}\right)~,
\end{align}
for appropriate functions $H_n,L_n$.
The expansion
\begin{align}
\det\cB =  \sum_{m_0,\dots,m_4\geq0} f_{m_0,\dots,m_4}(\psi) u^{m_0} (\phi^1)^{m_1} (\phi^2)^{m_2}(\phi^3)^{m_3}(\phi^4)^{m_4}
\end{align}
can be constructed explicitly from the expression of $\cB$ in appendix \ref{app:transflawoct}.
Considering the contribution from \eqref{eq:geninsertion} and integrating over $\arg(\phi^i)$
we obtain that the relevant contribution to the correlator is
\begin{align}
\alpha \det \cB = \sum_{m_0,\dots,m_4\geq0} f_{m_0,\dots,m_4}(\psi)  u^{m_0+t_0} (\phi^1)^{24}\left( \prod_{a=2}^4 (\phi^a)^{12}\right)
\delta_{m_1+t_1,24}\left(\prod_{a=2}^4 \delta_{m_a+t_a,12}\right)~,
\end{align}
where one can check explicitly that $m_0+t_0\leq48$.
A look at \eqref{eq:antiholmdepub} shows that the the integration with respect to $\arg(u)$ selects $m_0+t_0=24$.
Therefore we have determined that
\begin{align}
\label{eq:alphares}
\la \alpha \ra_{S^2} = f_{24-t_0,24-t_1,12-t_2,12-t_3,12-t_4}(\psi)~.
\end{align}
In particular, all such functions assume the form
\begin{align}
\la \alpha \ra_{S^2} &= \frac{\psi^{v_\alpha}}{1-\psi^8}~,	 &0\leq&v_\alpha \leq 8~,
\end{align}
where $v_\alpha$ is a non-negative integer that depends on the insertion $\alpha$.
The correlators diverge when $\psi^8=1$, which corresponds to a singularity in the conformal field theory.
This singularity appears in the hybrid theory as the potential condition $dW^{-1}(0)=B$ does not hold
for this choice of parameters.

As an example, we can consider the element of charge (3,3) in \eqref{eq:detHessW}, which can be interpreted as a three-point function as,
for example, $\alpha = \left(u^2 (\phi^1)^2(\phi^2)^2\right)\left(u^2 (\phi^1)^2(\phi^3)^2\right)\left(u^2 (\phi^1)^2(\phi^4)^2\right)$, which gives
\begin{align}
\label{eq:dethess22}
\la \det \Hess W \ra_{S^{2}} = \frac{1}{1-\psi^8}~.
\end{align}
As expected, this is the only possibility for a non-zero correlator in the undeformed theory at $\psi=0$.

\subsubsection{First row correlators}

Next, we tackle the case where $\alpha=\cO_1\cO_2\cO_3$, and at least one of the insertions arises from the first-row
of the spectral sequence. In particular, by inspecting the elements of the (c,c) ring computed above, it follows that
$\alpha = \alpha_1 \chib_1 \etab^{\overline0} + \alpha_0$, where $\alpha_0,\alpha_1\in PV^{0,0}(\bY)$.
Thus, the correlator $\la\alpha \ra_{S^2}$
splits as a sum of two integrals which we can compute separately. Although $\alpha_0 \notin H^0(\bY,\cO_{\bY})$,
its contribution to the correlator can be nonetheless computed by the methods of the previous section. Thus, the main novelty here
is the $\alpha_1$ contribution.

The correlator $\la \alpha_1\chib_1 \etab^{\overline0} \ra$ is determined in terms of the object $\cM^{\overline0}_1$, which we recall is
the determinant of the minor of the matrix $\cM=\pb \Jb$ obtained by removing the column corresponding
to the index $\alphab=\overline0$ (from the $\etab^{\overline0}$ insertion) and the row corresponding to the index $\beta=1$
(from the $\chib_1$ insertion).
Then, in this case \eqref{eq:Mminasdets} reads
\begin{align}
\cM_1^{\overline0}= \sum_{k=0}^4 \det_{0,k}(\pb \Tb) \det_{k,1} \cB~,
\end{align}
where $\det_{i,j}$ indicates the determinant of the minor obtained by removing the $i$-th row and the $j$-th column.
In particular, from \eqref{eq:octicpbTb} we have that $\det_{0,k}(\pb \Tb)=0$ for $k=2,3,4$. Thus, the sum
above reduces to just two terms, corresponding to
\begin{align}
\det_{0,0}(\pb \Tb) &=13182 (\ub^{26}+1)(\phib^1)^{11} (\phib^2)^{12}(\phib^3)^{12}(\phib^4)^{12}~,\nonumber\\
\det_{0,1}(\pb \Tb) &=28561 \ub^{25} (\phib^1)^{12} (\phib^2)^{12}(\phib^3)^{12}(\phib^4)^{12}~.
\end{align}
That is, putting all together the full correlator reads
\begin{align}
\label{eq:corrformH1}
\la \alpha \ra_{S^2} &=\int \Omega_{\bY}\wedge \Omegab_{\bY} \exp^{-\bv/4 ||T||^{2}} \alpha_1\left[ \det_{0,1} \cB \det_{0,0} (\pb \Tb) +  \det_{1,1} \cB \det_{0,1} (\pb \Tb)\right] \nonumber\\
&\quad+\int \Omega_{\bY}\wedge \Omegab_{\bY} \exp^{-\bv/4 ||T||^{2}} \alpha_0 \det \cB \det (\pb \Tb)~.
\end{align}

First, suppose there is only one first-row insertion, that is, say, $\cO_1$ is of the form \eqref{eq:firstrowins1} or \eqref{eq:firstrowins2}, while $\cO_{2,3}\in H^0_{\bQb}(\bY,\cO_{\bY})$
are bottom row insertions, again satisfying the charge condition $\bq(\cO_1)+\bq(\cO_2)+\bq(\cO_3)=3$.
The general expressions for $\alpha_{0,1}$ are
\begin{align}
\label{eq:alpha01exprs}
\alpha_0&= {u^7-\ub \over 1+ u\ub} u^{t_0} (\phi^1)^{t_1+3} \prod_{a=2}^4 (\phi^a)^{t_a} ~,  \nonumber\\
\alpha_1&= {1 \over (1+ u\ub)^2}u^{t_0} (\phi^1)^{t_1}  \prod_{a=2}^4 (\phi^a)^{t_a} ~,
\end{align}
such that
\begin{align}
t_\alpha&\geq0~,		&t_0&\leq 2t_1\leq 16~,		&t_1+t_2+t_3+t_4&=9~.
\end{align}
The relevant contribution to $\la \alpha_0 \ra_{S^2}$, determined again by integrating over the phases $\arg(\phi^i)$, is now given by
\begin{align}
\alpha_0 \det \cB = \sum_{m_0+t_0\geq-1}^{49} \sum_{m_1,\dots,m_4} f_{m_0,\dots,m_4}(\psi)  \frac{u^{m_0+t_0}}{1+u\ub} (\phi^1)^{24} \left(\prod_{a=2}^4 (\phi^a)^{12}\right)
\delta_{m_1+t_1,21}\left(\prod_{a=2}^4 \delta_{m_a+t_a,12}\right)~,
\end{align}
which, combined with \eqref{eq:antiholmdepub} and integrating over $\arg(u)$, yields
\begin{align}
\label{eq:alpha0res}
\la \alpha_0 \ra_{S^2} = f_{24-t_0,21-t_1,12-t_2,12-t_3,12-t_4}(\psi)~.
\end{align}
 For the $\la \alpha_1 \ra$ correlator, we need the $\ub$ dependence of the expansions
\begin{align}
e^{-\bv \Tb^\alpha T_\alpha} \det_{0,0} \pb \Tb &=  \sum_{n=-\infty}^\infty  \Ht_n(|\phi|,|u|) \ub^{26n}~,\nonumber\\
e^{-\bv \Tb^\alpha T_\alpha} \det_{0,1} \pb \Tb &=  \sum_{n=0}^\infty  \left(\Ft_n(|\phi|,|u|) \ub^{26n+25} +\Lt_n (|\phi|,|u|) u^{26n-25}\right)~,
\end{align}
for some appropriate phase-invariant functions $\Ht_n,\Ft_n,\Lt_n$. Similarly, we represent the relevant determinants as
\begin{align}
\det_{0,1}\cB =  \sum_{m_0,\dots,m_4\geq0} g_{m_0,\dots,m_4}(\psi) u^{m_0} (\phi^1)^{m_1} (\phi^2)^{m_2}(\phi^3)^{m_3}(\phi^4)^{m_4}~,\nonumber\\
\det_{1,1}\cB =  \sum_{m_0,\dots,m_4\geq0} h_{m_0,\dots,m_4}(\psi) u^{m_0} (\phi^1)^{m_1} (\phi^2)^{m_2}(\phi^3)^{m_3}(\phi^4)^{m_4}~.
\end{align}
Now, after integrating over the fiber $\arg(\phi^i)$, we are left with
\begin{align}
\alpha_1 \det_{0,1} \cB &= \sum_{m_0+t_0=0}^{38}  \sum_{m_1,\dots,m_4} g_{m_0,\dots,m_4}(\psi) u^{m_0+t_0} (\phi^1)^{11} (\phi^2)^{12}(\phi^3)^{12}(\phi^4)^{12}
\delta_{m_1+t_1,11}\prod_{a=2}^4 \delta_{m_a+t_a,12}~, \nonumber\\
\alpha_1 \det_{1,1} \cB &= \sum_{m_0+t_0=0}^{37}  \sum_{m_1,\dots,m_4} h_{m_0,\dots,m_4}(\psi)u^{m_0+t_0} (\phi^1)^{12} (\phi^2)^{12}(\phi^3)^{12}(\phi^4)^{12}
\prod_{i=1}^4 \delta_{m_i+t_i,12}~.
\end{align}
Thus, we obtain a contribution from three terms, which reads
\begin{align}
\label{eq:alpha1res}
\la \alpha_1 \ra_{S^2} &= I_1 g_{0,11-t_1,12-t_2,12-t_3,12-t_4}(\psi)+I_2 g_{26-t_0,11-t_1,12-t_2,12-t_3,12-t_4}(\psi) \nonumber\\
&\quad +I_3 h_{25-t_0,12-t_1,12-t_2,12-t_3,12-t_4}(\psi)~,
\end{align}
where the coefficients $I_{1,2,3}$ are represented by integrals over the magnitudes $|x^\alpha|$.
We see the difficulty with this expression, compared with \eqref{eq:alphares} and \eqref{eq:alpha0res},
where the unknown integral can be reabsorbed into a multiplicative constant.
In this case instead it appears that we cannot determine in principle the relative coefficients between the various contribution, and
thus we cannot determine the full dependence on the parameter $\psi$.
However, it turns out by inspection that all the three contributions to \eqref{eq:alpha1res}
vanish separately, thus $\la \alpha_1 \ra_{S^2}=0$. Finally, we conclude that for
one first-row insertion
\begin{align}
\la \alpha \ra_{S^2} = \la \alpha_0 \ra_{S^2}~.
\end{align}
It follows that we can choose a basis $A=\cO_1+\cO_1^{(0)}$, where\footnote{This notation emphasizes that $\cO_1$ has
both a $PV^{1,1}(\bY)$ and a $PV^{0,0}(\bY)$ component. Moreover, the spectral sequence degenerates at the second stage,
therefore $H^1_{\bQb}(\bY,\bullet)=H^\infty_{\bQb}(\bY,\bullet)$.}
$\cO_1\in H^1_{\bQb}(\bY,T_{\bY}\oplus\cO_{\bY})$,
$\cO_1^{(0)}\in H^1_{\bQb}(\bY,\cO_{\bY})$ and $\bq(\cO_1)=\bq(\cO_1^{(0)})$,
such that
\begin{align}
\label{eq:firstrowredef}
\la A \cO_2\cO_3 \ra_{S^{2}} = 0~,	\qquad\qquad	\forall \cO_{2,3}\in H^0_{\bQb}(\bY,\cO_{\bY})~.
\end{align}
Explicitly, let
\begin{align}
\label{eq:Fsfirstrow}
F^{(1)}_a&=\phi^a~,		&F^{(2)}_a&=\phi^a(\phi^b)^2(\phi^c)^2~,	\quad a\neq b\neq c~.
\end{align}
Then, there exists a unique $h\in\C$ such that
\begin{align}
\label{eq:firstrowelms}
A^{(\bq)}_a &= - {1 \over (1+u \ub)^2} F^{(\bq)}_{a} \chib_1 \etab^{\overline0} -\half {u^7-\ub \over  1+u\ub}(\phi^1)^3F^{(\bq)}_{a}  +\psi h F^{(\bq)}_{a} \phi^2\phi^3\phi^4
\end{align}
satisfy \eqref{eq:firstrowredef}. This simply follows from the fact that the following holds
\begin{align}
f_{24-t_0,21-t_1,12-t_2,12-t_3,12-t_4}(\psi)=\psi f_{24-t_0,24-t_1,11-t_2,11-t_3,11-t_4}(\psi)~.
\end{align}

Next, we turn to the case of two first row insertions of the form \eqref{eq:firstrowelms}, and one insertion
from the bottom row $\cO_3\in H^0_{\bQb}(\bY,\cO_{\bY})$, that is, we study the correlators
\begin{align}
\la A^{(\bq_1)}_a A^{(\bq_2)}_b \cO_3 \ra_{S^2} = \la F^{(\bq_1)}_a F^{(\bq_2)}_b \cA^2 \cO_3 \ra_{S^2} ~,
\end{align}
where we defined $\cA$ by $F_a^{(\bq)}\cA = A^{(\bq)}_a$.
The argument above can be
repeated almost identically in this case, the only difference being
in the expressions \eqref{eq:alpha01exprs}. Let us write $\cA = \cA_{1}\chib\etab + \cA_{0} + \psi \cC$, then the
components of the insertions are given by
\begin{align}
\alpha_1&= 2F^{(\bq_1)}_a F^{(\bq_2)}_b\cO_3 \cA_1(\cA_0+\psi\cC)~,		&\alpha_0&=F^{(\bq_1)}_a F^{(\bq_2)}_b\cO_3(\cA_0+\psi\cC)^2~.
\end{align}
The result is that there are only a few non-vanishing correlators, thus we can present the full list\footnote{Here we denote by $F_1F_2\cO_3$
the scalar product $F^{(\bq_1)}_a F^{(\bq_2)}_b\cO_3$ without specifying $a,b$ or $\bq$.
This completely determines the value of the correlator $\la A^{(\bq_1)}_a A^{(\bq_2)}_b \cO_3 \ra_{S^2}$,
but it is up to the reader to extract it from (\ref{eq:frfrcorrhyb}).
For example, $F_{1}F_{2}\cO_3=(\phi^2\phi^3\phi^4)^2$ determines the correlator $\la A^{(1)}_a, A^{(2)}_a, 1 \ra_{S^2}$ for any $a=2,3,4$.}
\begin{align}
\label{eq:frfrcorrhyb}
\xymatrix@R=0mm@C=15mm{
F_1F_2\cO_3					& \la A_1A_2\cO_3\ra_{S^2}\\
(\phi^2\phi^3\phi^4)^2			&\frac1{1-\psi^8}+I_0'\frac{\psi^8}{1-\psi^8}\\
u^2(\phi^1)^2(\phi^2)^4			&\frac{\psi^6}{1-\psi^8}\\
u^2(\phi^1)^2(\phi^3)^4			&\frac{\psi^6}{1-\psi^8}\\
u^2(\phi^1)^2(\phi^4)^4			&\frac{\psi^6}{1-\psi^8}\\
u^3(\phi^1)^3\phi^2\phi^3\phi^4		&\frac{\psi^4}{1-\psi^8}
}
\end{align}
We notice that for the first correlator we were not able to determine the relative coefficient $I_0'$ between the two contributions.

The last case to consider is for three first-row insertions. Here we find that all the operators must have $\bq=1$, which we can write
\begin{align}
\la A^{(1)}_aA^{(1)}_bA^{(1)}_c \ra_{S^{2}} = \la \phi^a\phi^b\phi^c \cA^3\ra_{S^{2}}~.
\end{align}
We can simplify the computation by expanding
\begin{align}
\cA^3 = (\cA_{1}\chib\eta + \cA_0)^3 + 3\psi  (\cA_{1}\chib\eta+\cA_0)^2 \cC+3\psi^2  (\cA_{1}\chib\eta+\cA_0) \cC^2  +  \psi^3 \cC^3~,
\end{align}
and compute the contribution from each term separately. Using \eqref{eq:firstrowredef} we have
\begin{align}
\la \phi^a\phi^b\phi^c(\cA_{1}\chib\eta+\cA_0) \cC^2 \ra_{S^2} &= -\psi \la\phi^a\phi^b\phi^c \cC^3 \ra_{S^2}~, 		\nonumber\\
\la\phi^a\phi^b\phi^c (\cA_{1}\chib\eta+\cA_0)^2 \cC \ra_{S^2} &=\la\phi^a\phi^b\phi^c \cA^2 \cC \ra_{S^2} -\psi^2\la\phi^a\phi^b\phi^c \cC^3 \ra_{S^2}~,
\end{align}
hence
\begin{align}
\la \phi^a\phi^b\phi^c \cA^3\ra_{S^{2}} = \la \phi^a\phi^b\phi^c(\cA_{1}\chib\eta + \cA_0)^3\ra_{S^{2}}
+ 3\psi \la \phi^a\phi^b\phi^c \cA^2 \cC\ra_{S^{2}}  -5\psi^3 \la \phi^a\phi^b\phi^c \cC^3\ra_{S^{2}}~.
\end{align}
The only choice for which the correlator does not automatically vanish is for $a\neq b\neq c$, where
\begin{align}
\label{eq:TTTcorr}
\la \phi^2\phi^3\phi^4(\cA_{1}\chib\eta + \cA_0)^3\ra_{S^{2}} &= I_1' \frac{\psi^9}{1-\psi^8} + I_2' \frac{\psi}{1-\psi^8}~,\nonumber\\
\la \phi^2\phi^3\phi^4\cA^2\cC\ra_{S^{2}} &=I_3' \frac1{1-\psi^8}+I_4'\frac{\psi^8}{1-\psi^8}~,\nonumber\\
\la \phi^a\phi^b\phi^c \cC^3\ra_{S^{2}} & = I_5' \frac{\psi^6}{1-\psi^8}~.
\end{align}
Thus, the full correlator reads
\begin{align}
\label{eq:H1H1H1corr}
\la \phi^a\phi^b\phi^c \cA^3\ra_{S^{2}} &= (I_1'+3I_4'-5I_5')\frac{\psi^9}{1-\psi^8} + (I_2'+3I_3') \frac{\psi}{1-\psi^8}~.
\end{align}
Again, we are not able to determine the various coefficients and therefore the full dependence of the correlator on the parameter $\psi$.

\subsection{GLSM and comparison with LG phase}

A non-trivial test for our formula is provided by the linear model. The B-twisted GLSM is independent of the K\"ahler parameters, and
the relations in the B ring can be evaluated at any point in the K\"ahler moduli space. Thus, when a hybrid model arises in a phase of a GLSM,
our formula must agree with the computations in other phases.

The hybrid model we have solved in this section arises as a phase of a (2,2) $\GU(1)^2$ GLSM \cite{Candelas:1993dm} with seven chiral superfields and gauge charges
\begin{align}
\label{eq:22octicgaugech}
\xymatrix@R=0mm@C=3mm{
		&X_1	&X_2	&X_3	&X_4	&X_5	&X_6	&P		&\text{F.I.}\\
\GU(1)_1	&1		&1		&1		&1		&0		&0		&-4		&r_1\\
\GU(1)_2	&-2		&0		&0		&0		&1		&1		&0		&r_2
}
\end{align}
We indicate as $x_{1,\dots,6}$ and $p$ the lowest components of the various superfields, and $r_{1,2}$ are the F.I.~parameters.
In the large radius phase, i.e., in the cone $r_1,r_2>0$, the model reduces to a NLSM with target space a K3-fibered CY${}_3$ obtained by resolving the singularities
of the hypersurface $W$ of degree 8 in the toric resolution of the weighted projective space $\CP^4_{11222}$.

The hybrid phase we have studied at length in this section arises in the cone $r_1<0, r_2>0$, where the D-terms force the field $p$ to acquire a non-zero vev,
as well as determine the irrelevant ideal to be $(x_5,x_6)$.
Upon the quotient by $\GU(1)_2$, this data determines the hybrid geometry $\bY=\tot\left(\cO(-2)\oplus\cO^{\oplus3}\rightarrow\CP^1\right)$.
Finally, $\GU(1)_1$ is broken by $<p>$ to a $\Z_4$ subgroup, which determines
the R-symmetry assignment and the orbifold quotient.

The Landau-Ginzburg orbifold phase arises instead in the cone $r_2<0, 2r_1+r_2<0$, where both $x_1$ and $p$ acquire non-zero vevs,
while the remaining fields are massless and interact through the superpotential
\begin{align}
W_{\text{LG}}&= \frac14 (x_2^4+x_3^4+x_4^4)+\frac18(x_5^8+x_6^8) -\psi x_2x_3x_4x_5x_6~,
\end{align}
which corresponds to our choice \eqref{eq:22supoct} for the superpotential in the hybrid phase.
The description for the observables in the chiral ring of LGO theories is well known:
for this example these are the elements in $R=\C[x_2,\dots,x_6]/\la \p W_{\text{LG}}\ra$ which are invariant under the $\Z_8$ orbifold.
Again, setting $\psi=0$ will generate good representatives for the cohomology classes.
Explicitly, a generic element in the ring is of the form
\begin{align}
\label{eq:LGchirringel}
\cO_{\text{LG}}=\left(\prod_{a=2}^4 x_a^{l_a}\right) x_5^{l_5}x_6^{l_6}~,		\qquad\quad 2\sum_{a=2}^4 l_a +l_5+l_6=8m~, \qquad m=0,\dots,4~.	
\end{align}
In particular, it follows that $l_5+l_6$ must be even.
The correspondence between hybrid and LG coordinates is quite straightforward.
The elements \eqref{eq:LGchirringel} lift in the GLSM to the operators
\begin{align}
\label{eq:cxstrGLSM}
\cO_{\text{GLSM}} = px_1^{l_1}\prod_{a=2}^4 x_a^{l_a} x_5^{l_5}x_6^{l_6}~,		\qquad\qquad l_1=\half(l_5+l_6)~,
\end{align}
which reduce in the hybrid theory, in the patch $U_1$ and identifyng without loss of generality $x_5=u$, to
\begin{align}
\cO_{\text{HY}} = u^{l_5}(\phi^1)^{l_1}\prod_{a=2}^4 (\phi^a)^{l_a} ~,\qquad\quad l_1+\sum_{a=2}^4 l_a=4m~, \qquad l_5\leq2l_1~.
\end{align}
These indeed coincide with the elements from the bottom row of the spectral sequence we described above.
It is possible to check explicitly that, up to a numerical factor which we are not able to determine, the following holds
\begin{align}
\la \cO^1_{\text{HY}} \cO^2_{\text{HY}} \cO^3_{\text{HY}} \ra_{S^2} = \la \cO^1_{\text{LG}}\cO^2_{\text{LG}}\cO^3_{\text{LG}}\ra
= \text{Res}\left\{  \frac{\cO^1_{\text{LG}}\cO^2_{\text{LG}}\cO^3_{\text{LG}}}{\p_2 W_{\text{LG}}\cdots \p_6W_{\text{LG}}} \right\}~.
\end{align}
In other words, we find a complete match of the correlators involving bottom row elements on the hybrid side and untwisted elements on the
LGO side.

From the linear model point of view, we can employ the elements \eqref{eq:cxstrGLSM} at $\bq=1$ (that is, $m=1$)
to deform the theory by
\begin{align}
W_{\text{GLSM}} &= W^0_{\text{GLSM}} + \psi \cO^{(\bq=1)}_{\text{GLSM}}~,
\end{align}
where $W^0_{\text{GLSM}}$ is the superpotential of the undeformed theory. For this reason
these deformations are dubbed {\it polynomial}.
On the other hand, there is no such interpretation for the elements which arise, in hybrid language,
from the first row of the spectral sequence.
There are no good representatives for these operators in the GLSM defined by \eqref{eq:22octicgaugech}
and thus no simple manner to correspondingly deform the GLSM action. In fact, as pointed out in
\cite{Aspinwall:2010ve}, these {\it non-polynomial} deformations, from the point of view of this GLSM, are obstructed as turning them on would prevent
the embedding of $W$ in the toric variety which is the toric resolution of $\CP^{4}_{11222}$.\footnote{The authors of \cite{Aspinwall:2010ve} provide
a different GLSM which describes the same moduli space of (2,2) SCFTs, but in which all 86 complex structure deformations are
realized polynomially. It is however unknown which 83 dimensional subspace corresponds to the polynomial complex structure deformations
of the original model. As the more general GLSM has a hybrid phase (but no LGO), the methods we are providing
in this work could be of help in answering this question.}

In the LGO phase, the non-polynomial representatives in the (c,c) ring appear in twisted sectors.
In this particular example \cite{Aspinwall:2010ve} they have the form
\begin{align}
\label{eq:LGtwistst}
T^{(1)}_a=&x_a|4\ra~,		&T^{(2)}_a=&x_ax_b^2x_c^2|4\ra~, \qquad a\neq b\neq c~,
\end{align}
where $|4\ra$ is the (NS,NS) vacuum state in the $k=4$ twisted sector.
Presently, to the knowledge of the authors, there is no
technique to evaluate correlators involving these states,
except when $W_{\text{LG}}$ is an invertible polynomial \cite{2016arXiv160808962B},
which, in our example corresponds to $\psi=0$.
However, at the LG point, the theory exhibits a $\Z_8$ quantum symmetry, which automatically yields
\begin{align}
\la T U U \ra = \la T T T \ra =0~,
\end{align}
where by $T$ we denote an element in \eqref{eq:LGtwistst} and by $U$ an element from the untwisted sector \eqref{eq:LGchirringel}.
The first equation above is reminiscent of
the structure in the hybrid theory in \eqref{eq:firstrowredef}.
Hence, it is natural to conjecture a correspondence between
\begin{align}
A^{(\bq)}_a \longleftrightarrow T^{(\bq)}_a~,
\end{align}
and given the identical structure of \eqref{eq:LGtwistst} and \eqref{eq:Fsfirstrow}, in particular $\cA \longleftrightarrow |4\ra$.
We can employ this conjecture to derive predictions on the structure of both theories.
On the one side, the correlators \eqref{eq:frfrcorrhyb} provide a prediction for correlators in the LGO theory involving twisted operators
\begin{align}
\label{eq:frfrcorrLGO}
\xymatrix@R=0mm@C=10mm{
F_1F_2U_3					& \la T_1T_2U_3\ra\\
x_2^2x_3^2x_4^2				&\frac1{1-\psi^8}+I_0'\frac{\psi^8}{1-\psi^8}\\
x_2^4x_5^2x_6^2				&\frac{\psi^6}{1-\psi^8}\\
x_3^4x_5^2x_6^2				&\frac{\psi^6}{1-\psi^8}\\
x_4^4x_5^2x_6^2				&\frac{\psi^6}{1-\psi^8}\\
x_2x_3x_4x_5^3x_6^3			&\frac{\psi^4}{1-\psi^8}
}
\end{align}
This agrees with the computation at the Fermat point ($\psi=0$) \cite{2016arXiv160808962B,Brunner:2013ota}, where
\begin{align}
\label{eq:twistedLGcorrFerm}
\la T_1 T_2 U_3 \ra= \text{Res}\left\{  \frac{F_1F_2U_3}{\p_2 \Wt_{\text{LG}}\p_3 \Wt_{\text{LG}} \p_4\Wt_{\text{LG}}} \right\}~,
\end{align}
and $\Wt_{\text{LG}}$ is obtained from $W_{\text{LG}}$ by setting to zero all non-invariant variables with respect to the orbifold action,
which in the $k=4$ twisted sector are $x_5$ and $x_6$.
For a general quasi-homogeneous $W_{\text{LG}}$, the generalization of \eqref{eq:twistedLGcorrFerm} is to our knowledge not known, thus our hybrid methods allow us to compute
the full list of correlators.

On the other side, the condition $\la TTT\ra=0$ predicts that the correlator \eqref{eq:H1H1H1corr} vanishes, that is
\begin{align}
\la\phi^2\phi^3\phi^4\cA^3 \ra_{S^2}=0~.
\end{align}

\section{$\mathcal{N}=(0,2)$ hybrid models and B/2 correlators}
\label{s:02hybrids}

We now turn to the analysis of hybrid theories which
flow in the IR to (0,2) SCFTs.
These models have been recently introduced in \cite{Bertolini:2017lcz},
and we begin this section by reviewing that construction.
A (0,2) hybrid model is defined by the quadruple $(\mathbf{Z},\mathcal{E},V,J)$ where $\mathbf{Z}$ is a K\"ahler
manifold and $\mathcal{E}\rightarrow \mathbf{Z}$ is a rank-$R$ holomorphic vector bundle.
As before, we take $\mathbf{Z}$ to be the total space of a holomorphic vector
bundle $\mathbf{Z}=\text{tot}\left(X\xrightarrow{\pi} B\right)$, and $B$ to be a smooth compact manifold.
Let $x^\alpha$, $\alpha=1,\dots,d=\dim \bZ$, be local coordinates on $\bZ$, which we split according to the fiber/base
decomposition\footnote{For ease of exposition, we assume that $X$ splits as a sum of line bundles, although the general case
can be treated at the price of a more involved notation.} as $(y^{\mu},\phi^i)$, $\mu=1,\dots,b=\dim B$ and $i=1,\dots,n=\rank\ X$.
Similarly to the (2,2) case, $V$ is a $\GU(1)$-action on $\mathbf{Z}$
determined by a holomorphic Killing vector on $\bZ$.
We again assume it acts vertically on $\bZ$, thus defining a good (0,2) hybrid,
and that it induces the decomposition $X=\oplus_i X_i$, where $\phi^i$ is the
coordinate along $X_i$, into eigenspaces of positive eingenvalues, that is
\begin{align}
V(B)&=0~,				&V(X_i)&=q_i X_i~,	\qquad q_i \in \mathbb{Q}_{>0}~.	
\end{align}
The bundle $\cE$ must respect the bundle structure of $\bZ$ and must admit a lift of the $V$-action.
Let $\lambda^A$, $A=1,\dots,R$, be a section of $\cE$, and let $G^{A}_{ \ B}$ be the transition functions
for $\cE$. If $\{U_a\}$ is a cover of $B$, we indicate as
$\lambda^A_a$ the restriction of $\lambda^A$ to the patch $\pi^{-1}U_a$.
Then, the bundle $\cE$ admits a lift of the $V$-action if, on the intersection $U_a\cap U_b\neq\emptyset$, the following holds
\begin{align}
\label{eq:chargelocglob}
V(\lambda^A_a) &= Q_A\lambda^A_a~,		&V(\lambda^A_b=(G_{ba})^{A}_{ \ B}\lambda^B_a) &= Q_A\lambda^A_b~,
\end{align}
that is, the charge assignment holds globally. We take $-1\leq Q_A <0$ and $0<q_i<1$.
A class of bundles that satisfy this property are classified by extensions of the form
\begin{align}
\label{eq:defbundleE02}
\xymatrix@R=0mm@C=10mm{
0\ar[r] &\oplus_{I=1}^N \pi^\ast \cO(L_I^a)    \ar[r] &\cE   \ar[r] & \pi^\ast \cE_B \ar[r] &0~,
}
\end{align}
where $\cE_B$ is a rank-$(R-N)$ bundle and $\cO(L_I^a)$, $I=1,\dots,N$, $a=1,\dots,\dim\text{Pic}B$,
are a collection of line bundles over $B$ such that\footnote{Again, the restriction to line bundles is not essential but simplifies notation.}
\begin{align}
V(\cE_B)&=-\cE_B~,		&V(\cO(L_I^a))=Q_I\cO(L_I^a), \qquad Q_I>-1~.
\end{align}
Finally, the (0,2) superpotential is specified by a holomorphic section $J\in \Gamma(\mathcal{E}^{*})$, such that
\begin{align}
V(J_A) &= \sum_\alpha V(x^\alpha) \p_\alpha J_A~, &V(x^{\alpha})&=q_{\alpha}x^{\alpha}~,
\end{align}
where the $V$-action on a section $J\in\Gamma(\cE^*)$ is determined by the $V$-action on sections of $\cE$ in \eqref{eq:chargelocglob}.
Given the property \eqref{eq:chargelocglob} this is a globally well-defined condition.
This construction defines a nonsingular model when the potential condition $J^{-1}(0)=B$ is satisfied.

In order to write the action for the corresponding NLSM we introduce the (0,2) superfields
\begin{align}
\cX^\alpha & = x^\alpha + \sqrt{2}\theta^+ \psi_+^\alpha -i \theta^+\thetab^+ \pb_{\bar{z}} x^{\alpha}~,&
\cXb^{\alphab} & = {\xb}^{\alphab} - \sqrt{2}\thetab^+ \psib_+^{\alphab} +i\theta^+\thetab^+ \pb_{\bar{z}} \xb^{\alphab}~, \nonumber\\
\Psi^A & = \psi_-^A - \sqrt{2}\theta^+ F^A -i\theta^+\thetab^+ \pb_{\bar{z}} \psi_{-}^{A}~,&
\Psib^{\Ab} & = \psib_-^{\Ab} - \sqrt{2}\thetab^+ \Fb^{\Ab} +i\theta^+\thetab^+ \pb_{\bar{z}} \psib_-^{\Ab}~.
\end{align}
These satisfy the same chirality conditions as in \eqref{eq:chiralconds}.

Let $K$ be the K\"ahler potential on $\bZ$ and $g_{\alpha\bar{\beta}}$ the associated K\"ahler metric,
and let $\mathcal{H}_{A\overline{B}}$ be a Hermitian metric on $\cE\rightarrow\bZ$.
The action in components reads
\begin{align}
\label{eq:actioncomps02}
\cL_K &= -g_{\alpha\betab}\partial_{\mu} x^\alpha \partial^{\mu} \xb^{\betab} - 2ig_{\alphab\beta}\psib_+^{\alphab} D_{z} \psi_+^\beta
+2i\mathcal{H}_{\Ab B}\psib_-^{\Ab} \Db_{\bar{z}} \psi_-^B + R_{\alpha\Bb A\bar{\beta}} \psi_{+}^{\alpha}\psi_{-}^A\psib_{-}^{\Bb}\bar{\psi}_{+}^{\bar{\beta}}~,  \nonumber\\
\cL_J&= -\frac{1}{2}\psi_+^\alpha\psi_-^A D_\alpha J_A + \frac{1}{2}\psib_+^{\alphab}\psib_-^{\Ab} \Db_{\alphab} \Jb_{\Ab} -\frac{1}{4} \mathcal{H}^{A\Bb} \Jb_{\Bb}J_A~,
\end{align}
where the covariant derivatives
\begin{align}
D_{z}\psi_+^{\alpha}&=(\partial_{z}+\frac{1}{2}\omega_{z})\psi_+^{\alpha}+\partial_{z} x^{\beta}\Gamma^{\alpha}_{\beta\delta}\psi_+^{\delta}~,
&\Db_{\bar{z}} \psi_-^A&=(\pb_{\bar{z}}+\frac{1}{2}\omega_{\bar{z}})\psi_-^A+\pb_{\bar{z}} x^{\beta}\Gamma^A_{\beta B}\psi_-^B~,
\end{align}
are constructed with the K\"ahler and Hermitian connection, respectively, and
$R_{A\Bb \alphab\beta}$ is the curvature constructed from the Hermitian connection.
In writing \eqref{eq:actioncomps02} we have already imposed the equations of motion for the auxiliary field.

\subsection{Anomalies and the low-energy limit}
\label{ss:anomalies}

Given the construction outlined above, the action \eqref{eq:actioncomps02} admits
an unbroken $\GU(1)_L\times\GU(1)_R^0$ symmetry with charges
\begin{align}
\label{eq:naivecharges}
\xymatrix@R=0mm@C=10mm{
\text{fields}		&\theta^+		&\cX^\alpha		&\Psi^A\\
\GU(1)_R^0		&1			&0				&1\\
\GU(1)_L			&0			&q_\alpha			&Q_A
}
\end{align}
However, anomaly cancellation plays a much more predominant role in the context of (0,2) hybrids.
In order to probe this IR theory, it is useful again to construct a left-moving algebra
\begin{align}
\label{eq:leftcurrents}
J_L&:= Q_A \psi_-^A  \psib_{-,A} - q_\alpha x^\alpha \rho_\alpha~, \nonumber\\
T&:= -\p_{z} x^\alpha\rho_\alpha - \psi_-^A \p_{z} \psib_{-,A} -\half \p_{z} J_L~,
\end{align}
corresponding to the generators of the global $\GU(1)_L$ symmetry and of the energy-momentum tensor.
Again, we have introduced the field $\rho_\alpha \equiv g_{\alpha\alphab} \p_{z} \xb^{\alphab} + \Gamma^A_{\alpha B}\psib_{-,A} \psi_-^B$.
Using this structure and free-fields OPEs, it follows that $\GU(1)_L$ is non-anomalous, that is,
\begin{eqnarray}
T(z)J_{L}(w)\sim \frac{J_L(w)}{(z-w)^{2}}+\frac{\p_w J_L(w)}{(z-w)}~,
\end{eqnarray}
when
\begin{align}
\label{eq:02LGfiban}
\sum_A Q_A^2-\sum_\alpha q_{\alpha}^{2}=-\sum_A Q_A - \sum_\alpha q_\alpha~,
\end{align}
which corresponds to the condition that the anomaly of the (0,2) LG fiber theory vanishes \cite{Kawai:1994np,Distler:1993mk}.
Note that this fixes the normalization (and the sign) of the charges $\GU(1)_L$ in \eqref{eq:naivecharges}.
The remaining anomaly cancellation conditions \cite{Bertolini:2017lcz} impose constraints on the allowed geometric structure and are given by
\begin{align}
\label{eq:anomaly02goem}
&\sum_I Q_Ic_1(\cO(L_I^a))-c_1(\cE_B) -\sum_i q_ic_1(X_i) =0~,
 &c_1(\cE) + c_1(T_{\bY})&=0~.
\end{align}
When these are satisfied, it is argued in \cite{Bertolini:2017lcz} that the theory flows to a non-trivial IR fixed point
characterized by the left-moving central charge
\begin{align}
\label{eq:cc02}
c&=2(d-R)+3\mathfrak{r}~,
&\mathfrak{r}&= -\sum_A Q_A - \sum_\alpha q_\alpha~,
\end{align}
where $\mathfrak{r}$ is the level of the $\mathfrak{u}(1)_L$ Kac-Moody (KM) algebra.

Next, we need to determine the IR right-moving R-current. Barring accidental symmetries \cite{Bertolini:2014ela},
this must be given as a linear combination of the symmetries \eqref{eq:naivecharges},
which we take to be $\GU(1)_R^0+\lambda \GU(1)_L$.
The parameter $\lambda$ can be determined by c-extremization \cite{Benini:2013cda} of the right-moving central charge
\begin{eqnarray}
\cb(\lambda)=\sum_{\alpha}(\lambda q_{\alpha}-1)^{2}-\sum_{A}(\lambda Q_{A}+1)^{2}~,
\end{eqnarray}
which yields
\begin{align}
\lambda =\frac{\sum_{\alpha}q_{\alpha}+\sum_{A} Q_{A}}{\sum_{\alpha}q_{\alpha}^{2}-\sum_{A} Q_{A}^{2}}=1~,
\end{align}
where we used \eqref{eq:02LGfiban}.
Thus, the IR $\GU(1)_L\times\GU(1)_R$ symmetries are given by
\begin{align}
\label{eq:02LRcharges}
\xymatrix@R=0mm@C=10mm{
\text{fields}		&\theta		&\cX^\alpha		&\Psi^A\\
\GU(1)_R			&1			&q_\alpha			&Q_A+1\\
\GU(1)_L			&0			&q_\alpha			&Q_A
}
\end{align}
In particular, this yields the right-moving central charge
\begin{align}
\cb=3(d-R+\mathfrak{r})~.
\end{align}
Finally, there is one additional anomaly condition we need to impose, which reads
\begin{align}
\ch_2(\cE) &= \ch_2(T_\bZ)~.
\end{align}
This is the condition that the NLSM constructed above is well-defined.

A comment might be useful at this point. The anomaly cancellation \eqref{eq:anomaly02goem} does not require $c_1(T_{\bZ})=0$.
However, as familiar from GLSM \cite{Distler:1995mi}
and NLSM \cite{Guffin:2008pi} constructions, it is always possible
to recover the condition that the target space is CY by adding spectators fields to the theory.
These are a massive pair, consisting of a bosonic field $S$ and a Fermi field $\Xi$ such that $V(S)=-V(\Xi)=q_S<1$.
Let $X_S\rightarrow B$ be a line bundle such that $\bZ':=\tot\left( X\oplus X_S \rightarrow B\right)$ has vanishing first Chern class, i.e., $c_1(T_{\bZ'})=0$.
Then, we take the spectator fields to transform according to
\begin{align}
S&\in \Gamma(X_S)~,		&\Xi&\in \Gamma(X_S^\ast)~.
\end{align}
By \eqref{eq:anomaly02goem} we have that $\cE':=\cE\oplus \pi^\ast X_S^\ast$ satisfies $c_1(\cE')=0$.
Now, these fields interact through the potential $\int d^2z d\thetab^+ \Xi S$, which is consistent with the charge assignment and the bundle geometry.
Thus, the theory with spectators
defines a UV geometry satisfying $c_1(T_{\bZ'})=c_1(\cE')=0$, but since these fields are massive
in the IR and can be simply integrated out,
the model with spectators flows to the same
IR theory as the model without them. We will therefore consider models without spectators
satisfying the weaker topological condition.

\subsection{The heterotic topological ring}

While in (2,2) SCFTs the B ring is defined by the cohomology of the supercharge $\bQb_{\text{(c,c)}}=\bQb_-+\bQb_+$, as considered in the
first part of this work, in general (0,2) SCFTs we do not have this definition at our disposal, as there is no left-moving supersymmetry
and therefore no operator which can assume the role of $\bQb_-$. The cohomology of $\bQb_+$, which we denote $\cH_{\bQb_+}$, still defines the (infinite dimensional) ring
of right-moving chiral operators, which from here on we will refer to as the chiral ring.
In a large class of theories,
it is possible \cite{Adams:2005tc} to define a subring of the chiral ring, where we take a projection within $\bQb_+$-cohomology
onto elements $\cO$ that satisfy $2h(\cO)=\bq(\cO)$,
where $h(\cO)$ and $\bq(\cO)$ are the left-moving weight and $\GU(1)_L$ charge, respectively. We denote this subset $\cH_{\text{B}/2}$.
It is not hard to show that our theories fall into this category. Let $\cO_{1,2}\in\cH_{\text{B}/2}$, then their OPE
takes the general form
\begin{align}
\label{eq:OPE02B2}
\cO_1(z)\cdot \cO_2(0) = \sum_s c_{12}^s \cO_s z^{h_s-q_s/2}~,
\end{align}
up to $\bQb_+$-exact terms, where $h_s=h(\cO_s)$ and $q_s=\bq(\cO_s)$. Here the sum on the RHS is over elements $\cO_s\in\cH_{\bQb_+}$.
The charges of the allowed operators in the OPE is fixed by $q_s=q_1+q_2$. By applying the standard
Sugawara decomposition for the level $\mathfrak{r}$ $\mathfrak{u}(1)$ KM algebra \eqref{eq:leftcurrents}, we can write any $\cO_s$ as
\begin{align}
\cO_s(z)= e^{iq_s\Phi/\sqrt{\mathfrak{r}}}(z) \widehat{\cO}_s(z)~,
\end{align}
where we bosonized the current $J_L=i\sqrt{\mathfrak{r}}\p \Phi$ and $\Phi$ is a free chiral boson, and
\begin{align}
\bq(\widehat{\cO}_s)&=0~, 		&h(\widehat{\cO}_s)&=h_s-{q_s^2\over2\mathfrak{r}}\geq0~.
\end{align}
This unitary bound, together with the charge integrality requirement of the (NS,NS) spectrum, yield the desired bound $h_s\geq q_s/2$.
Therefore, taking the limit $z\rightarrow0$ in \eqref{eq:OPE02B2} defines a finite subring
of the chiral ring known as the heterotic topological ring.\footnote{In a geometric phase these rings are also denoted by the term quantum sheaf cohomology rings.}

At a conceptual level, the computation of the full heterotic topological ring is a
only a slight generalization of the methods in section \ref{s:chiralring},
thus our discussion will be brief and focused on highlighting the differences which arise in the (0,2) setting.
Again, we will restrict our attention to the untwisted (NS,NS) sector.
The supersymmetry transformation of
the relevant supercharge again splits as a sum of two terms $\bQb_+=\bQb_0+\bQb_J$ and are given by
\begin{align}
\label{eq:chirringcohmnewfermi02}
[\bQb_0,\xb^{\alphab}] &=\psib_+^{\alphab}~,		&\{\bQb_0,\psi_+^\alpha\} &=i \pb_{\zb} x^\alpha~, 	&\{\bQb_J,\psib_{-,A} \}&=J_A~.
\end{align}
Then, a generic $\bQb_+$-closed element takes the form
\begin{align}
\label{eq:02states}
\cO(\omega) = \omega(x,\xb)_{\betab_1\dots\betab_s,B_1\dots B_t}^{A_1\dots A_r}\psib_+^{\betab_1}\cdots\psib_+^{\betab_s}\psi_-^{B_1}\cdots\psi_-^{B_t}\psib_{-,A_1}\cdots\psib_{-,A_r}~.
\end{align}
We can compute the corresponding weight and the left-moving charge using the operators \eqref{eq:leftcurrents}, and we find
\begin{align}
2h(\cO(\omega)) &= \bq(\omega) + \sum_{\mu=1}^t(1+Q_{B_{\mu}}) - \sum_{\nu=1}^r Q_{A_{\nu}}~,	&\bq(\cO(\omega)) = \bq(\omega) + \sum_{\mu=1}^tQ_{B_{\mu}} - \sum_{\nu=1}^r Q_{A_{\nu}}~.
\end{align}
It follows that the condition $2h(\cO)=\bq(\cO)$ can only be satisfied if $t=0$.
In this case, we can interpret $\omega$ in \eqref{eq:02states} as a $(0,s)$-horizontal form valued in $\wedge^r \cE$.
Extending the notation from the tangent bundle case we define
\begin{align}
\wedge^r_s \cE&:=\Omega^{0,s}(\bZ,\wedge^r \cE)~,
\end{align}
where we assume at most polynomial growth along the fiber directions. On these the action of the supercharges is
\begin{align}
\label{eq:Qaction02}
\bQb_0 &: \cO(\omega) \mapsto (\pb \omega)(x,\xb)_{\betab_1\dots\betab_s\gammab}^{A_1\cdots A_r}
\psib_+^{\gammab}\psib_+^{\betab_1}\cdots\psib_+^{\betab_s}\psib_{-,A_1}\cdots\psib_{-,A_r}~,\nonumber\\
\bQb_J &: \cO(\omega) \mapsto (-1)^s \omega(x,\xb)_{\betab_1\dots\betab_s}^{A_1\dots A_r}J_{A_1}
\psib_+^{\betab_1}\cdots\psib_+^{\betab_s}\psib_{-,A_2}\cdots\chib_{-,A_r}~.
\end{align}
In particular, the relevant states are double graded with respect to $r$ and $s$, and the supercharge still acts as expected
\begin{align}
\bQb_0 &: \wedge^r_s \cE \rightarrow \wedge^r_{s+1} \cE~,		&\bQb_J &: \wedge^r_s \cE \rightarrow \wedge^{r-1}_s \cE~.
\end{align}
Therefore, the elements of the topological heterotic ring are computed by a spectral sequence, as in \eqref{eq:22spetrseq}, where $\wedge^r_s \cE$
replaces $\wedge^r_s T_{\bY}$.

At a technical level instead, the explicit computation of the cohomology groups $H^\bullet_{\bQb_+}(\bZ,\wedge^r_s\cE)$ can be quite daunting.
The class of theories reviewed here are amenable to such computations, as shown in $\cite{Bertolini:2017lcz}$ together with
a method to compute their cohomology. Although the formal formula for the $S^2$ correlators we are going to derive next in this section
is valid for a generic hybrid model, its applicability, and hence its usefulness, relies on our ability to compute the heterotic topological ring.
Thus, it is natural to restrict our attention to the class of theories studied in $\cite{Bertolini:2017lcz}$.

\subsection{B/2-twisted hybrid models and $S^2$ localization}

The B/2-twist is defined as in the $(2,2)$ case by
\begin{eqnarray}
\GU(1)_{\text{B/2}}:=\GU(1)_{R}-\GU(1)_{L}~,
\end{eqnarray}
where in this setting
the role of the left-moving R-symmetry is taken by the flavor symmetry $\GU(1)_{L}$,
which is guaranteed to exist by construction in our models.
This does not lead to a contradiction, since we have shown in section \ref{ss:anomalies} that anomaly cancellation fixes the sign
of the IR flavor symmetry.\footnote{If the sign of the flavor symmetry had not been fixed,
we would not have been able to distinguish between the A/2 and the B/2 twists.}

We make again the choice of working with the $B_{(+)}$-twist. Under this choice the supercharge $\overline{\bQ}_{+}$ becomes a scalar,
and the Fermi fields become $C^{\infty}$ sections of the following bundles
\begin{eqnarray}
\psi^{A}_{-}&\in&\Gamma(K_{\Sigma}\otimes x^{*}\mathcal{E})~,\nonumber\\
{\psib}^{\overline{A}}_{-}&\in&\Gamma(x^{*}\overline{\mathcal{E}})~,\nonumber\\
{\psib}^{\bar{\alpha}}_{+}&\in&\Gamma(x^{*}\overline{T}_{\mathbf{Z}})~,\nonumber\\
\psi^{\alpha}_{+}&\in&\Gamma(\overline{K}_{\Sigma}\otimes x^{*}T_{\mathbf{Z}})~.
\end{eqnarray}
The action \eqref{eq:chirringcohmnewfermi02} of $\overline{\mathbf{Q}}_{+}$ on the component fields can be identified in the B/2-twisted theory
with a cohomology problem
via the map
\begin{align}
\psib_+^{\alphab}&\rightarrow d\xb^{\alphab}~,		 &\psib_{-,A}&\rightarrow \partial_{A}~,
\end{align}
where we write $\partial_{A}$ for a local basis of $\mathcal{E}$. It then follows that the operators (\ref{eq:02states}) (for $t=0$)
are mapped to $(0,\bullet)$ differential horizontal forms valued in $\wedge^{\bullet}\mathcal{E}$
\begin{align}\label{identops02}
\mathcal{O}(\omega)\rightarrow \omega\in \Omega^{0,\bullet}(\mathbf{Z},\wedge^{\bullet}\mathcal{E})~.
\end{align}
Under this identification, we map the supercharge $\overline{\mathbf{Q}}_{+}$ to a differential operator
\begin{align}
\overline{\mathbf{Q}}_{+}\rightarrow \delta:=\overline{\partial}-\iota_{J}~,
\end{align}
where $\iota_{J}$ acts as
\begin{align}
\iota_{J}\circ\omega^{s}_{r}=s(-1)^{r}J_{A}\omega^{A,A_{1},\ldots,A_{s-1}}_{r}\partial_{A_{1}}\wedge\ldots\wedge\partial_{A_{s-1}}~,
\end{align}
for any $\omega^{s}_{r}\in \Omega^{(0,r)}(\mathbf{Z},\wedge^{s}\mathcal{E})$.

Now we turn to the computation of the $S^2$ correlators via localization. The procedure is analogous to the (2,2) case, that is,
we localize with respect to a subset of dynamical fields $\{ x^{\alpha},\xb^{\alphab},\psi_{+}^{\alpha},\psib_+^{\alphab}\}$,
and we keep the remaining ones as background fields. The saddle point equations in this case are just
\begin{eqnarray}
\delta \psi^\alpha_+=i\overline{\partial}_{\bar{z}}x^{\alpha}=0~.
\end{eqnarray}
In contrast to the (2,2) case, this does not imply $x=\text{const.}$, but $x$ is allowed more generically to be a holomorphic map.
This in fact implies that worldsheet instanton corrections to B/2 model correlators cannot be ruled out.
For the purpose of deriving our formula we will ignore such corrections, that is, we assume that in fact the stronger condition $x=\text{const}$ holds.
In section \ref{ss:instcorrs} we will address the question of whether instanton corrections do appear in our models.
However, let us point out here that there are several known classes of examples in the literature where
it has been shown that such corrections are absent in their B/2-twisted versions \cite{McOrist:2008ji}.
With our assumption, $\psi_{+}^{\alpha}$ and $\psi^{A}_{-}$ have no zero modes, and integrating over the auxiliary fields $F,\Fb$ we obtain
\begin{align}
\label{eq:02formula}
\langle\mathcal{O}(\omega)\rangle_{S^{2}}:=\int_{\bZ}d^{2}x\int(\prod_{\alphab} d\psib_+^{\alphab}) (\prod_{\Ab}d \psib_-^{\Ab}) \exp\left(-\frac{\mathbf{v}}{4}\| J \|^{2}
+\frac{\mathbf{v}}{4}\Db_{\alphab}\Jb_{\Ab} \psib_+^{\alphab} \psib_-^{\Ab}\right)\mathcal{O}(\omega)~.
\end{align}
Some remarks about the Fermi measure are in order. The two contributions are identified with sections of the bundles
\begin{align}
\prod_{\alphab} d\psib_+^{\alphab}&\in \Gamma(\Kb_{\bZ})~,	 	&\prod_{\Ab}d \psib_-^{\Ab}&\in \Gamma(\wedge^R\cEb^\ast)~,
\end{align}
where $\Kb_{\bZ}:=\wedge^d \Tb_{\bZ}^\ast$ is the anti-holomorphic anti-canonical bundle of $\bZ$.
The requirement that the measure is well-defined
now reads
\begin{align}
\label{eq:02Fermimescond}
\Kb_{\bZ}\otimes(\wedge^{R}\overline{\mathcal{E}})^\ast \cong \cOb_{\bZ}~,
\end{align}
since $\cOb_{\bZ}$ always admits a nowhere vanishing global section. This condition was obtained also in \cite{Sharpe:2006qd} and can be rewritten,
upon an overall conjugation of \eqref{eq:02Fermimescond}, as
\begin{eqnarray}\label{ccondbundle02}
\wedge^{R}\mathcal{E}\cong K_{\bZ}^\ast~.
\end{eqnarray}
Since $\wedge^{R}\mathcal{E}$ and $K_{\bZ}^{*}$ are line bundles over $\bZ$, topologically they are determined by their first Chern class.
Thus (\ref{ccondbundle02}) is equivalent to the condition $c_{1}(T\bY)=0$, where $\bY=\mathrm{tot}(\mathcal{E}\rightarrow\bZ)$, and in particular
it coincides with the second anomaly cancellation condition in \eqref{eq:anomaly02goem}. Thus, we do not need any additional constraints
on our models in order for the B/2-twisted theory to be well-defined.

Another assumption we implicitly made in writing the above, following \cite{McOrist:2008ji} and \cite{Katz:2004nn}, is that the ratio of 1-loop determinants
\begin{eqnarray}
\frac{\mathrm{det}\overline{\partial}_{x^{*}\mathcal{E}}}{\mathrm{det}\overline{\partial}_{x^{*}\mathcal{T_{\mathbf{Z}}}}}
\end{eqnarray}
is a number, which we can just ignore.

Finally, we integrate over the Fermi zero modes and we obtain
\begin{eqnarray}
\label{eq:02corr}
\la \cO(\omega) \ra_{S^2}=\int_{\mathbf{Z}}\Omega_{\mathcal{E}}\lrcorner e^{\widehat{L}} \omega~,
\end{eqnarray}
where, by a slight abuse of notation, we define
\begin{align}
\widehat{L}:=-\frac{\mathbf{v}}{4}\| J \|^{2}+\frac{\mathbf{v}}{4}\overline{\partial}\overline{J}^{A}\partial_{A}~,
\end{align}
and $\Omega_{\mathcal{E}}$ is the nowhere vanishing section
\begin{align}
\Omega_{\mathcal{E}}\in \Gamma(K_{\bZ}\otimes  \wedge^{R}\mathcal{E})=\Gamma( \mathcal{O}_{\bZ})~.
\end{align}
In particular, $\Omega_{\mathcal{E}}$  is unique up to rescaling by a non-vanishing holomorphic section of $\mathcal{O}_{\bZ}$.
In deriving \eqref{eq:02corr} we used the condition that, if $s:=\mathcal{H}\oplus g$ is a hermitian K\"ahler metric on the total space
$\mathcal{E}\rightarrow \mathbf{Z}$, then the unique compatible connection is given by $s^{-1}\partial s$, which implies that
\begin{eqnarray}
\Gamma_{\bar{\alpha} C}^{A}=\Gamma_{\alpha \overline{C}}^{\overline{A}}=0~.
\end{eqnarray}
Thus, if we define
\begin{eqnarray}
\overline{J}^{A}:=\mathcal{H}^{A\overline{B}}\overline{J}_{\overline{B}}~,
\end{eqnarray}
it follows that
\begin{eqnarray}
\overline{D}_{\bar{\alpha}}\overline{J}^{A}=\overline{\partial}_{\bar{\alpha}}\overline{J}^{A}~.
\end{eqnarray}
The operator $\Omega_{\mathcal{E}}\lrcorner$ is defined as
\begin{align}
\Omega_{\mathcal{E}}\lrcorner\omega^{s}_{r}&:=f_{\bZ}\epsilon_{\alpha_{1}\cdots \alpha_{d}A_{1}\cdots A_{s}A_{s+1},\cdots A_{R}}\omega^{A_{1}\cdots A_{s}}_{\bar{\beta}_{1}\cdots \bar{\beta}_{r}}d\bar{x}^{\bar{\beta}_{1}}\wedge\cdots\wedge d\bar{x}^{\bar{\beta}_{r}}\nonumber\\
&\qquad\qquad\qquad\qquad\otimes e^{A_{s+1}}\wedge\cdots\wedge e^{A_{R}} \otimes dx^{\alpha_{1}}\wedge\cdots\wedge dx^{\alpha_{d}}~,
\end{align}
where $f_{\bZ}$ is a nowhere vanishing function and $\{ e^{A}\}$ is the dual basis to $\partial_{A}$.

Let $\mathcal{O}(\omega)\in \Omega^{0,p}(\mathbf{Z},\wedge^q\mathcal{E})$.
Then, necessary conditions to have a non-vanishing correlator are
\begin{align}
p-q&=\dim\mathbf{Z}-\mathrm{rank} \mathcal{E}~,		&\bq(\mathcal{O}(\omega)) &= \mathfrak{r}~,
\end{align}
which follow by requiring \eqref{eq:02formula} to be invariant under the chiral symmetries $\GU(1)_L\times\GU(1)_R$.

As a final remark, for the case of a $(0,2)$ LG model, we have $\bZ=\mathbb{C}^{d}$ and $\mathcal{E}$
a trivial bundle of rank $R$. Then, $\Omega_{\mathcal{E}}=\Omega_{\mathbb{C}^{d}}\wedge\Omega_{\mathbb{C}^{R}}$
and therefore the correlator (\ref{eq:02corr}) reproduces exactly the one derived in \cite{Melnikov:2009nh},
where the action of $\Omega_{\mathbb{C}^{R}}\lrcorner$ becomes simply the contraction with the Levi-Civita symbol of rank $R$.

\subsection{Properties of (0,2) correlators}

In this section we derive some formal properties for our formula.
We start by showing that \eqref{eq:02corr} does not depend on the representatives in $\cH_{\text{B/2}}$.
Let us define
\begin{align}
\label{primedd02}
\langle \beta\rangle'&:=\int_{\mathbf{Z}}\Omega_{\mathcal{E}}\lrcorner\beta~,		 &\beta&\in \Gamma(\Omega^{0,\bullet}(\bZ,\wedge^{\bullet}\mathcal{E}))~.
\end{align}
In order for the integral to be well-defined, we assume $\beta$ to be compactly supported along the fiber $X$.
In particular, the quantity $\la \delta \beta \ra'$ is identically zero unless $\delta \beta \in\Gamma(\wedge^R_d\cE)$. Thus,
$\beta\in \Gamma(\wedge^R_{d-1}\cE)$ and $\delta \beta = \pb \beta$.
Since $\Omega_{\mathcal{E}}$ is a section of a holomorphic bundle, we have
\begin{align}
\langle \delta \beta\rangle'=\int_{\mathbf{Z}}\Omega_{\mathcal{E}}\lrcorner\pb\beta=\int_{\mathbf{Z}}d(\Omega_{\mathcal{E}}\lrcorner\beta)=0~,
\end{align}
where we assumed that $\beta$ has no poles.
Let us now take a look at $\delta e^{\hat{L}}$. By a direct computation it is possible to show that
\begin{eqnarray}
\label{eq:deltaexpL02}
\delta e^{\hat{L}}\sim J_{A}\overline{\partial}\overline{J}^{A}\wedge\overline{\partial}\overline{J}^{A_{1}}\wedge\ldots\wedge\overline{\partial}\overline{J}^{A_{R}}\partial_{A_{1}}\wedge\ldots\wedge\partial_{A_{R}}
\in \Gamma\left(\Omega^{(0,R+1)}\times\wedge^{R}\mathcal{E}\right)~.
\end{eqnarray}
If $\rank\ \mathcal{E}\geq \dim \mathbf{Z}$,\footnote{We remark that almost all the examples in the literature, among which deformations of (2,2) hybrids and (0,2) LG models,
satisfy this condition, but there are examples that elude it. For instance, the authors of \cite{Andreas:2011cx} construct a class of stable rank-2 bundles over CY${}_3$
satisfying the heterotic anomaly conditions.}
$\delta e^{\hat{L}}$ vanishes identically because it is a $(0,R+1)$ form. If $\rank\ \mathcal{E}< \dim \mathbf{Z}$ instead, we note that it is possible to write (\ref{eq:deltaexpL02}) as
\begin{eqnarray}
\delta e^{\hat{L}}\sim J_{A}\overline{\partial}\overline{J}^{A}\wedge\overline{\partial}\overline{J}^{1}\wedge\ldots\wedge\overline{\partial}\overline{J}^{R}\epsilon^{A_{1},\ldots,A_{R}}\partial_{A_{1}}\wedge\ldots\wedge\partial_{A_{R}}~.
\end{eqnarray}
In particular, this expression always involves the product $\overline{\partial}\overline{J}^{A}\wedge\overline{\partial}\overline{J}^{A}=0$ for some $A$. Hence, we conclude that
\begin{eqnarray}
\delta e^{\hat{L}}=0~,
\end{eqnarray}
and that
\begin{eqnarray}
\langle \delta\mathcal{O}(\omega)\rangle_{S^{2}}=\langle \delta(e^{\hat{L}}\omega)\rangle'=0~,
\end{eqnarray}
which proves that B/2-correlators do not depend on the representatives of the $\delta$-cohomology classes.

With this result at our disposal, we can proceed further and show that our formula (\ref{eq:02corr})
is invariant under variations of the following parameters:\footnote{In the following
we use the notation $\delta\overline{J}$, $\delta\mathcal{H}_{\overline{A}B}$ and $\delta\bv$ for the variations of
the parameters and $\delta(\bullet)$ for the differential acting on $(\bullet)$. We hope this is not too confusing for the reader.}
\begin{enumerate}
\item anti-holomorphic parameters $\overline{J}\rightarrow \overline{J}+\delta\overline{J}$;
\item worldsheet volume $\mathbf{v}\rightarrow\mathbf{v}+\delta\mathbf{v}$;
\item metric $\mathcal{H}_{A\overline{B}}\rightarrow \mathcal{H}_{A\overline{B}}+\delta\mathcal{H}_{\overline{A}B}$
such that $\delta\mathcal{H}_{AB}=\delta\mathcal{H}_{\Ab\Bb}=0$.
\end{enumerate}
It is possible to show that under the above variations, the exponential term in \eqref{eq:02corr}
transforms, respectively, as
\begin{eqnarray}
\label{eq:expLvarpars}
\delta_{J}e^{\hat{L}}&=&\delta\left(\frac{\mathbf{v}}{4}\mathcal{H}^{A\overline{B}}\delta\overline{J}_{\overline{B}}\partial_{A} \right)\wedge e^{\hat{L}}~, \nonumber\\
\delta_{\bv}e^{\hat{L}}&=& \delta\left(\frac{\delta\mathbf{v}}{4}\overline{J}^{A}\partial_{A} \right)\wedge e^{\hat{L}}~,\nonumber\\
\delta_{\mathcal{H}}e^{\hat{L}}&=& \delta\left(\frac{\mathbf{v}}{4}\delta\mathcal{H}^{A\overline{B}}\overline{J}_{\overline{B}}\partial_{A} \right)\wedge e^{\hat{L}}~.
\end{eqnarray}
Hence, given any operator $\mathcal{O}(\omega)$ satisfying $\delta\mathcal{O}(\omega)=0$, we have
\begin{eqnarray}
\delta_{J}\langle \mathcal{O}(\omega)\rangle_{S^{2}}=\delta_{\mathcal{H}}\langle \mathcal{O}(\omega)\rangle_{S^{2}}=\delta_{\bv}\langle \mathcal{O}(\omega)\rangle_{S^{2}}=0~,
\end{eqnarray}
as claimed, where we used the property
\begin{eqnarray}
\langle \delta(\alpha)\wedge e^{\hat L} \mathcal{O}(\omega)\rangle' = \la \delta\left( \alpha\wedge e^{\hat L} \cO(\omega)  \right)\ra'=0~,
\end{eqnarray}
where $\alpha\in\Gamma(\Omega^{0,\bullet}(\bZ,\wedge^{\bullet}\mathcal{E}))$ is determined in each case in \eqref{eq:expLvarpars}.

\subsection{The transformation law for (0,2) hybrid integrals}
\label{ss:02transflaw}

We now turn to a proposal for a generalized transformation law for the integrals (\ref{eq:02corr}) that arise in the B/2-twisted hybrid correlators.
Our derivation can be regarded as a natural generalization of the argument presented in section \ref{ss:transflaw} for (2,2) correlators.
The key assumption is that there exists a section $\mathcal{B}$ of $\mathrm{Hom}(\mathcal{E}^{*},\mathcal{E}^{*})$ such that
\begin{align}
\mathcal{B}J&=T~, 	&\partial_{\psi}T&=0~,
\end{align}
where $T\in\Gamma(\cE^\ast)$ satisfies the potential condition $T^{-1}(0)=B$
and it does not depend on the parameters of $J$, which we denoted collectively as $\psi$.
Recall that $\cE\rightarrow\bZ$ is equipped with a hermitian metric $\mathcal{H}_{A\overline{B}}$.
Let us define a new metric $h$ by
\begin{eqnarray}
\label{transfmetric}
\mathcal{H}^{-1}=\mathcal{B}^{t}h^{-1}\mathcal{B}~.
\end{eqnarray}
Note that (\ref{transfmetric}) is not well-defined in all of $\mathbf{Z}$ since $\mathcal{B}$, in general, is not invertible at every point $x^\alpha\in\mathbf{Z}$.
Nonetheless, we assume that the argument can be extended to those points as well.
Then, by substituting (\ref{transfmetric}) into (\ref{eq:02corr}) we obtain the transformed correlator, which reads
\begin{eqnarray}\label{inttransformed}
\langle\mathcal{O}(\omega)\rangle= \int_{\mathbf{Z}}\Omega_{\mathcal{E}}\lrcorner e^{\hat{L}'}\omega~,
\end{eqnarray}
where
\begin{eqnarray}
\label{eq:hatLpr}
\hat{L}'=-\frac{\mathbf{v}}{4}\|T\|^{2}_{h}+\frac{\mathbf{v}}{4}(\mathcal{B}^{t})^{A}_{ \ B }\overline{\partial}\overline{T}^{B}\partial_{A}~.
\end{eqnarray}
Although $\hat{L}'=\hat L$, we wish to emphasize that in \eqref{eq:hatLpr} the indices are contracted with $h$ instead of with $\mathcal{H}$, that is,
\begin{align}
\|T\|^{2}_{h}&:=T_{A}h^{A\overline{A}}\overline{T}_{\overline{A}}~,	&\overline{T}^{B}&:=h^{B\overline{B}}\overline{T}_{\overline{B}}~.
\end{align}
At this point we invoke the property, shown above, that the correlators are independent of variations $\delta h^{A\overline{A}}$, and thus
we argue that $h$ can be considered independent of the parameters $\psi$. In order to obtain a more explicit formula,
let $\omega$ be a section of $\wedge^{q}_{p}\mathcal{E}$ satisfying $p-q=\dim\mathbf{Z}-R$, and let
\begin{eqnarray}
\label{eq:alphadef02}
\alpha:=(\mathcal{B}^{t})^{[A_{1}}_{ \ B_{1} }\overline{\partial}\overline{T}^{B_{1}}\cdots(\mathcal{B}^{t})^{A_{k}}_{ \ B_{k} }\overline{\partial}\overline{T}^{B_{k}}\omega^{A_{k+1},\ldots,A_{R}]}
\in \Gamma\left(\Omega^{0,d}(\mathbf{Z},\wedge^{R}\mathcal{E})\right)~,
\end{eqnarray}
where the antisymmetrization $[\cdots]$ is over the $A_{i}$ indices only, and $k:=R-q$.
For ease of notation, in \eqref{eq:alphadef02} we omitted the indices corresponding to the form degree.
Then, (\ref{inttransformed}) can be written as
\begin{eqnarray}\label{alphaform}
\langle\mathcal{O}(\omega)\rangle= \int_{\mathbf{Z}}e^{\frac{-\mathbf{v}}{4}\|T\|^{2}}\Omega_{\mathcal{E}}\lrcorner\alpha~.
\end{eqnarray}
Let us define\footnote{This is the analogue of the matrix $\mathcal{M}$ we defined in section \ref{ss:transflaw} for the (2,2) case.} 
$M$ to be the $R\times d$ matrix whose components are given by $M^{A}_{\bar{\alpha}}:= (\mathcal{B}^{t} \overline{\partial}\overline{T})^{A}_{\bar{\alpha}}$
and let $M^{\bar{\alpha}_{1},\ldots,\bar{\alpha}_{r}}_{A_{1},\ldots,A_{s}}$ be the completely skew-symmetric tensor where each component is given by the determinant
of the $(R-s)\times (d-r)$ minor of $M$ obtained by removing the columns $\bar{\alpha}_{1},\ldots,\bar{\alpha}_{r}$ and the rows $A_{1},\ldots,A_{s}$.
With this notation, (\ref{alphaform}) finally reads
\begin{eqnarray}
\label{eq:transfloafin}
\langle\mathcal{O}(\omega)\rangle= \int_{\mathbf{Z}}d^{2d}xe^{\frac{\mathbf{v}}{4}\|T\|^{2}}f_{\bZ}(x)
M^{\bar{\alpha}_{1},\ldots,\bar{\alpha}_{p}}_{A_{1},\ldots,A_{q}}\omega^{A_{1},\ldots,A_{q}}_{\bar{\alpha}_{1},\ldots,\bar{\alpha}_{p}}~.
\end{eqnarray}
The functions $M^{\bar{\alpha}_{1},\ldots,\bar{\alpha}_{p}}_{A_{1},\ldots,A_{q}}\omega^{A_{1},\ldots,A_{q}}_{\bar{\alpha}_{1},\ldots,\bar{\alpha}_{p}}$
are $C^{\infty}$ sections of $\mathcal{O}_{\mathbf{Z}}$ with coefficients given by rational functions of the parameters $\psi$.
Thus, the integral \eqref{eq:transfloafin} can be expanded as a sum of integrals where the $\psi$ dependence can be simply factored out.
Unfortunately, it is not possible to derive a more explicit form for the correlators, as in (\ref{xidepen}),
due to the fact that the property \eqref{eq:Mminasdets} does not hold for a generic $\cE$, since in general $\rank\ \mathcal{E}\neq \dim \mathbf{Z}$.

\subsection{Instanton corrections}
\label{ss:instcorrs}

In the case of $(0,2)$ models, B/2 correlators may be subject to instanton corrections.
In this section, we analyze the structure of Fermi zero modes in the background of a non-trivial instanton,
and we derive simple selection rules for the absence of such corrections.

In hybrid models, worldsheet instantons are associated to holomorphic maps from the worldsheet into $B$.
These are characterized by homology classes in $H_2(B,\Z)$.
Picking a basis $\xi^a$ for $H_2(B,\Z)$, a given homology class
is determined by a set of integers $n_a$, which we denote instanton numbers.

The Fermi fields of the B/2-twisted theory couple over the worldsheet to the following bundles
\begin{align}
\label{eq:fermizmbundles}
\xymatrix@C=5mm@R=0mm{
				&\psi_+^\alpha					&\psib_+^{\alphab}				&\psi_-^A					&\psib_-^{\Ab}\\
\text{bundle}		&\Kb_\Sigma\otimes x^\ast T_{\bZ}	&x^\ast \overline{T}_{\bZ}			&K_\Sigma\otimes x^\ast\cE	&x^\ast\overline{\cE}\\
\text{\# z.m.}		&h^1(\Sigma,x^\ast \Tb_{\bZ})		&h^0(\Sigma,x^\ast \Tb_{\bZ})		&h^1(\Sigma,x^\ast \cE^\ast)	&h^0(\Sigma,x^\ast \cE^\ast)		~,
}
\end{align}
and the number of zero modes is computed by the appropriate cohomology group.\footnote{We emphasized in \eqref{eq:fermizmbundles} that
the zero modes of $\psi_+$ and $\psib_+$ are counted by anti-holomorphic sections of the appropriate bundles, while $\psi_-$ and $\psib_-$
by holomorphic ones.
Of course, this does not affect the dimensions of the cohomology groups.}
The computation is fairly treatable given the fact that these bundles split over $\Sigma=\CP^1$ as sums of line bundles
\begin{align}
\label{eq:pullbacksplit}
x^\ast(T_{\bZ})&=\oplus_{\alpha=1}^d \cO(d_\alpha)~,		&x^\ast(\cE)&=\oplus_{A=1}^R \cO(D_A)~,
\end{align}
where the degrees $d_\alpha,D_A$ depend on the instanton numbers $n_a$.
The cohomology of $x^\ast(T_{\bZ})$ is given by
\begin{align}
h^0(\Sigma,x^\ast T_{\bZ}) &=\sum_{\alpha|d_\alpha\geq0}(d_\alpha+1) ~,
&h^1(\Sigma,x^\ast T_{\bZ}) &= \sum_{\alpha|d_\alpha<0}(-d_\alpha-1)~,
\end{align}
and for $\cE^\ast$ similarly
\begin{align}
h^0(\Sigma,x^\ast \cE^\ast)&=\sum_{A|D_A\leq0}(-D_A+1)~,
&h^1(\Sigma,x^\ast \cE^\ast)&=\sum_{A|D_A>0}(D_A-1)~.
\end{align}

Given a specific instanton background, there will be a non-trivial contribution to the correlator
if we can absorb all the Fermi zero modes from the measure.
From the general expression of the insertions \eqref{eq:02states},
the only possibility to soak up zero modes of $\psi_{\pm}$ is by bringing down appropriate terms from the action.
There are only two terms in the action that contain these fields:
the Yukawa coupling $D_\alpha J_A \psi_+^\alpha \psi_-^A$ and
the curvature four-fermi term. It then follows that the contribution can only be in the form of the product $\psi_-\psi_+$.
Hence, the instanton contribution must vanish unless these zero modes appear in equal number in the measure, that is, if
\begin{align}
\label{eq:psipsiinst}
\sum_{\alpha|d_\alpha<0}(-d_\alpha-1) =\sum_{A|D_A>0}(D_A-1)~.
\end{align}
For the zero modes of $\psib_{\pm}$, in addition to the four-fermi term and the Yukawa coupling $\Db_{\alphab} \Jb_{\Ab} \psib_+^{\alphab} \psib_-^{\Ab}$,
there are in general contributions from the insertion itself.
Thus, the instanton contribution to the correlator $\la \alpha \ra_{S^2}$, where $\alpha\in\Gamma(\wedge^r_s\cE)$
vanishes unless
\begin{align}
r-s&=\sum_{A|D_A\leq0}(-D_A+1)-\sum_{\alpha|d_\alpha\geq0}(d_\alpha+1)~.
\end{align}
Combining this with \eqref{eq:psipsiinst}, and using the fact that $c_1(T_{\bZ})+c_1(\cE)=0$, we obtain the expected relation
\begin{align}
r-s&=\rank\ \cE-\dim\bZ~.
\end{align}
If $\cE=T_{\bZ}$, which describes a subset of deformations of a (2,2) model, these formulae simplify considerably. Trivially $r=s$, as
$\rank\ T_{\bZ}=\dim\bZ$, while \eqref{eq:psipsiinst} becomes
\begin{align}
\sum_{\alpha|d_\alpha\geq 1} (d_\alpha-1) = \sum_{\alpha|d_\alpha \leq -1} (-d_\alpha-1)~,
\end{align}
which, using the fact that $c_1(T_{\bZ})=0$, gives a simple selection rule, which reads
\begin{align}
\label{eq:psipsicondTZ}
\big| \big\{ \alpha |d_\alpha \geq 1\big\}\big| = \big|\big\{ \alpha|d_\alpha \leq -1\big\}\big|~.
\end{align}
As a final comment, we would like to point out that these selection rules, although simple, are rather mild, and
instanton corrections in a given model can still be absent even if the conditions above are satisfied.
This is somewhat expected, as the discussion in this section only involves the hybrid geometry, and it is insensitive to the remaining structure
of the model. In the example we are going to study next, we will see that in order to fully exclude instanton corrections
we need to employ more sophisticated techniques.

\subsection{A (0,2) example}

We conclude this section with an example of a (0,2) hybrid. In particular, we choose a model which is obtained as a (0,2) deformation
of the octic model we studied in section \ref{s:22example}.
For the geometric data we choose again $\bZ=\tot\left(\cO(-2)\oplus\cO^{\oplus3}\rightarrow \CP^1\right)/\mathbb{Z}_{4}$ and $\cE=T_{\bZ}$,
and $\GU(1)_V$ acts with charges $q_i=\frac14$, $i=1,\dots,4$ on the fiber coordinates.
We choose the (0,2) superpotential
\begin{align}
\label{eq:02supoctu02u}
J_0^u&=u^7(\phi^1_u)^4-\psi_1 \phi_u^1\phi_u^2\phi_u^3\phi_u^4~,		&J_1^u&=\half(u^8+1)(\phi_u^1)^3-\psi_1 u \phi_u^2\phi_u^3\phi_u^4~,\nonumber\\
J_2^u&= (\phi_u^2)^3 - \psi_2 u\phi_u^1\phi_u^3\phi_u^4~,			&J_3^u&= (\phi_u^3)^3 - \psi_3 u\phi_u^1\phi_u^2\phi_u^4~,	
&J_4^u&= (\phi_u^4)^3 - \psi_4 u\phi_u^1\phi_u^2\phi_u^3~,
\end{align}
in the patch $U_1$, and
\begin{align}
\label{eq:02supoctu02v}
J_0^u&=v^7(\phi^1_v)^4-\psi_1 \phi_v^1\phi_v^2\phi_v^3\phi_v^4~,		&J_1^v&=\half(v^8+1)(\phi_v^1)^3-\psi_1 v \phi_v^2\phi_v^3\phi_v^4~,\nonumber\\
J_2^v&= (\phi_v^2)^3 - \psi_2 v\phi_v^1\phi_u^3\phi_u^4~,			&J_3^v&= (\phi_v^3)^3 - \psi_3 v\phi_v^1\phi_v^2\phi_v^4~,	
&J_4^v&= (\phi_v^4)^3 - \psi_4 v\phi_v^1\phi_u^2\phi_u^3~,
\end{align}
in the patch $U_2$.
For generic values of the parameters, \eqref{eq:02supoctu02u} and \eqref{eq:02supoctu02v} are not integrable to a single function, thus defining a (0,2) model.
When $\psi_1=\cdots=\psi_4$ we recover the (2,2) superpotential \eqref{eq:22supoct}.

\subsubsection*{Instanton corrections?}

The geometry of this example is fairly simple, which allows us to be explicit. Here $B=\CP^1$, thus
instantons are classified by an integer $n\in\Z_{\geq0}$, and since we are interested in probing
non-trivial instanton corrections we restrict to the case $n>0$. The splitting \eqref{eq:pullbacksplit} is given by
\begin{align}
x^\ast(T_{\bZ})&=\cO(-2n)\oplus\cO^{\oplus3}\oplus\cO(2n)~.
\end{align}
Thus, \eqref{eq:psipsicondTZ} is satisfied for each value of $n$, and it appears that the correlators do receive instanton corrections,
in disagreement with \cite{McOrist:2008ji}.
In order to resolve this apparent puzzle, we implement the same approach as in \cite{Sharpe:2006qd}.
That is, we look for a suitable compactification of the space of worldsheet instantons, as well as a suitable extension of the sheaves \eqref{eq:fermizmbundles} over the moduli space.

Following \cite{Bertolini:2017lcz}, we construct a linear model ($V^+$ model in the terminology of \cite{Morrison:1994fr}) with target space $\cO(-2)\oplus\cO^{\oplus3}\rightarrow \CP^1$
by introducing the (0,2) chiral matter superfields
\begin{align}
\label{eq:gauchar02GLSMex}
\xymatrix@R=0mm@C=3mm{
			&X_1	&X_2	&P_1	&P_2	&P_3	&P_4	&\text{F.I.}\\
\GU(1)		&1		&1		&-2		&0		&0		&0		&r
}
\end{align}
together with a neutral chiral field $\Sigma$, with lowest component $\sigma$.
We indicate by $x_{1,2}$ and $p_{1,2,3,4}$ the lowest components of the superfields $X_{1,2}$ and $P_{1,2,3,4}$, respectively,
and by $r$ the F.I.~parameter.
We introduce a collection of (0,2) Fermi fields with the same gauge charge assignments
\begin{align}
\xymatrix@R=0mm@C=3mm{
			&\Gamma^1		&\Gamma^2		&\Lambda^1		&\Lambda^2		&\Lambda^3		&\Lambda^4\\
\GU(1)		&1				&1				&-2				&0				&0				&0		
}
\end{align}
which satisfy the chirality conditions
\begin{align}
\label{eq:Fermi02GLSM}
\cDb\Gamma^1&=\Sigma x_1~,	&\cDb\Gamma^2&=\Sigma x_2~,	&\cDb\Lambda^1&=-2\Sigma p_1~,	&\cDb\Lambda^2&=\cDb\Lambda^3=\cDb\Lambda^4=0~.
\end{align}
We choose the following assignment for the chiral symmetries
\begin{align}
\xymatrix@R=0mm@C=5mm{
			&X_{1,2}		&P_{1,2,3,4}			&\Gamma^{1,2}		&\Lambda^{1,2,3,4}		&\Sigma		\\
\GU(1)_L		&0			&\frac14				&-1					&-\frac34				&-1			\\
\GU(1)_R		&0			&\frac14				&0					&\frac14				&1
}
\end{align}
This allows us to introduce potential terms, which we take of the form $\Gamma\cdot J + \Lambda\cdot H$, where
\begin{align}
\label{eq:GLSMpot02}
J_1 &=x_1^7 p_1^4-\psi_1 x_2 p_1p_2p_3p_4~,				&J_2 &=x_2^7p_1^4-\psi_1 x_1p_1p_2p_3p_4~,		\nonumber\\
H_1&= \half(x_1^8+x_2^8)p_1^3 - \psi_3 x_1x_2p_2p_3p_4~,		&H_2&= p_2^3 - \psi_4 x_1x_2p_1p_3p_4~,\nonumber\\
H_3&= p_3^3 - \psi_3 x_1x_2p_1p_2p_4~,					&H_4&= p_4^3 - \psi_4 x_1x_2p_1p_2p_3~.
\end{align}
For $r>0$, $x_{1,2}$ cannot simultaneously vanish,
and $\sigma$ is instead forced to vanish,
while the F-term constraints from \eqref{eq:GLSMpot02} require $p_1=\cdots=p_4=0$. Thus, the classical vacuum of the theory is $B=\CP^1$.
In order for (0,2) supersymmetry to be unbroken, these need to satisfy the constraint\cite{Witten:1993yc}
\begin{align}
0= x_1J_1+x_2J_2-2p_1H_1~,
\end{align}
which holds for the superpotential \eqref{eq:GLSMpot02}.
The coordinates $p_{1,\dots,4}$ transform as sections of line bundles over $B$ specified by their gauge charges,
$p_1\in\cO(-2)$ and $p_{2,3,4}\in\cO$, while the massless left-moving
fermions are described by the SES
\begin{align}
\label{eq:SES02exTZ}
\xymatrix@R=0mm@C=8mm{
0\ar[r]	&\pi^\ast\cO		\ar[r]^-{g}
&\pi^\ast \cO(1)^{\oplus2}\oplus\pi^\ast\cO(-2) \oplus \pi^\ast \cO^{\oplus3} \ar[r]	&\cE	\ar[r]&0~,
}
\end{align}
where $g=\begin{pmatrix} x_1 &x_2 &-2p_1 &0&0&0 \end{pmatrix}$ is determined by \eqref{eq:Fermi02GLSM}.
This determines $\cE=T_{\bZ}$, and we have recovered our (0,2) hybrid model.

The gauge instantons for this model in the phase $r>0$
are characterized by an integer $n\in\Z_{\geq0}$,
and again we restrict to the case $n>0$.
In this case, $x_{1,2}\in\Gamma(\cO(n))$ have $n+1$ zero modes each, $p_1\in\Gamma(\cO(-2n))$ has no zero modes,
while $p_{2,3,4}\in\Gamma(\cO)$ have one zero mode each.
It appears that the moduli space of instantons is therefore non-compact.
However, the localization conditions
need to be supplemented by $J=H=0$, which imply $p_{1,2,3,4}=0$.
Thus, taking into account the quotient by the gauge action \eqref{eq:gauchar02GLSMex}
we find that the moduli space is $B_n=\CP^{2n+1}$, which is indeed compact.

The strategy, following \cite{Katz:2004nn}, to determine the extension of the sheaves \eqref{eq:fermizmbundles} over the moduli space $B_n$, is to expand
the various Fermi fields into zero modes, and then interpret the coefficients of the expansion
as line bundles over the moduli space.\footnote{If one denotes the moduli space of instantons $x:\Sigma\rightarrow \bZ$ by $\mathcal{M}$,
the universal instanton map by $\alpha:\Sigma\times \mathcal{M}\rightarrow\bZ$
and the projection $\pi:\Sigma\times\mathcal{M}\rightarrow\mathcal{M}$,
then the sheaves where the zero modes of the fermions belong are given by
(possibly dual or conjugates of) $\mathcal{F}_{i}=R^{i}\pi_{*}\alpha^{*}\mathcal{E}^{*}$
and $\mathcal{G}_{i}=R^{i}\pi_{*}\alpha^{*}T_{\bZ}$. It is crucial for this analysis to choose
a compactification of $\mathcal{M}$ and an extension of these bundles over it.
While there is no systematic method to do this in general, when a GLSM model is available this is indeed possible,
as pointed out in \cite{Katz:2004nn,Sharpe:2006qd}.} In particular, we are interested in the extensions of the bundles
$\Kb_\Sigma\otimes x^\ast \Tb_{\bZ}^\ast$ and $K_\Sigma\otimes x^\ast\cE$ over $B_n$.
The former is determined by the zero modes of the massless right-moving Fermi fields in $X_{1,2}$ and $P_{1,2,3,4}$,
which we denote collectively as $\rho_+^{\alpha'}$, $\alpha'=1,\dots,6$.
The latter is parametrized by the zero modes of the massless lowest components of left-moving fields $\Gamma^{1,2}$ and $\Lambda^{1,2,3,4}$,
which similarly we denote collectively as $\rho_-^{\alpha'}$.

Applying this procedure to the LES\footnote{In the following all the sheaves are over $B_{n}$.}
\begin{align}
\xy {\ar(0.05,-13)*{};(0.05,-15)*{}};
\xymatrix@R=2mm@C=8mm{
0 \ar[r]	&\cO \ar[r]	&{\begin{matrix} \cO(1)^{\oplus h^0(\Sigma,\cO(n))}\\\oplus\\  \cO(-2)^{\oplus h^0(\Sigma,\cO(-2n))}
\oplus \cO^{\oplus3} \end{matrix}}	\ar[r]	& {\begin{matrix} \\ \cG_0 \end{matrix}}
\ar@{-} `d[l]`[llld]  \\
 0 \ar[r]	&{\begin{matrix} \cO(1)^{\oplus h^1(\Sigma,\cO(n))}\\\oplus\\  \cO(-2)^{\oplus h^1(\Sigma,\cO(-2n))}\end{matrix}}	\ar[r]	& \cG_1		\ar[r]&0
}
\endxy
\end{align}
associated to the SES \eqref{eq:SES02exTZ} and using the fact that
\begin{align}
H^0(\Sigma,\Kb_\Sigma \otimes x^\ast \Tb_{\bZ}^\ast) = H^1(\Sigma, \Tb_{\bZ})^\ast ~,
\end{align}
we have that the zero modes of $\rho_+^{\alpha'}$ couple to
\begin{align}
\overline{\cG}_1^\ast&=\overline{\cO}(2)^{\oplus (2n-1)}~.
\end{align}
It is often more convenient to work with holomorphic bundles, thus we can make use of a Hermitian fiber metric
to dualize the bundle to $\cG_1 = \cO(-2)^{\oplus(2n-1)}$.

For the extension sheaves of zero modes of $\rho_-^{\alpha'}$,
we instead consider the LES induced by
the dual of \eqref{eq:SES02exTZ}, which simplifies to
\begin{align}
\label{eq:LESEastinst}
\xymatrix@R=0mm@C=10mm{
0 \ar[r]	& \cF_0	\ar[r]	& \cO(2)^{\oplus(2n+1)}	\ar[r]^-{\widetilde{g}=0}	&\cO	\ar[r] &
 \cF_1	\ar[r]	& \cO(-1)^{2(n-1)}	\ar[r]	&0~.
}
\end{align}
The map $\widetilde{g}$, induced from $g$, vanishes identically because $p_1$ has no zero modes. Therefore we obtain
\begin{align}
\cF_0&=\cO(2)^{\oplus(2n+1)}~,
&\cF_1&=\cO\oplus\cO(-1)^{\oplus2(n-1)}~.
\end{align}
Now, since we have
\begin{align}
H^0(\Sigma,K_\Sigma \otimes x^\ast T_{\bZ}) = H^1\left(\Sigma, T_{\bZ}^\ast\right)^\ast~,
\end{align}
it follows that the zero modes of $\rho_-$ couple to $\cF_1^\ast = \cO\oplus\cO(1)^{\oplus2(n-1)}$, 
while the zero modes of $\rhob_-$ couple to $\cF_0$.

Now, to solve the aforementioned puzzle, let us have a closer look at the term in the action $D_\alpha J_A \psi_+^\alpha \psi_-^A$, which
appears it can be used to soak up $\psi_\pm$ zero modes.
In our GLSM interpretation, this corresponds to a term in the action of the form $D_{\alpha'} J_{\beta'} \rho_+^{\alpha'} \rho_-^{\beta'}$, where
for simplicity we have grouped the various superpotential terms $J_{\alpha'}=(J,H)$.
Following the same logic as in \cite{Sharpe:2006qd},
we interpret $D_\alpha J_A$ as an element of $H^0(B_n,\cG_1\otimes \cF_1^\ast)$, and
\begin{align}
H^0(B_n,\cG_1\otimes \cF_1^\ast) &= H^0\left(\CP^{2n+1},\cO(-2)^{\oplus(2n-1)} \otimes \left(\cO\oplus\cO(1)^{\oplus2(n-1)}\right)\right)\nonumber\\
&=H^0\left(\CP^{2n+1},\cO(-2)\right)^{\oplus(2n-1)}  \oplus H^0\left(\CP^{2n+1},\cO(-1)\right)^{\oplus2(n-1)(2n-1)}=0~.
\end{align}
The other term that can be used to soak up $\psi_{\pm}$ zero modes is the 4-Fermi term 
$R_{\alpha\overline{B}A\bar{\beta}}\rho^{\alpha}_{+}\rho^{A}_{-}\rhob^{\overline{B}}_{-}\rhob^{\bar{\beta}}_{+}$. 
Again, following \cite{Sharpe:2006qd} we can interpret this term as an element in $H^1(B_n,\cG_1\otimes \cF_1^\ast\otimes \mathcal{F}_{0})$. 
Therefore we compute:
\begin{align}
H^1(B_n,\cG_1\otimes \cF_1^\ast\otimes \mathcal{F}_{0})=H^1\left(\CP^{2n+1},\cO\right)^{\oplus(2n-1)(2n+1)}  \oplus H^1\left(\CP^{2n+1},\cO(1)\right)^{\oplus2(n-1)(4n^{2}-1)}=0~.
\end{align}
Thus, the terms that we are required to bring down from the action in order to soak up the relevant Fermi zero modes are trivial in cohomology,
and they cannot contribute to the correlator. Hence, we conclude that no instanton corrections are possible for our example,
in agreement with \cite{McOrist:2008ji}.

Although in our (0,2) octic hybrid example a simple zero mode counting could not rule out instanton contributions,
our selection rules prove themselves useful in some non-trivial model. A nice example \cite{Aspinwall:2010ve} is the hybrid geometry
$\bZ=\tot\left(\cO(-1)^{\oplus2}\oplus\cO^{\oplus2}\rightarrow \CP^1\right)$ and $\cE=T_{\bZ}$, with $\GU(1)_V$ charge assignment
$q_i=1/4$ for the fiber coordinates.
The analysis of the corresponding linear model \cite{McOrist:2008ji} does not rule out instanton corrections.
However, from our hybrid perspective, it is obvious that \eqref{eq:psipsicondTZ} cannot be satisfied and thus instanton corrections are absent
in the B/2 model correlators.

\subsubsection*{B/2 correlators}

Let us now turn to the computation of the correlators, where for simplicity we restrict our attention to bottom row insertions.
A choice of representatives for the cohomology classes of the
B/2 chiral ring is still described by \eqref{eq:geninsertion}.
It has been shown in \cite{Bertolini:2013xga} that for this class of $J$-deformations, for which $\cE=T_{\bZ}$,
the dimension of the heterotic topological ring does not jump, and it agrees with its value at the (2,2) locus.
Thus, we can carry out the computation for any choice of non-singular superpotential $J$, in particular at $\psi_{1,2,3,4}=0$,
where it becomes isomorphic to our computation in section \ref{s:22example}.
That is, $\alpha = \cO_1\cO_2\cO_3$, where $\cO_{1,2,3}\in H_{\bQb_+}(\bZ,\cO_{\bZ})$, assumes again the form
\begin{align}
\label{eq:02exins}
\alpha = u^{t_0}(\phi^1)^{t_1}(\phi^2)^{t_2}(\phi^3)^{t_3}(\phi^4)^{t_4}~,		\qquad\qquad t_\alpha\geq0~.
\end{align}
Here the charge condition implies $t_1+\cdots+t_4=12$, and the condition that this is a restriction to the patch $U_1$ of the
product of sections of the trivial bundle $\cO_{\bZ}$ forces $t_0\leq 2t_1$.

Moreover, there exists $\cB\in \Gamma(T_{\bZ}\otimes T^\ast_{\bZ})$ such that
\begin{align}
\label{eq:02translawTJ}
T_I=\cB_{I}^{ \ J}J_J = \begin{pmatrix}  u^{25}\phi_1^{13}	&\half(u^{26}+1)\phi_1^{12}	&\phi_2^{13}	&\phi_3^{13}	&\phi_4^{13}  \end{pmatrix}^\top~.
\end{align}
Note that although this is a section of the same bundle $\cE=T_{\bZ}$, and $T$ is given by the same expression as in the (2,2) case,
$\cB$ is necessarily a different section, as it depends on
the parameters $\psi_1,\dots,\psi_4$. We report it in appendix \ref{app:transflawoct}, and we represent
its determinant as
\begin{align}
\det\cB = \sum_{m_0,\dots,m_4\geq0} \ft_{m_0,\dots,m_4}(\psi_1,\dots,\psi_4)u^{m_0} (\phi^1)^{m_1}(\phi^2)^{m_2}(\phi^3)^{m_3}(\phi^4)^{m_4}~.
\end{align}
Integrating over the fiber phases $\arg(\phi^i)$ we have that the only contribution to the integral is
\begin{align}
\alpha \det\cB = \sum_{m_0,\dots,m_4\geq0} \ft_{m_0,\dots,m_4}(\psi_1,\dots,\psi_4)u^{m_0+t_0}(\phi^1)^{24}\left(\prod_{a=2}^4(\phi^a)^{12}\right)
\delta_{m_1+t_1,24}\left(\prod_{a=2}^4\delta_{m_a+t_a,12}\right)~.
\end{align}
Putting all together, and integrating over $\arg(u)$, where we make use of \eqref{eq:antiholmdepub}, we obtain
\begin{align}
\la \alpha \ra_{S^2} = \ft_{24-t_0,24-t_1,12-t_2,12-t_3,12-t_4}(\psi_1,\dots,\psi_4)~.
\end{align}
As an example, we see that
\begin{align}
\la \det \Hess W\ra_{S^2} = \frac{1}{1-\psi_1^2\psi_2^2\psi_3^2\psi_4^2}~.
\end{align}
When $\psi_1=\cdots=\psi_4=\psi$ we recover the (2,2) correlator \eqref{eq:dethess22}.
More generally, for each insertion of the form \eqref{eq:02exins} we obtain the structure
\begin{align}
\la \alpha \ra_{S^2} = \frac{\psi_1^{v^{\alpha}_1}\psi_2^{v^{\alpha}_2}\psi_3^{v^{\alpha}_3}\psi_4^{v^{\alpha}_4}}{1-\psi_1^2\psi_2^2\psi_3^2\psi_4^2}~,
\end{align}
where $v^\alpha_i$ are non-negative integers that depend on the form of the insertion $\alpha$. In particular, we have that for this class of
(0,2) deformations the discriminant is the locus $\psi_1^2\psi_2^2\psi_3^2\psi_4^2=1$, which determines the locus where the
condition $J^{-1}(0)=B$ fails to be satisfied.

To check our results we invoke again the GLSM. In particular, our (0,2) hybrid arises in a phase of a linear model which is obtained as a (0,2) deformation
of the (2,2) GLSM we studied in section \ref{s:22example}. In particular,
the phase structure is unaltered by the class of deformations considered, and in the cone $r_2<0,2r_1+r_2<0$
the model exhibits a LG phase, where the hybrid
(0,2) superpotential \eqref{eq:02supoctu02u} corresponds to the LG (0,2) superpotential
\begin{align}
\label{eq:02supoctuLG}
J^{\text{LG}}_2&= x_2^3 - \psi_2 x_3x_4x_5x_6~,			&J^{\text{LG}}_3&= x_3^3 - \psi_3 x_2x_4x_5x_6~,	
&J^{\text{LG}}_4&= x_4^3 - \psi_4x_2x_3x_5x_6~,\nonumber\\
J^{\text{LG}}_5&=x_5^7 - \psi_1 x_2x_3x_4x_6		&J^{\text{LG}}_6&=x_6^7 - \psi_1 x_2x_3x_4x_5~.
\end{align}
The B/2 ring in this phase is described by
$R=\C[x_2,\dots,x_6]/\la J_{\text{LG}}\ra$. Invariance under the $\Z_8$ orbifold action
implies that good representatives for the ring are given again by \eqref{eq:LGchirringel} and
the correspondence
between hybrid and LG operators is quite straightforward
\begin{align}
\label{eq:cxstrGLSM02}
\cO_{\text{LG}}=\prod_{a=2}^4 x_a^{l_a} x_5^{l_5}x_6^{l_6}\quad	\longleftrightarrow\quad 	\cO_{\text{HY}} = u^{l_5}(\phi^1)^{(l_5+l_6)/2}\prod_{a=2}^4 (\phi^a)^{l_a} ~.
\end{align}
By applying the methods of \cite{McOrist:2007kp} we verify explicitly that the following holds
\begin{align}
\la \cO^1_{\text{HY}} \cO^2_{\text{HY}} \cO^3_{\text{HY}} \ra_{S^2} = \la \cO^1_{\text{LG}}\cO^2_{\text{LG}}\cO^3_{\text{LG}}\ra
= \text{Res}\left\{  \frac{\cO^1_{\text{LG}}\cO^2_{\text{LG}}\cO^3_{\text{LG}}}{J^{\text{LG}}_2 \cdots J^{\text{LG}}_6} \right\}~.
\end{align}
Again, up to a numerical factor, we find a complete agreement between these sets of correlators.

\subsubsection{Bundle deformations and $E$-parameters dependence}

The hybrid model offers another set of deformations off the (2,2) locus. Namely, we can take $\cE$ to be a deformation of $T_{\bZ}$.
As an example, we can take $\cE_\ep=\cO^{\oplus3}\oplus \cE'_\ep$, where $\cE'_\ep$ is a one-parameter family of rank 2 bundles with
transition functions given by
\begin{align}
G_{uv}=\begin{pmatrix}
-v^{-2} & 2v\ep \phi_v^1\\
0 & v^{2}
\end{pmatrix}~.
\end{align}
In particular $\cE_{\ep=1}=T_{\bZ}$. Of course, $J\in\Gamma(\cE^\ast_\ep)$ depends on the parameter $\ep$.
However, one can show that
\begin{enumerate}
\item the dimension of $\cH_{\text{B/2}}$ is independent of $\ep$;
\item the number of parameters of $J$, which we collectively denote $\psi$, is independent of $\ep$.
\end{enumerate}
A natural question is then the following: is $\ep$ actually a parameter of the B/2-twisted theory?
From the hybrid perspective the answer seems to be yes, as any correlation function $\la\alpha\ra_{S^{2}}$ 
will be a function of $\psi$ as well as of $\ep$.

This is somewhat puzzling when we interpret our result from the GLSM perspective. In fact, parameters defining the bundle $\cE\rightarrow\bZ$
arise as $E$-parameters in the GLSM construction\cite{Witten:1993yc}. These are the natural parameters that appear in A/2-model computations, and their
appearance in B/2-model correlators was ruled out in a class of theories, including our example, in \cite{McOrist:2008ji}.
It then appears that from the hybrid perspective, B/2-model correlators do depend on these.
The resolution of the apparent puzzle resides in taking into consideration the action of field redefinitions in the linear model.
In fact, at the level of the GLSM, it is always possible to perform a field redefinition absorbing the parameter $\ep$.
Hence, in the hybrid phase of the GLSM obtained after performing such a field redefinition
we have $\ep=1$, i.e., $\cE=T_{\bZ}$. In other words, the hybrid model defined by $(\cE_\ep,J_{\ep})$ is
equivalent to the model $(\cE_{\ep=1},\Jt)$, where $\Jt$ is not necessarily equal to $J_{\ep=1}$.
While this equivalence is evident at the level of the linear model, it would be interesting to
investigate it in the hybrid model directly.

\section{Discussion}
\label{s:outlook}

In this work we have started an analysis of the $S^{2}$ B-type (and B/2-type) correlators in (2,2) and (0,2) hybrid models.
Some features of these correlators
have been already studied for a different class of (2,2) hybrid models in \cite{Guffin:2008kt}.
In fact, the treatment there applies only when $dW^{-1}(0)=0$ is a complete intersection and,
moreover, their explicit formulae require the points $dW^{-1}(0)=0$ to be non-degenerate.
Both of these conditions are not satisfied generically in the class of hybrid theories studied in this work.
For instance, the octic example studied at length in this work does not satisfy this second criterion.
More relevant to us are the properties of the correlators studied in \cite{Garavuso:2013zoa}.
There, the authors propose Mathai-Quillen like forms for (0,2) and (2,2) correlators,
for both B- and A-twist.
To make connection with our notation, the Mathai-Quillen like form allows to write integrals
over $B$ of products of elements in $\mathbf{\overline{Q}}$-cohomology as integrals over $\bY$ (of appropriately defined lifts of the insertions).
From our perspective, this can be achieved by integrating along the fibers.
It would be interesting to give a more thorough connection between their results and ours.

We also find an interesting connection with the work \cite{2015arXiv150204872S}.
There, the author defines the Koszul-De Rham complex, which is an extension of the usual
De Rham complex by auxiliary commuting and anti-commuting variables.
This is a bi-complex with differentials $(d_{DR},\partial_{\mathcal{K}})$ that can be identified as 
$d_{DR}\rightarrow\mathbf{Q}_{-}+\mathbf{\overline{Q}}_{0}$ and $\partial_{\mathcal{K}}\rightarrow \overline{\mathbf{Q}}_{W}$
acting on $\rho_{\alpha},x^{\alpha},\bar{x}^{\bar{\alpha}}, \bar{\psi}_{-,\alpha}$ and $J_{\alpha}$.

Another connections with recent mathematical work can be found in \cite{2014arXiv1409.5996K}.
In this work, the authors study hybrid models in the context of
homological mirror symmetry of Fano manifolds and define different classes of Hodge numbers associated to a hybrid.
It would be interesting to elucidate the physical interpretation of these invariants.

A natural extension of this work would be to consider the cases when $\partial \Sigma\neq \emptyset$ and the theory admits boundary conditions corresponding to B-branes.
This situation is considered, in the (2,2) case, in \cite{Lazaroiu:2003zi,Herbst:2004ax,Babalic:2016mbw}.
It would be interesting to extend the constructions of this paper to correlators involving B-branes, or more in general defects,
in particular to the situation where the orbifold action is non-trivial.
We hope to return to these cases in a sequel.

On a more technical note, our results fall short in a number of ways. First, it would be
important to give a formal proof of the transformation law we proposed for non-trivial hybrid models.
Second, it would be illuminating to posses a residue formula where the integration is over a holomorphic cycle over $\bY \backslash B$.
Third, even without a residue formula, it would be important nonetheless to have a technique to evaluate the integrals that arise in our correlators.
In fact, while for simple examples we managed to elude this, in order to compute the full dependence in
more complicated situations (see for instance \eqref{eq:TTTcorr}) this seems to be a unavoidable.  \\\\

\textbf{Acknowledgments}. We would like to thank P.~Aspinwall, F.~ Benini, K.~Hori, M.~Kapranov, S.~Lee,
I.~Melnikov, R.~Plesser, D.~Pomerleano, E.~Sharpe and T.~Spencer for enlightening discussions.
We would also like to thank I.~Melnikov, R.~Plesser and E.~Sharpe for helpful comments on the manuscript.
MB and MR would like to thank the Perimeter Institute for Theoretical Physics and the respective institutions for hospitality while part of this work has been performed.
We thank the physics department at Duke University
for allowing MB to present a preliminary version of this work. We also thank KIAS, Steklov Mathematical Institute,
the mathematics department of Higher School of Economics and YMSC at Tsinghua University for their hospitality during the final stages of this work.
MB is supported by NSF Grant PHY-1521053.
MR gratefully acknowledges the support of the Institute for Advanced Study, DOE grant DE-SC0009988 and the Adler Family Fund.
This work was supported by World Premier International Research Center Initiative (WPI Initiative), MEXT, Japan.

\appendix

\section{Conventions}
\label{app:convents}

In this appendix we collect our conventions for superspace. Our choice of Euclidean signature is obtained from the Minkowski metric
\begin{align}
d^2s=-(dx^0)^2+(dx^1)^2
\end{align}
by performing the Wick rotation
\begin{align}
x^0&=-iy^2~,	&x^1&=y^1~.
\end{align}
With this choice we have
\begin{align}
x^+&=x^0+x^1=y^1-iy^2=\zb~,		&x^-&=x^0-x^1=-y^1-iy^2=-z~,
\end{align}
and
\begin{align}
\p_+&\equiv{\p\over\p x^+}=\p_{\zb}~,		&\p_-&\equiv{\p\over\p x^-}=-\p_{z}~.
\end{align}
We follow the convention of \cite{Hori:2013ika} for Dirac spinors
\begin{align}
\label{eq:spinorconv}
\begin{pmatrix}
\psi_-\sqrt{dz} \\
\psi_+\sqrt{d\bar{z}}
\end{pmatrix}~,
\end{align}
with the following product structure
\begin{align}
\label{eq:productspin}
\langle \psi_{1},\psi_{2}\rangle=\psi_{1\alpha}\varepsilon^{\alpha\beta}\psi_{2\beta}=-\psi_{1-}\psi_{2+}+\psi_{1+}\psi_{2-}~,
\end{align}
where we defined $\varepsilon^{-+}=-1$. In our (0,2) application we will work with left- and right-moving Weyl spinors, which are obtained from
\eqref{eq:spinorconv} by setting the bottom or the top component to zero, respectively.
Conjugation acts on bosonic and fermionic fields alike, by exchanging barred and unbarred components, that is
\begin{align}
\overline{\langle \psi^1,\psi^2\rangle}=-\langle \psib^1,\psib^2\rangle~.
\end{align}
The product \eqref{eq:productspin} also satisfies
\begin{eqnarray}
\langle \psi^1,\gamma^{\mu}\psi^2\rangle=-\langle \gamma^{\mu}\psi^1,\psi^2\rangle~,
\end{eqnarray}
as well as the Fierz identities
\begin{align}
\langle \bar{\epsilon},\epsilon\rangle\lambda+\langle \bar{\epsilon},\lambda\rangle\epsilon+\langle\epsilon,\lambda\rangle\bar{\epsilon}&=0~,\nonumber\\
\langle \bar{\epsilon},\gamma^{m}\epsilon\rangle \gamma_{m}\lambda+\langle \epsilon,\bar{\epsilon}\rangle\lambda+2\langle\bar{\epsilon},\lambda\rangle\epsilon&=0~.
\end{align}
The Dirac operator in flat space in Euclidean signature reads
\begin{align}
\gamma^{\mu}D_{\mu}=\gamma^{2}\partial_{2}+\gamma^{1}\partial_{1}=\begin{pmatrix}
0& 2\partial_{z} \\
 2\overline{\partial}_{\bar{z}} & 0
\end{pmatrix}~,
\end{align}
where we have used the following choice for the $\gamma$ matrices
\begin{align}
\gamma^{1}&=\sigma_{1}~,		&\gamma^{2}&=\sigma_{2}~.
\end{align}

\section{A note on cohomology of polyvector fields}
\label{app:fibcohom}

In Proposition 3.7 of \cite{Babalic:2016mbw} is proved that the inclusion $\imath:PV_{c}(\bY)\hookrightarrow PV(\bY)$ 
induces a quasi-isomorphism on the cohomologies $(PV_{c}(\bY),\delta)$ and $(PV(\bY),\delta)$, where $PV_{c}(\bY)$ 
are compactly supported polyvector fields on $\bY$ and $PV(\bY)$ are $C^{\infty}$ polyvector fields.
The proof of the quasi-isomorphism involves two operators
\begin{align}
\pi&:=\rho\cdot \mathrm{id}+(\overline{\partial}\rho)\widehat{R}~, 	&\mathbf{\mathcal{R}}&:=(1_{\bY}-\rho)\widehat{R}~,
\end{align}
where $1_{\bY}$ is the unit function on $\bY$, $\rho$ is a function with compact support on a open set of $\bY$
containing $B=dW^{-1}(0)$ and $\widehat{R}$ is constructed out of the section
$s=\frac{i}{\|dW\|^{2}}(\overline{\partial}_{\bar{\alpha}}\overline{W})g^{\bar{\alpha}\beta}\partial_{\beta}$ and its $\delta$-derivatives.
These operators maps the spaces as
 \begin{align}
\pi&:PV(\bY)\rightarrow PV_{c}(\bY)~, 	&\mathbf{\mathcal{R}}&:PV(\bY)\rightarrow PV(\bY)~.
\end{align}
The quasi-isomorphism is proved by showing that $\imath \circ \pi$ and $\pi\circ \imath$ are homotopically equivalent to the identity (acting on the respective spaces).
The proof requires that $\mathcal{R}$ preserves the subspace $PV_{c}(\bY)$ inside $PV(\bY)$.
Since the operators $\pi$ and $\mathbf{\mathcal{R}}$ have at most polynomial growth because $dW$ is a regular section of $T^{*}_{\bY}$,
the proof goes through by replacing $PV_{c}(\bY)$ with $PV_{\mathrm{pol}}(\bY)$,
the space of polyvector fields on $\bY$ with at most polynomial growth along the fiber $X$,
where $\bY=\mathrm{Tot}(X\rightarrow B)$ and $dW^{-1}(0)=B$.
Therefore the cohomology rings $(PV(\bY),\delta)$, $(PV_{c}(\bY),\delta)$ and $(PV_{\mathrm{pol}}(\bY),\delta)$ are isomorphic.

\section{Sections of various bundles}
\label{app:sectbundles}

In this appendix we provide explicit expressions for various geometrical quantities we need in solving the examples in the body of the work.
The geometric set-up is given by $\bY=\cO^{\oplus3}\oplus\bY'$, where $\bY'=\tot\left(\cO(-2)\rightarrow\CP^1\right)$. We denote by $u=v^{-1}$
the local coordinates on the standard cover $U_1=U|_u$ and $U_2=U|_v$ of $B=\CP^1$.

\subsubsection*{The tangent bundle}

The tangent bundle splits as
$T_{\bY}=\cO^{\oplus3}\oplus T_{\bY'}$, therefore we can restrict our attention to the non-trivial summand $\bY'$.
A generic section of $T_{\bY'}$ at grade $d$, which here is simply determined by the overall power of $\phi^1$, satisfies
\begin{align}
\begin{pmatrix} (\phi_u^1)^d T_u &  (\phi^1_u)^{d+1}Y_u
\end{pmatrix} &=
 \begin{pmatrix} (\phi_v^1)^d T_v &  (\phi^1_v)^{d+1}Y_v
\end{pmatrix}
\begin{pmatrix}
-v^{2} & 2v \phi_v\\
0 & v^{-2}
\end{pmatrix}~,
\end{align}
which implies
\begin{align}
T_u &= -\Sigma_{2d+2}\big|_u~,		&Y_u&=\Sigma_{2d}\big|_u-2u^{-1}(T_u(u)-T_u(0))~,\nonumber\\
T_v &= \Sigma_{2d+2}\big|_v~,			&Y_v&=\Sigma_{2d}\big|_v+2v^{2d+1}T_u(0)~.
\end{align}
Here $\Sigma_{m}\in H^0(\CP^1,\cO(m))$. In particular, in the patch $U_1$, these have the form
\begin{align}
T_u&=a_0+a_1u+a_2u^2+\cdots+a_{2d+2}u^{2d+2}~,\nonumber\\
Y_u&=(b_0-2a_1) + (b_1-2a_2)u +\cdots+(b_{2d}-2a_{2d+1})u^{2d}-2a_{2d+2} u^{2d+1}~.
\end{align}

\subsubsection*{The cotangent bundle}

The cotangent bundle has a similar splitting $T^\ast_{\bY}=\cO^{\oplus3}\oplus T^\ast_{\bY'}$
and again the only non-trivial component is given by sections of $T^\ast_{\bY'}$. At grade $d$ we have
\begin{align}
\begin{pmatrix} S_u (\phi^1_u)^{d+1}\\ Z_u (\phi^1_u)^d \end{pmatrix} &=
\begin{pmatrix}
-v^{2} & 2v \phi_v\\
0 & v^{-2}
\end{pmatrix}
\begin{pmatrix} S_v (\phi^1_v)^{d+1}\\ Z_v (\phi^1_v)^d \end{pmatrix}~,
\end{align}
which means that $Z_{u,v}$ are restriction of a section of $\cO(2d+2)$ and we can take
\begin{align}
S_u&=- \Sigma_{2d}\big|_{u}+2u^{-1}(Z_u-Z_u(0))~,		&S_v&=\Sigma_{2d}\big|_{v}+2v^{2d+1}Z_u(0)~,
\end{align}
where as before $\Sigma_{2d}\in H^0(\CP^1,\cO_{\CP}(2d))$.
Explicitly, in the $U_1$ patch we have
\begin{align}
Z_u &= a_0 + a_1 u +\cdots+ a_{2d+2}u^{2d+2}~,\nonumber\\
S_u&=(2a_1-b_0)+(2a_2-b_1)u + (2a_3-b_2)u^2 + \cdots + (2a_{2d+1}-b_{2d})u^{2d}+2a_{2d+2}u^{2d+1}~.
\end{align}

\subsubsection*{The tensor product $T_{\bY}\otimes T_{\bY}^\ast$}

For our hybrid geometry, the tensor product splits as $T_{\bY}\otimes T^\ast_{\bY}=\cO^{\oplus9}\oplus (T_{\bY'})^{\oplus3}\oplus (T^\ast_{\bY'})^{\oplus3}\oplus (T_{\bY'}\otimes T^\ast_{\bY'})$.
The only novelty here is the section of $T_{\bY'}\otimes T^\ast_{\bY'}$, which, at degree $d$, satisfies
\begin{align}
\begin{pmatrix}
-v^{2} & 2v \phi_v\\
0 & v^{-2}
\end{pmatrix}
\begin{pmatrix}
A_v \phi_v^{d+1} & B_v \phi_v^{d+2} \\ C_v \phi_v^d& D_v \phi_v^{d+1}
\end{pmatrix}
\begin{pmatrix}
-v^{-2} & 2v \phi_v\\
0 & v^{2}
\end{pmatrix} =
\begin{pmatrix}
A_u \phi_u^{d+1}& B_u \phi_u^{d+2} \\ C_u \phi_u^d & D_u \phi_u^{d+1}
\end{pmatrix}~.
\end{align}
With a bit of algebra one shows that the most general solution is given by
\begin{align}
C_u &= \Sigma_{2d+4}\big|_{u}~,		&C_v &=- \Sigma_{2d+4}\big|_{v}~,\nonumber\\
D_u&=\Sigma_{2d+2}\big|_u-2u^{-1}(C_u-C_u(0))~,	&D_v&=\Sigma_{2d+2}\big|_v+2v^{2d+3}C_u(0)~,\nonumber\\
A_u&=\Sigmat_{2d+2}\big|_u+2u^{-1}(C_u-C_u(0))~,	&A_v&=\Sigmat_{2d+2}\big|_v-2v^{2d+3}C_u(0)~,
\end{align}
where $\Sigma_{2d+2},\Sigmat_{2d+2}$ are two distinct elements of $H^0(\CP^1,\cO_{\CP^1}(2d+2))$, as well as
\begin{align}
B_u&=\Sigma_{2d}\big|_{u} - 2u^{-1}(A_u-A_u(0)) + 2u^{-1}(D_u-D_u(0))+4u^{-2}C_u''~, \nonumber\\
B_v&= -\Sigma_{2d}\big|_{v} -2v^{2d+1}A_u(0)+2v^{2d+1}D_v(0) + 4v^{2d+2} C_u(0)+4v^{2d+1}C_u'(0) ~,
\end{align}
where, if $C_u=c_0+c_1u+c_2u^2+\cdots$, then
\begin{align}
C_u'&= u^{-1}\left(C_u-C_u(0)\right) = c_1+c_2u^2+\cdots~,\nonumber\\
C_u''&= \left(C_u-C_u(0)-C_u'(0)\right) = c_2u^2+\cdots~.
\end{align}
Finally, in the ${U}_1$ patch these quantities assume the following form
\begin{align}
C_u&=c_0+c_1u+\cdots+c_{2d+4}u^{2d+4}~,\nonumber\\
D_u&=(d_0-2c_1)+(d_1-2c_2)u + \cdots + (d_{2d+2}-2c_{2d+3})u^{2d+2}-2c_{2d+4}u^{2d+3}~,\nonumber\\
A_u&=(a_0+2c_1)+(a_1+2c_2)u + \cdots + (a_{2d+2}+2c_{2d+3})u^{2d+2}+2c_{2d+4}u^{2d+3}~,\nonumber\\
B_u&=(b_0-2a_1+2d_1-4c_2)+(b_1-2a_2+2d_2-4c_3)u + \cdots + (b_{2d}-2a_{2d+1}+2d_{2d+1}-4c_{2d+2})u^{2d}\nonumber\\
&\quad+(-2a_{2d+2}+2d_{2d+2}-4c_{2d+3})u^{2d+1}-4_{2d+4}u^{2d+2}~.
\end{align}

\section{Transformation law for the octic hybrid}
\label{app:transflawoct}

In this appendix we collect the details concerning the section $\cB\in\Gamma(T_{\bY}\otimes T_{\bY}^\ast)$
which we implement in the transformation law to solve our examples. We work in the patch $U_1$ and we recall that
\begin{align}
\cB J = T = \begin{pmatrix} u^{25} \phi_1^{13} & (u^{26}+1)\phi_1^{12} &  \phi_2^{13} & \phi_3^{13}& \phi_4^{13}  \end{pmatrix}^\top~.
\end{align}
For ease of notation, in this appendix we write the fiber coordinates with lower indices.

\subsection{(2,2) model}

In this case we have
\begin{align}
u^{25} \phi_1^{13}&=\frac1{1-\psi^8}\Big[ \left[ -\psi^8 u^{18}\phi_1^9-\psi^9 u^{11}\phi_1^6\phi_2\phi_3\phi_4 -2\psi^{10}u^4\phi_1^3\phi_2^2\phi_3^2\phi_4^2 \right. \nonumber\\
&\quad \left.+3\psi u^{11}\phi_1^6\phi_2\phi_3\phi_4-u^{10}\phi_1^9
+\psi^6\phi_1^3\phi_2^2\phi_3^2\phi_4^2 +\psi^2 u^4\phi_1^3\phi_2^2\phi_3^2\phi_4^2 \right] J_0 \nonumber\\
&\quad+ \left[2u^{17}\phi_1^{10}+2\psi^{10}u^3\phi_1^4\phi_2^2\phi_3^2\phi_4^2-2\psi u^{10}\phi_1^7 \phi_2\phi_3\phi_4\right] J_1\nonumber\\
&\quad +\left[ \psi^7\phi_1^4\phi_3^3\phi_4^3+\psi^3 u^4\phi_1^4\phi_3^3\phi_4^3\right]J_2+\left[ \psi^8 u\phi_1^5\phi_3\phi_4^4+\psi^4 u^5\phi_1^5\phi_3\phi_4^4\right] J_3 \nonumber\\
&\quad+\left[\psi^9 u^2\phi_1^6\phi_2\phi_3\phi_4^2+\psi^5 u^6\phi_1^6\phi_2\phi_3\phi_4^2\right]J_4\Big]~, \nonumber\\
(u^{26}+1)\phi_1^{12}&= \frac1{1-\psi^8}\Big[ \left[- u\phi_1^8 + u^{19} \phi_1^8 -\psi u^2\phi_1^5 \phi_2\phi_3\phi_4  +\psi u^{12} \phi_1^5  \phi_2\phi_3\phi_4 -\psi^2 u^3 \phi_1^2\phi_2^2\phi_3^2\phi_4^2\right. \nonumber\\
&\quad +\psi^2u^5\phi_1^2 \phi_2^2\phi_3^2\phi_4^2
+\psi^6u\phi_1^2 \phi_2^2\phi_3^2\phi_4^2-\psi^6u^7\phi_1^2  \phi_2^2\phi_3^2\phi_4^2+2\psi^{10}u^3\phi_1^2 \phi_2^2\phi_3^2\phi_4^2\nonumber\\
&\quad\left.-2\psi^{10}u^5\phi_1^2 \phi_2^2\phi_3^2\phi_4^2
+\psi^9u^2\phi_1^5  \phi_2\phi_3\phi_4-\psi^9u^{12}\phi_1^5 \phi_2\phi_3\phi_4+\psi^8 u\phi_1^8-\psi^8 u^{19}\phi_1^8 \right] J_0\nonumber\\
&\quad +2\left[ \psi^{10} u^4\phi_1^3\phi_2^2\phi_3^2\phi_4^2-\psi^{10} u^2\phi_1^3\phi_2^2\phi_3^2\phi_4^2+\psi^6 u^6\phi_1^3\phi_2^2\phi_3^2\phi_4^2-\psi^9 u\phi_1^6\phi_2\phi_3\phi_4\right. \nonumber\\
&\quad \left.-\psi^8\phi_1^9 +\psi^2 u^2\phi_1^3\phi_2^2\phi_3^2\phi_4^2+\psi u\phi_1^6\phi_2\phi_3\phi_4+\phi_1^9 \right] J_1 \nonumber\\
&\quad +\left[ \psi^3 u^5\phi_1^3\phi_3^3\phi_4^3+\psi^3 u^3\phi_1^3\phi_3^3\phi_4^3+\psi^7u^7\phi_1^3\phi_3^3\phi_4^3+\psi^7u\phi_1^3\phi_3^3\phi_4^3\right] J_2 \nonumber\\
&\quad+\left[\psi^4u^6\phi_1^4\phi_3\phi_4^4+\psi^4u^4\phi_1^4\phi_3\phi_4^4+\psi^8u^8\phi_1^4\phi_3\phi_4^4+\psi^8u^2\phi_1^4\phi_3\phi_4^4\right]J_3\nonumber\\
&\quad+\left[ \psi^5u^7\phi_1^5\phi_2\phi_3\phi_4^2+\psi^5u^5\phi_1^5\phi_2\phi_3\phi_4^2+\psi^9u^9\phi_1^5\phi_2\phi_3\phi_4^2+\psi^9u^3\phi_1^5\phi_2\phi_3\phi_4^2\right]J_3 \Big]~.
\end{align}
The remaining entries are
\begin{align}
 \phi_2^{13}&=\frac1{1-\psi^8}\Big[\psi^7\phi_1^3\phi_3^3\phi_4^3 \begin{pmatrix}  -u^8 & 2u^7\phi_1  \end{pmatrix} \begin{pmatrix} J_0 \\ J_1  \end{pmatrix}
+\psi^8\phi_2\phi_3^4\phi_4^4 \begin{pmatrix}  u & 0  \end{pmatrix} \begin{pmatrix} J_0 \\ J_1  \end{pmatrix}  \nonumber\\
 &\quad+\left[ \left(1- \psi^8 \right) \phi_2^{10} + \left(\psi- \psi^9 \right) u\phi_2^7\phi_1 \phi_3\phi_4 + \left(\psi^2- \psi^{10} \right) u^2\phi_1^2\phi_2^4\phi_3^2\phi_4^2 \right. \nonumber\\
&\quad \left. +\psi^3 u^3\phi_1^3\phi_2\phi_3^3\phi_4^3 + \psi^6u^6\phi_1^6 \phi_3^2\phi_4^2 \right] J_2 + \left(\psi^4 u^4\phi_1^4 \phi_2\phi_3\phi_4^4 +  \psi^9 u \phi_1 \phi_2^2\phi_3^2\phi_4^5\right) J_3 \nonumber\\
&\quad+ \left(\psi^5 u^5\phi_1^5 \phi_2^2\phi_3\phi_4^2 +  \psi^{10} u^2\phi_1^2 \phi_2^3\phi_3^2\phi_4^3 \right) J_4 \Big]~.
\end{align}
Similarly, by permuting the indices $a=2,3,4$ we obtain the expressions for $\phi_3^{13}$ and $\phi_4^{13}$.

\subsection{(0,2) model}

In this case we have instead
\begin{align}
u^{25} \phi_1^{13}&=\frac1{1-\psi_1^2\psi_2^2\psi_3^2\psi_4^2}\Big[ [ -\psi_1^2\psi_2^2 \psi_3^2 \psi_4^2 u^{18} \phi_1^9-\psi_1^3\psi_2^2 \psi_3^2 \psi_4^2  u^{11}\phi_1^6 \phi_2 \phi_3 \phi_4 \nonumber\\
&\quad-2 \psi_1^4\psi_2^2 \psi_3^2 \psi_4^2  u^4 \phi_1^3\phi_2^2 \phi_3^2 \phi_4^2 +3 \psi_1 u^{11}\phi_1^6 \phi_2 \phi_3 \phi_4 -u^{10} \phi_1^9+ \psi_1^3\psi_2 \psi_3 \psi_4 \phi_1^3\phi_2^2 \phi_3^2 \phi_4^2 \nonumber\\
&\quad+\psi_1^2 u^4\phi_1^3 \phi_2^2 \phi_3^2 \phi_4^2 ]J_0+[ 2 u^{17} \phi_1^{10}+2\psi_1^4 \psi_2^2 \psi_3^2 \psi_4^2  u^3\phi_1^4 \phi_2^2 \phi_3^2 \phi_4^2 -2 \psi_1 u^{10} \phi_1^7\phi_2 \phi_3 \phi_4 ]J_1 \nonumber\\
&\quad +[\psi_1^4\psi_2 \psi_3 \psi_4 \phi_1^4 \phi_3^3 \phi_4^3 +\psi_1^3 u^4 \phi_1^4 \phi_3^3 \phi_4^3] J_2+ [\psi_1^4\psi_2^2 \psi_3 \psi_4  u\phi_1^5 \phi_3 \phi_4^4 +\psi_1^3\psi_2  u^5 \phi_1^5\phi_3 \phi_4^4 ]J_3\nonumber\\
&\quad+ [\psi_1^4 \psi_2^2 \psi_3^2 \psi_4  u^2 \phi_1^6 \phi_2 \phi_3 \phi_4^2+\psi_1^3\psi_2 \psi_3  u^6 \phi_1^6\phi_2 \phi_3 \phi_4^2 ]J_4 \Big] ~,\nonumber\\
 \frac{u^{26}+1}2 \phi_1^{12}&=\frac1{1-\psi_1^2\psi_2^2\psi_3^2\psi_4^2}\Big[[\half u^{19} \phi_1^8 -\half u \phi_1^8 +\half \psi_1 u^{12} \phi_1^5\phi_2 \phi_3 \phi_4 -\half \psi_1u^2 \phi_1^5 \phi_2 \phi_3 \phi_4 \nonumber\\
&\quad+\half \psi_1^2 u^5 \phi_1^2 \phi_2^2 \phi_3^2 \phi_4^2- \half \psi_1^2u^3\phi_1^2\phi_2^2 \phi_3^2 \phi_4^2 -\half  \psi_1^3\psi_2 \psi_3 \psi_4 u^7\phi_1^2 \phi_2^2 \phi_3^2 \phi_4^2 \nonumber\\
&\quad+\half \psi_1^3 \psi_2 \psi_3 \psi_4u \phi_1^2 \phi_2^2 \phi_3^2 \phi_4^2 - \psi_1^4\psi_2^2 \psi_3^2 \psi_4^2 u^5\phi_1^2 \phi_2^2 \phi_3^2 \phi_4^2 +\psi_1^4 \psi_2^2 \psi_3^2 \psi_4^2 u^3\phi_1^2\phi_2^2 \phi_3^2 \phi_4^2  \nonumber\\
&\quad-\half \psi_1^3\psi_2^2 \psi_3^2 \psi_4^2  u^{12} \phi_1^5 \phi_2 \phi_3 \phi_4+\half \psi_1^3 \psi_2^2 \psi_3^2 \psi_4^2  u^2 \phi_1^5\phi_2 \phi_3 \phi_4 -\half \psi_1^2\psi_2^2 \psi_3^2 \psi_4^2  u^{19} \phi_1^8\nonumber\\
&\quad+\half \psi_1^2 \psi_2^2 \psi_3^2 \psi_4^2 u \phi_1^8]J_0 +  [\psi_1^4\psi_2^2 \psi_3^2 \psi_4^2 u^4\phi_1^3 \phi_2^2 \phi_3^2 \phi_4^2 - \psi_1^4\psi_2^2 \psi_3^2 \psi_4^2 u^2\phi_1^3 \phi_2^2 \phi_3^2 \phi_4^2 \nonumber\\
&\quad+\psi_1^3\psi_2 \psi_3 \psi_4 u^6 \phi_1^3 \phi_2^2 \phi_3^2 \phi_4^2 -\psi_1^3 \psi_2^2 \psi_3^2 \psi_4^2  u\phi_1^6 \phi_2 \phi_3 \phi_4 -\psi_1^2 \psi_2^2 \psi_3^2 \psi_4^2 \phi_1^9\nonumber\\
&\quad+\psi_1^2 u^2\phi_1^3 \phi_2^2 \phi_3^2 \phi_4^2 +\psi_1 u \phi_1^6 \phi_2 \phi_3 \phi_4+\phi_1^9]J_1
[\half \psi_1^3 (u^5+u^3) \phi_1^3\phi_3^3 \phi_4^3 \nonumber\\
&\quad+\half\psi_1^4 \psi_2 \psi_3 \psi_4 (u^7+u) \phi_1^3 \phi_3^3 \phi_4^3 ]J_2 +[\half\psi_1^3 \psi_2 (u^6+u^4) \phi_1^4 \phi_3 \phi_4^4  \nonumber\\
&\quad +\half\psi_1^4 \psi_2^2 \psi_3 \psi_4 (u^8+u^2) \phi_1^4 \phi_3 \phi_4^4 ]J_3 +[\half \psi_1^3 \psi_2 \psi_3 (u^7+u^5) \phi_1^5 \phi_2 \phi_3 \phi_4^2\nonumber\\
&\quad+\half  \psi_1^4 \psi_2^2 \psi_3^2 \psi_4 (u^9+u^3) \phi_1^5\phi_2 \phi_3 \phi_4^2]J_4 \Big]~,
\end{align}
and
\begin{align}
\label{eq:02phi213B}
\phi_2^{13}&=\frac1{1-\psi_1^2\psi_2^2\psi_3^2\psi_4^2}\Big[ (-\psi_2^5 \psi_3 \psi_4 u^8\phi_1^3 \phi_3^3 \phi_4^3 +\psi_1\psi_2^5 \psi_3 \psi_4  u \phi_2 \phi_3^4 \phi_4^4) J_0+2 \psi_2^5 \psi_3 \psi_4 u^7\phi_1^4 \phi_3^3 \phi_4^3  J_1 \nonumber\\
&\quad + ((1-\psi_1^2\psi_2^2 \psi_3^2 \psi_4^2 ) \phi_2^{10}+\psi_2 (1-\psi_1^2\psi_2^2 \psi_3^2 \psi_4^2) u\phi_1 \phi_2^7 \phi_3 \phi_4 \nonumber\\
&\quad+\psi_2^2 (1-\psi_1^2\psi_2^2 \psi_3^2 \psi_4^2) u^2\phi_1^2 \phi_2^4 \phi_3^2 \phi_4^2 +\psi_2^3 u^3\phi_1^3 \phi_2 \phi_3^3 \phi_4^3 +\psi_2^4 \psi_3 \psi_4 u^6\phi_1^6 \phi_3^2 \phi_4^2) J_2\nonumber\\
&\quad+(\psi_2^4 u^4 \phi_1^4 \phi_2  \phi_3 \phi_4^4+\psi_1^2\psi_2^5 \psi_3 \psi_4 u\phi_1 \phi_2^2 \phi_3^2 \phi_4^5)J_3 +(\psi_1^2\psi_2^5 \psi_3^2 \psi_4 u^2 \phi_1^2 \phi_2^3 \phi_3^2 \phi_4^3+\psi_2^4 \psi_3 u^5\phi_1^5 \phi_2^2 \phi_3 \phi_4^2 ) J_4 \Big]~.
\end{align}
As before, the remaining relations can be obtained from \eqref{eq:02phi213B} by permutations of the indices $a=2,3,4$.

\section{A (0,2) LG model}

In this appendix we apply the hybrid techniques to solve a (0,2) LG model, for which a residue form is not attainable \cite{Melnikov:2009nh}.
Let us consider the model with $n=2$, $N=3$ and superpotential
\begin{align}
J_1&=\phi_1^2\phi_2~,	&J_2&=\phi_2(\phi_1^2-\psi \phi_2)~,	&J_3&=\phi_1(\phi_1^2-\psi\phi_2)~.
\end{align}
If $\psi\neq0$ the model is nonsingular, and it has the property that there does not exist a two-dimensional subset of the $J$'s such that the model is nonsingular.
This is precisely the condition \cite{Melnikov:2009nh} for the existence of a residue formula for the correlators.
Anomaly cancellation fixes the normalization of the charges to be
\begin{align}
q_1&=\frac29~,		&q_2&=\frac49~,	&Q_1&=-\frac89~,	&Q_2&=-\frac89~,	&Q_3&=-\frac69~.
\end{align}
These lead to $c=10/3$, $\cb=5/3$ and $r=16/9$.
The chiral ring is computed as the cohomology of the Koszul complex
\begin{align}
\xymatrix@R=0mm@C=10mm{
\bK:=\quad	0 \ar[r]	&\wedge^3 \cE	\ar[r]^-{J}	&\wedge^2 \cE	\ar[r]^-{J}		&\cE	\ar[r]^-{J}	&R	\ar[r]	&0~,
}
\end{align}
where we interpret $\cE=R^3$ as a module over the ring $R=\C[\phi_1,\phi_2]$. 
In particular, we find that the only non-zero elements are
\begin{align}
H^0(\bK) &= R/J\cong\C^5~,
&H^1(\bK)&=\{\phi_{1}^{2}\phi_{2}\psib^1_{-},\phi_{1}^{2}\phi_{2}\psib^2_{-},\phi_{1}^{4}\psib^2_{-},\phi_{1}^{3}\phi_{2}\psib^3_{-},\phi_{1}\phi_{2}^{2}\psib^3_{-} \}~.
\end{align}
In order to solve this model we apply the transformation law, which for this example is fairly straightforward.
Let $\cB: R^3\rightarrow R^3$ given by
\begin{align}
\cB\cdot J=\begin{pmatrix}
1		&0			&0\\
\psi		&0			&\phi_1\\
\psi^{-1}	&-\psi^{-1}		&0
\end{pmatrix}
\begin{pmatrix}
J_1 \\ J_2 \\ J_3
\end{pmatrix}=
\begin{pmatrix}
\phi_1^2\phi_2 \\ \phi_1^4 \\ \phi_2^2
\end{pmatrix} \equiv T~.
\end{align}
The quantity $T$ does not depend on the parameter $\psi$ and satisfies the condition $T^{-1}(0)=\{\phi=0\}$.
Next, we construct the matrix
\begin{align}
M^A_{\alphab}=\begin{pmatrix}
2\phib_1\phib_2+4\psi\phib_1^3	&0				&4\phib_1^3\phi_1\\
\phib_1^2+2\psi^{-1}\phib_2		&-2\psi^{-1}\phib_2	&0
\end{pmatrix}~.
\end{align}
The determinants relevant for the correlators are obtained from the appropriate minors, as prescribed in section \ref{ss:02transflaw}. We obtain
\begin{align}
M_1&=8\psi^{-1}\phib_1^3\phib_2\phi_1~,\nonumber\\
M_2&=4\phib_1^3\phi_1(\phib_1^2+2\psi^{-1}\phib_2)~,\nonumber\\
M_3&=-4\phib_1\phib_2(2\phib_1^2+\psi^{-1}\phib_2)~,
\end{align}
where $M_A$ denotes the determinant of the minor of $M$ given by ignoring the $i$-th column. A general correlator is then given by
\begin{align}
\la \alpha \ra_{S^2} = \int_{\C^2} d^2\phi_1d^2\phi_2 e^{-\frac{\bv}4 ||T||^2} M_A \alpha^A~,
\end{align}
where $\alpha^A \in H^1(\bK)$ has the general form $\alpha^A = \phi_1^k\phi_2^l\psib_-^A$.
It is then straightforward to compute, up to overall constants, the full list of correlators
\begin{align}
\la\phi_1^k \phi_2^l \psib_-^1\ra &= 2c_{1,3}\psi^{-1} \delta_{l,1}\delta_{k,2}~,\nonumber\\
\la\phi_1^k \phi_2^l \psib_-^2\ra &=c_{5,0}\delta_{l,0}\delta_{k,4}+ 2c_{3,1}\psi^{-1} \delta_{l,1}\delta_{k,2}~,\nonumber\\
\la\phi_1^k \phi_2^l \psib_-^3\ra &=-2c_{3,1}\delta_{l,1}\delta_{k,3}-c_{1,2} \psi^{-1} \delta_{l,2}\delta_{k,1}~.
\end{align}
The coefficients $c_{a,b}\in\C$ are given as integrals of the form
\begin{align}
c_{a,b}=4\int_{\mathbb{R}^2_{\geq 0}} d|\phi_1|d|\phi_2| e^{-\frac{\bv}4 ||T||^2} |\phi_1|^{2a} |\phi_2|^{2b} ~.
\end{align}

\bibliographystyle{fullsort}
\bibliography{bibliography}
\end{document}